\documentclass[10pt,journal,compsoc]{IEEEtran}

\ifCLASSOPTIONcompsoc
  \usepackage[nocompress]{cite}
\else
  \usepackage{cite}
\fi

\usepackage[pdftex]{graphicx}
\graphicspath{{../pdf/}{../jpeg/}}
\DeclareGraphicsExtensions{.pdf,.jpeg,.png, .jpg, .eps}

\usepackage[precent]{overpic}

\usepackage{amsmath}
\usepackage{amssymb}
\usepackage[utf8]{inputenc}		
\usepackage{cite}                      
\usepackage{tabu}
\usepackage[html]{xcolor}
\usepackage{tikz} 
\usepackage{subcaption}
\usepackage{tabularx}
\usepackage{pifont}
\usepackage{colortbl}
\usepackage{url}
\usepackage{breakurl}
\usepackage[breaklinks]{hyperref}

\usepackage{stfloats}
\hypersetup{
	colorlinks,
	linkcolor={blue!80!black},
	citecolor={blue!80!black},
	urlcolor={blue!80!black}
}

\definecolor{dgreen}{rgb}{0.0, 0.42, 0.24}
\definecolor{solar_red}{HTML}{AF0000}
\definecolor{solar_green}{HTML}{5F8700}
\newcommand{\cmark}{\textcolor{dgreen}{\ding{52}}}%
\newcommand{\xmark}{\textcolor{solar_red}{\ding{56}}}%
\newcommand{\pmark}{\textcolor{dgreen}{\textbf{+}}}%
\newcommand{\mmark}{\textcolor{solar_red}{\textbf{--}}}%
\newcommand{\semimark}[2]{{#1/#2}}%

\definecolor{MarkBlack}{HTML}{252525} 
\definecolor{MarkCyan}{HTML}{225EA8} 
\definecolor{MarkYellow}{HTML}{FFFF33} 
\definecolor{MarkPurple}{HTML}{54278F} 
\definecolor{MarkOrange}{HTML}{EC7014} 

\begin{document}
%
\title{A Comparison of Rendering Techniques \\ for 3D Line Sets with Transparency}

%
%


\author{Michael Kern,
		Christoph Neuhauser,
		Torben Maack,
		Mengjiao Han,
		Will Usher,
		R\"udiger Westermann
\IEEEcompsocitemizethanks{
	\IEEEcompsocthanksitem M. Kern, C. Neuhauser, T. Maack, and R. Westermann are with the Computer Graphics \& Visualization Group, Technische Universit{\"a}t M{\"u}nchen, Garching, Germany\protect\\
	E-mail: \{michi.kern, christoph.neuhauser, westermann\}@tum.de, torben.maack@mytum.de\protect\\
	\IEEEcompsocthanksitem M. Han and W. Usher are with the Scientific Computing and Imaging Institute, University of Utah, U.S.\protect\\
	E-mail: \{mengjiao, will\}@sci.utah.edu}
	\thanks{\textbf{Preprint to appear in IEEE Transactions on Visualization and Computer Graphics, 2020. Manuscript revised on Feb 20, 2020.}}
}

\IEEEtitleabstractindextext{%
\begin{abstract}
	This paper presents a comprehensive study of rendering techniques for 3D line sets with transparency.
	The rendering of transparent lines is widely used for visualizing trajectories of tracer particles in flow fields. 
	Transparency is then used to fade out lines deemed unimportant, based on, for instance, geometric properties or attributes defined along with them. 
	Accurate blending of transparent lines requires rendering the lines in back-to-front or front-to-back order, yet enforcing this order for space-filling 3D line sets with extremely high-depth complexity becomes challenging. 
	In this paper, we study CPU and GPU rendering techniques for transparent 3D line sets. 
	We compare accurate and approximate techniques using optimized implementations and several benchmark data sets.
	We discuss the effects of data size and transparency on quality, performance, and memory consumption. 
	Based on our study, we propose two improvements to per-pixel fragment lists and multi-layer alpha blending. 
	The first improves the rendering speed via an improved GPU sorting operation, and the second improves rendering quality via transparency-based bucketing.
\end{abstract}
	
	\begin{IEEEkeywords}
		Scientific visualization, line rendering, order-independent transparency.
	\end{IEEEkeywords}
}

\maketitle


\IEEEdisplaynontitleabstractindextext

%


\IEEEpeerreviewmaketitle


%
%
%
%

\label{Teaser Image}
\begin{figure*}[ht!]
	\centering
	\begin{subfigure}{0.245\linewidth}
		\setlength{\lineskip}{0pt}
		\begin{tikzpicture}
		\node[anchor=south west,inner sep=0] at (0,0) {\includegraphics[width=1\textwidth]{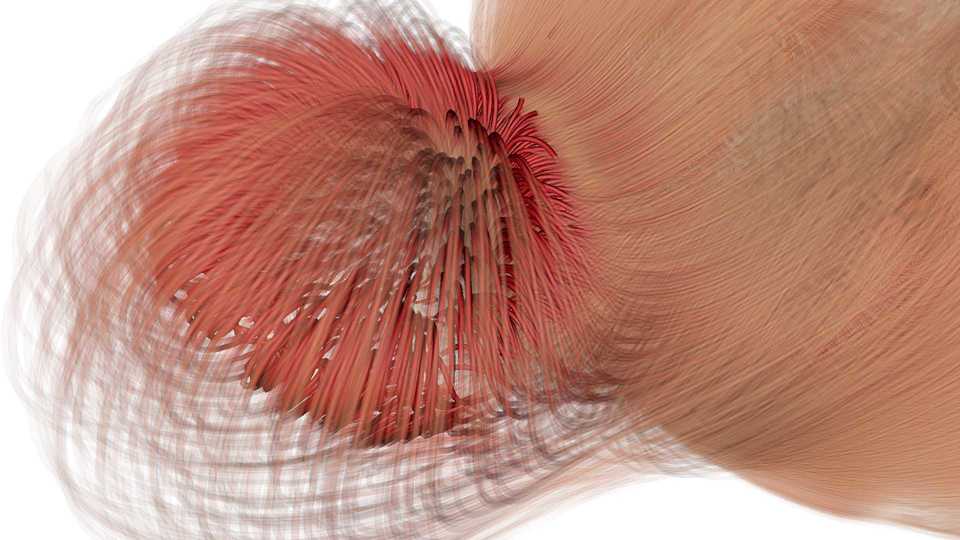}};
		\draw[MarkYellow,thick] (3.2,0.7) rectangle (4.4,1.45);
		\node[scale=1, anchor=west] at (0, 0.3) {\textbf{(a)}};
		\end{tikzpicture}
		\vspace*{-9pt}
	\end{subfigure}%
	\hspace{1mm}%
	\begin{subfigure}{0.245\linewidth}
		\setlength{\lineskip}{0pt}
		\begin{tikzpicture}
		\node[anchor=south west,inner sep=0] at (0,0) {\includegraphics[width=1\textwidth]{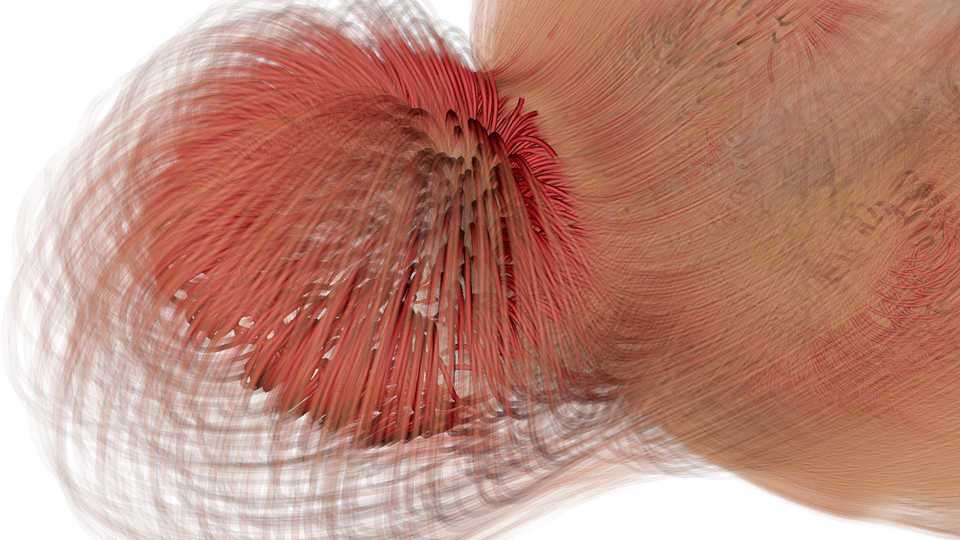}};
		\draw[MarkYellow,thick] (3.2,0.7) rectangle (4.4,1.45);
		\end{tikzpicture}
		\vspace*{-9pt}
	\end{subfigure}%
	\hspace{1mm}%
	\begin{subfigure}{0.245\linewidth}
		\setlength{\lineskip}{0pt}
		\begin{tikzpicture}
		\node[anchor=south west,inner sep=0] at (0,0) {\includegraphics[width=1\textwidth]{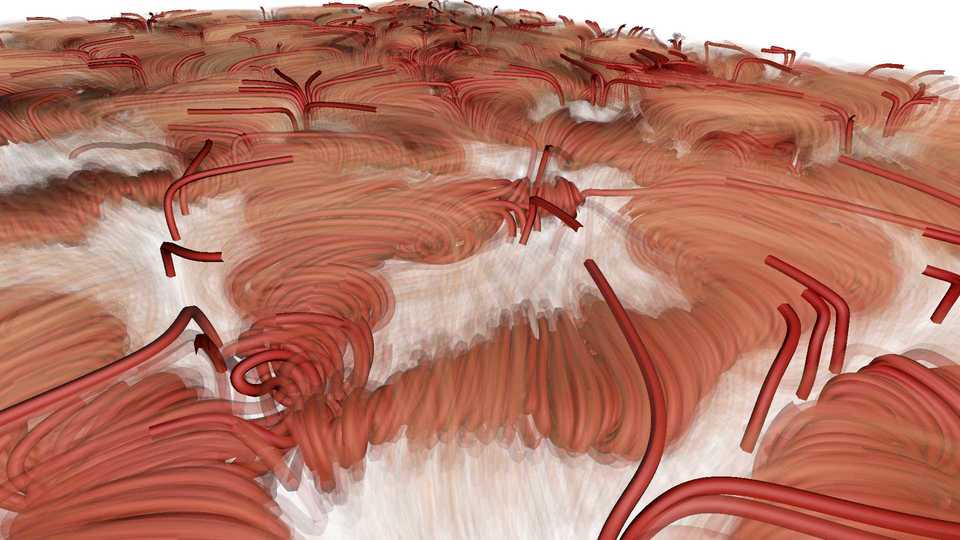}};
		\draw[MarkPurple,thick] (3.0,0.9) rectangle (4.2,1.65);
		\node[scale=1, anchor=west, text=white] at (0, 0.3) {\textbf{(b)}};
		\end{tikzpicture}
		\vspace*{-9pt}
	\end{subfigure}%
	\hspace{1mm}%
	\begin{subfigure}{0.245\linewidth}
		\setlength{\lineskip}{0pt}
		\begin{tikzpicture}
		\node[anchor=south west,inner sep=0] at (0,0) {\includegraphics[width=1\textwidth]{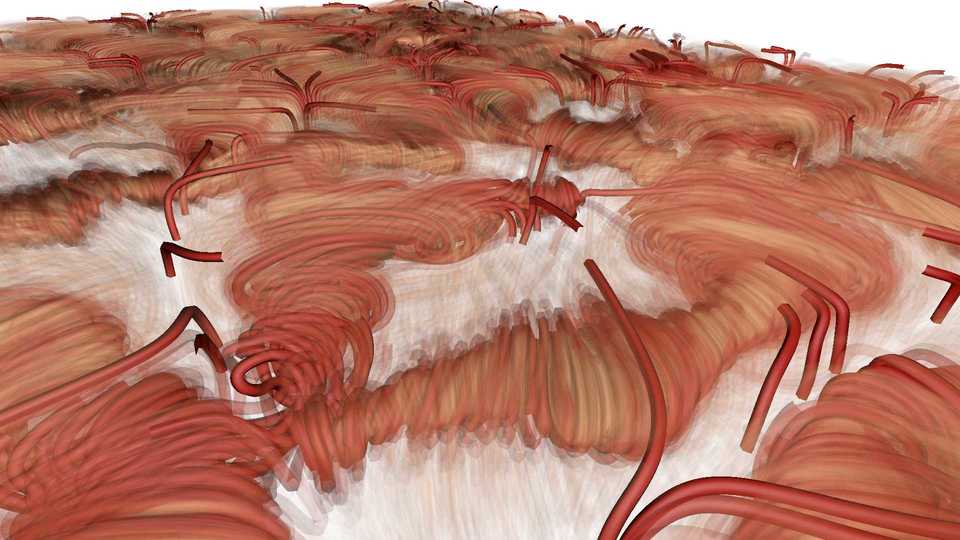}};
		\draw[MarkPurple,thick] (3.0,0.9) rectangle (4.2,1.65);
		\end{tikzpicture}
		\vspace*{-9pt}
	\end{subfigure}
	\begin{subfigure}{0.245\linewidth}
		\setlength{\lineskip}{0pt}
		\begin{tikzpicture}
		\node[anchor=south west,inner sep=0] at (0,0) {\includegraphics[width=1\textwidth]{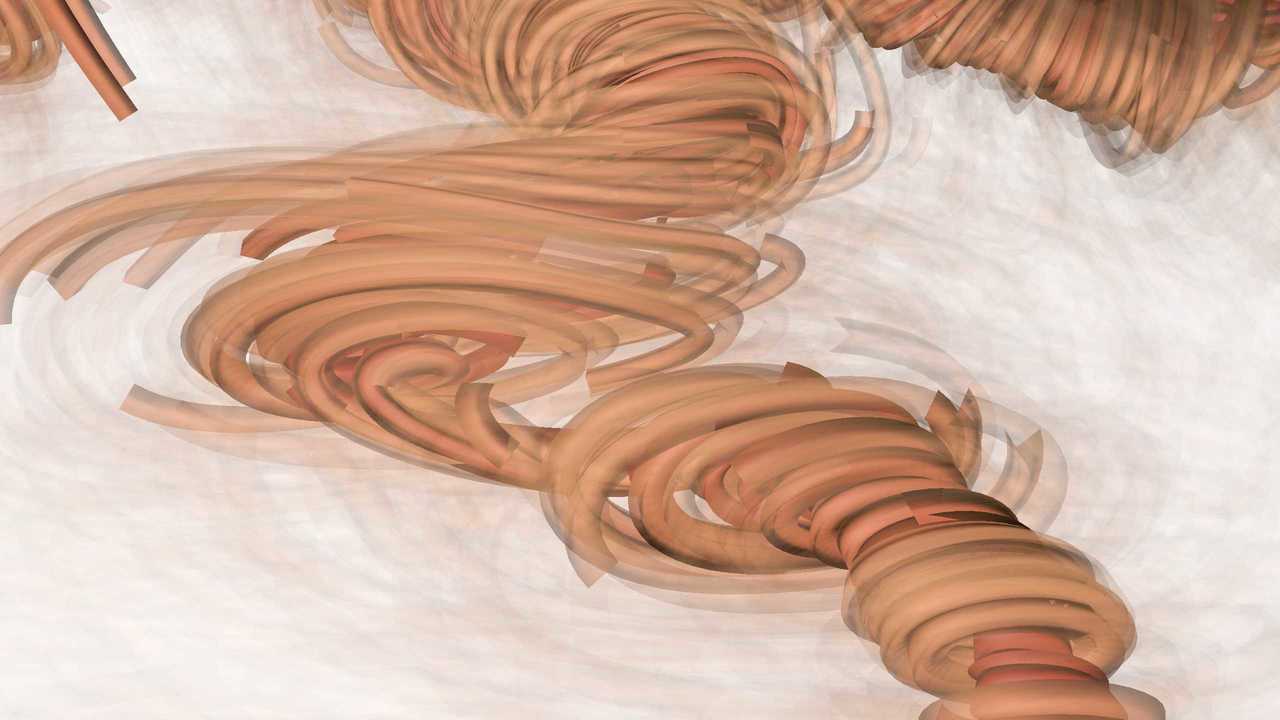}};
		\draw[MarkBlack,thick] (2.1,0.5) rectangle (3.3,1.25);
		\node[scale=1, anchor=west] at (0, 0.3) {\textbf{(c)}};
		\end{tikzpicture}
		\includegraphics[width=0.5\linewidth]{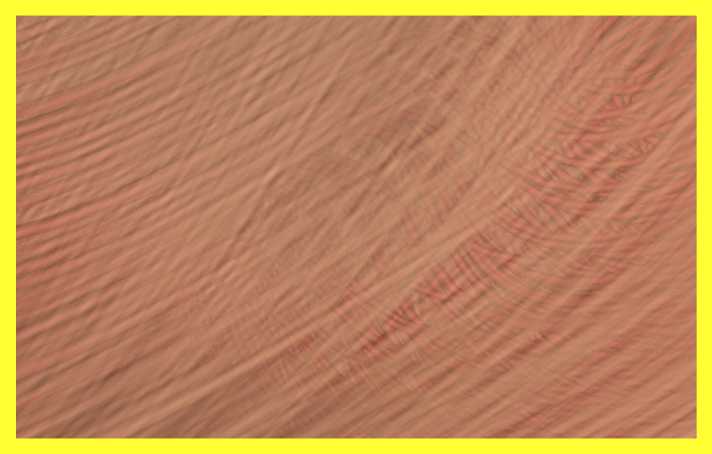}%
		\includegraphics[width=0.5\linewidth]{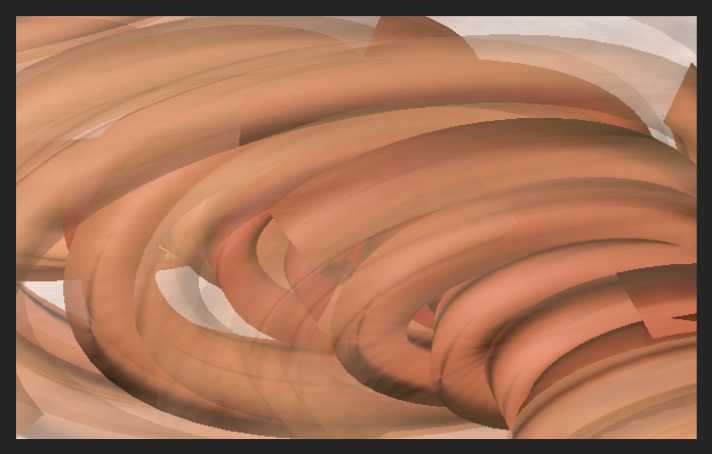}%
		\caption*{\textbf{(a)} PSNR = 33.41, SSIM = 0.907}
	\end{subfigure}%
	\hspace{1mm}%
	\begin{subfigure}{0.245\linewidth}
		\setlength{\lineskip}{0pt}
		\begin{tikzpicture}
		\node[anchor=south west,inner sep=0] at (0,0) {\includegraphics[width=1\textwidth]{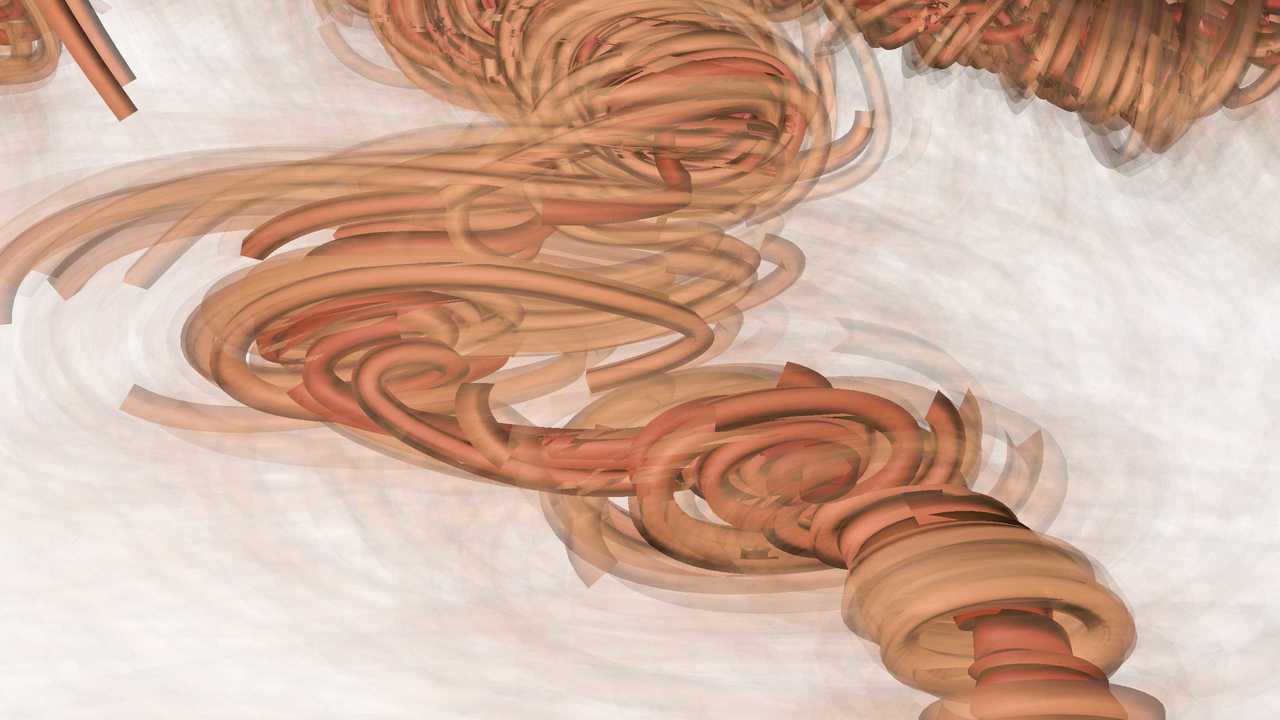}};
		\draw[MarkBlack,thick] (2.1,0.5) rectangle (3.3,1.25);
		\end{tikzpicture}
		\includegraphics[width=0.5\linewidth]{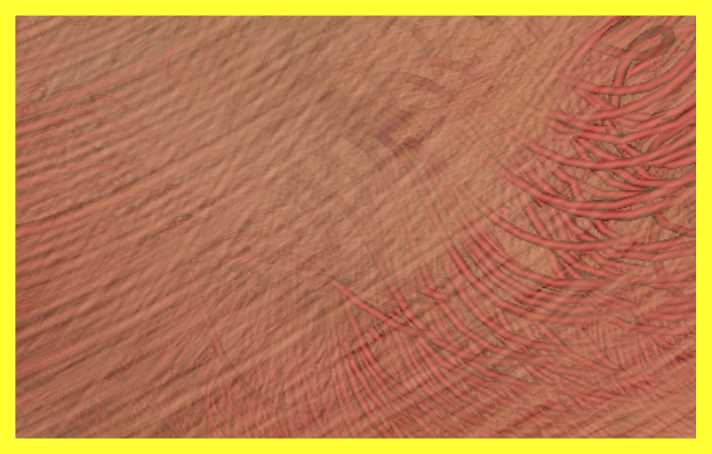}%
		\includegraphics[width=0.5\linewidth]{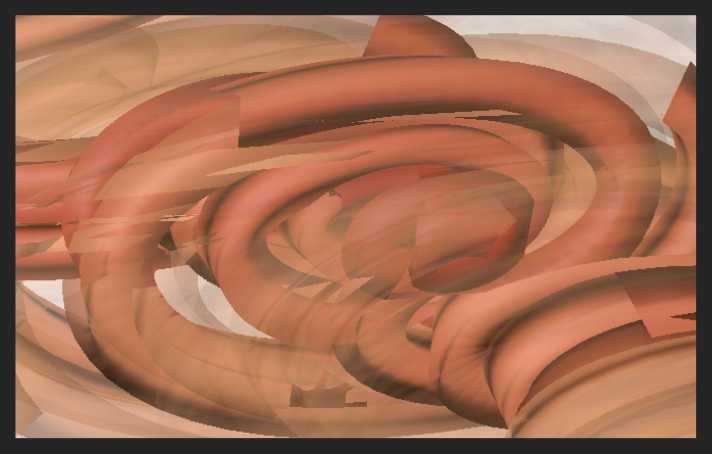}
		\caption*{\textbf{(b)} PSNR = 31.98, SSIM = 0.901}
	\end{subfigure}%
	\hspace{1mm}%
	\begin{subfigure}{0.245\linewidth}
		\setlength{\lineskip}{0pt}
		\begin{tikzpicture}
		\node[anchor=south west,inner sep=0] at (0,0) {\includegraphics[width=1\textwidth]{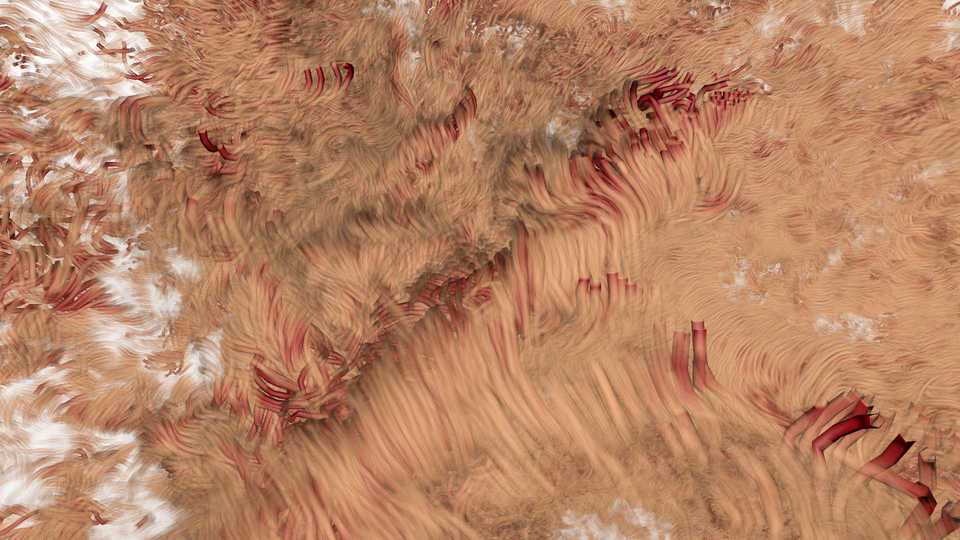}};
		\draw[MarkCyan,thick] (1.4,0.5) rectangle (2.6,1.25);
		\node[scale=1, anchor=west, text=black] at (0, 0.3) {\textbf{(d)}};
		\end{tikzpicture}
		\includegraphics[width=0.5\linewidth]{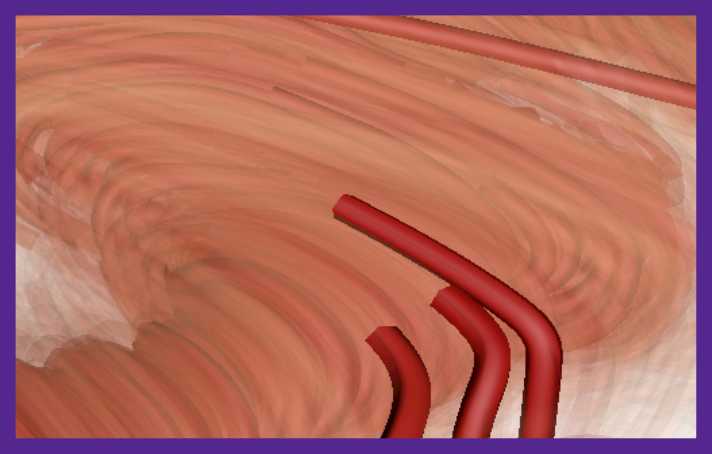}%
		\includegraphics[width=0.5\linewidth]{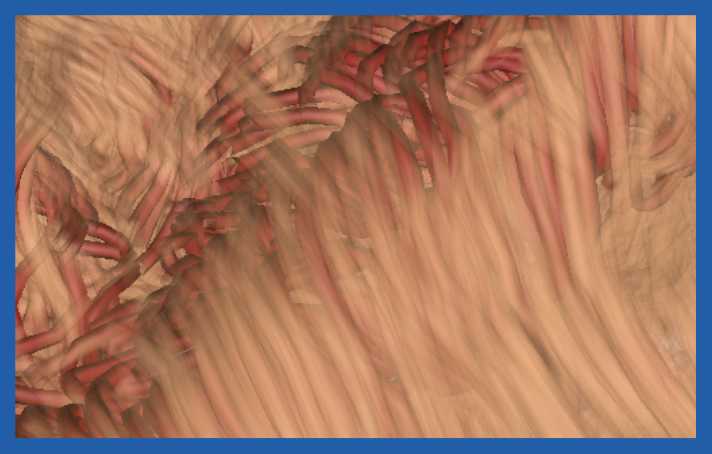}
		\caption*{\textbf{(c)} PSNR = 34.42, SSIM = 0.913}
	\end{subfigure}%
	\hspace{1mm}%
	\begin{subfigure}{0.245\linewidth}
		\setlength{\lineskip}{0pt}
		\begin{tikzpicture}
		\node[anchor=south west,inner sep=0] at (0,0) {\includegraphics[width=1\textwidth]{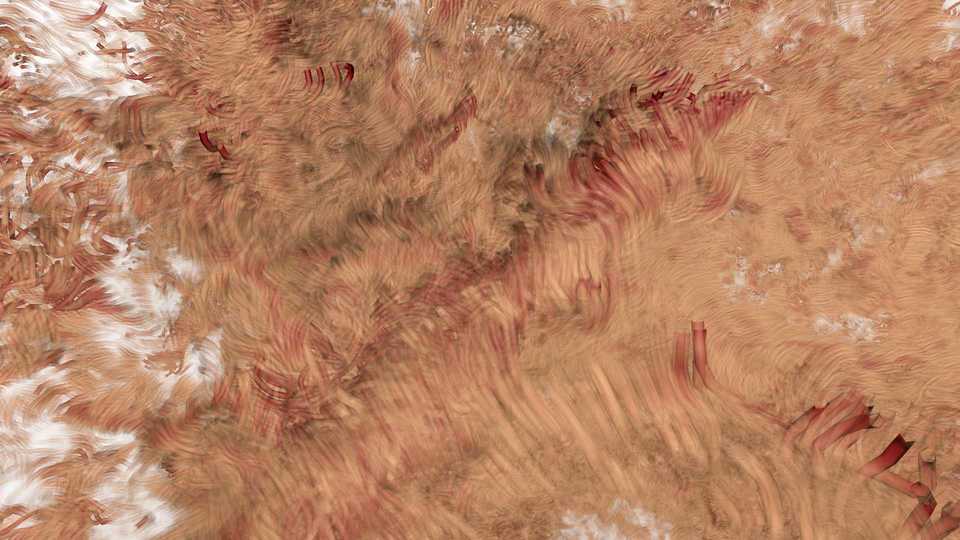}};
		\draw[MarkCyan,thick] (1.4,0.5) rectangle (2.6,1.25);
		\end{tikzpicture}
		\includegraphics[width=0.5\linewidth]{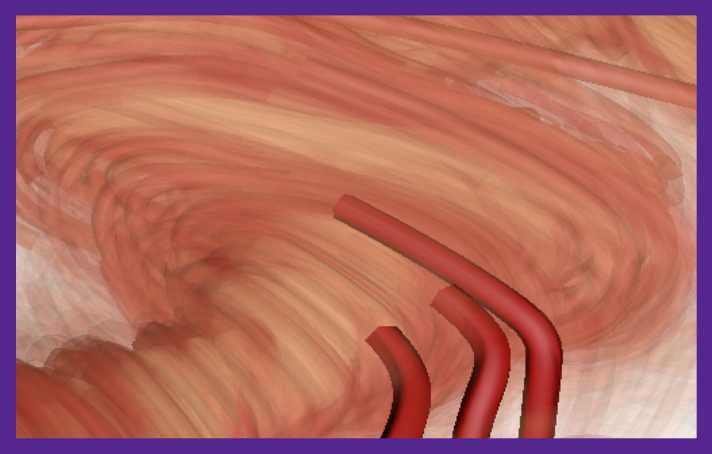}%
		\includegraphics[width=0.5\linewidth]{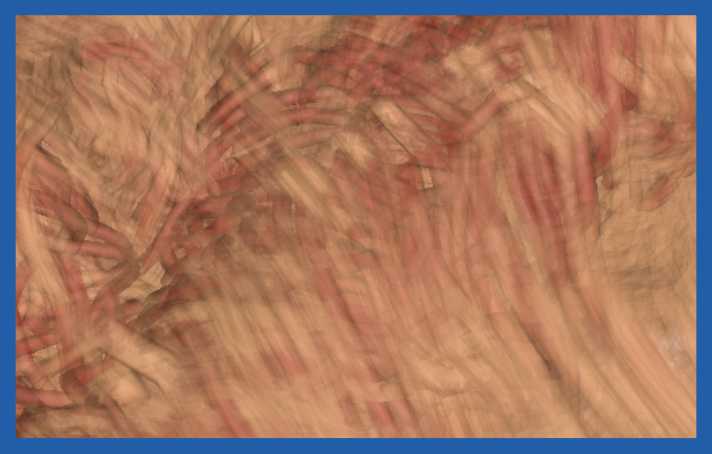}
		\caption*{\textbf{(d)} PSNR = 31.10, SSIM = 0.842}
	\end{subfigure}
	\caption{Strengths and weaknesses of transparent line rendering techniques. For each pair, the left image shows the ground truth (GT). Right images show \textbf{(a)} approximate blending using MLABDB, \textbf{(b)} opacity over-estimation of MBOIT, \textbf{(c)} reverse blending order of MLABDB, \textbf{(d)} blur effect of MBOIT. Speed-ups to GT rendering technique: (a) 7, (b) 2, (c) 3.5, (d) 4.5.}
	\label{fig:teaser}
 	\vspace{-1em}
\end{figure*}
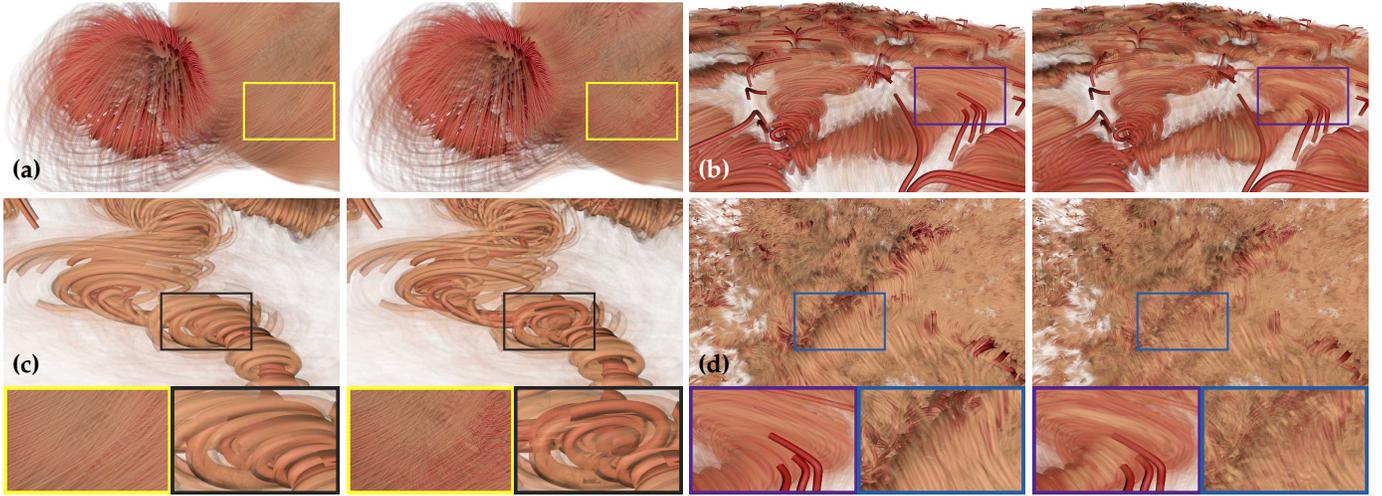
\IEEEraisesectionheading{\section{Introduction}\label{sec:introduction}}
In many visualization tasks, the need to efficiently display sets of 3D lines is paramount. 
Applications range from the visualization of pathways of particle tracers in flow  fields or over moving vehicles for smart transportation and urban planning, to exploring neural connections in the brain or relations encoded in large graphs and network structures. 
Prior work such as~\cite{Guenther2013, Lawonn2014, Oeltze2014, Rojo2019} has shown that transparency, when used carefully to avoid overblurring, can be used effectively to relieve occlusions and to accent important structures while maintaining less important context information. It is particularly useful for exploratory visualization tasks, where users interactively select the strength of transparency and the mapping of data values to transparency.

Rendering transparency, however, introduces a performance penalty. When using transparency, the per-pixel color and opacity contributions need to be blended in correct visibility order, i.e.,
by using $\alpha$-blending (where $\alpha$ represents a point's opacity) in either front-to-back or back-to-front order.
Rendering techniques can be distinguished as to whether they compute the visibility order exactly or approximately, and how this order is established. 
Especially for line sets, which have a significantly higher depth complexity than surface or point models, maintaining the visibility order during rendering can become a severe performance bottleneck. 

In this study, we evaluate exact and approximate object- and image-order transparency rendering techniques, with intending to analyze the performance of such techniques when used to render line sets with an extremely high depth complexity. Our evaluation includes an in-depth evaluation of model-specific acceleration schemes. We further demonstrate the use of approximate transparency rendering techniques for surface and point models with high depth complexity, though refrain from a detailed performance evaluation on these cases. The latter would require considering specific acceleration structures for such surface or point models, which is  beyond the scope of a single paper.

Object-order techniques make use of GPU rasterization. We consider Depth Peeling (DP)~\cite{Everitt2001} and Per-Pixel Linked Lists (LL)~\cite{Yang2010}, both of which can render transparency accurately at the cost of computing or memory. 
Other object-order techniques use (stochastic) transmittance approximations, where transmittance refers to the multiplicative accumulation of per-fragment transparencies. 
Of the many different variants of approximate techniques, we selected Multi-Layer Alpha Blending (MLAB)~\cite{Salvi2014} and the most recent Moment-Based Blending Technique (MBOIT)~\cite{Muenstermann2018} (see Fig.~\ref{fig:teaser} for example images). 
Both approximate techniques 
use only small and constant additional buffer resources. 

We also evaluate four image-order techniques based on ray tracing. We consider the Generalized Tubes method~\cite{han_ray_2019} as well as Embree’s built-in Bezier curve primitives~\cite{Wald2014} implemented in Intel's OSPRay CPU ray tracing framework (OSP)~\cite{OSP}, a GPU ray-tracer using NVIDIA's RTX ray tracing interface~\cite{NVIDIA2018} through the Vulkan API (RTX), and voxel-based GPU line ray tracing (VRC)~\cite{Kanzler2018}. 
All techniques utilize dedicated data structures to facilitate efficient ray traversal as well as empty space skipping and thus provide effective means to evaluate the capabilities of image-order line rendering.

OSP, RTX, VRC, DP, and LL, despite their algorithmic differences, are all accurate methods and yield the same rendering result.
Performance-wise, on the other hand, these techniques differ substantially, and for large data sets some of them even turn out to be impractical. The main goal of our evaluation study is to shed light on the differences between these techniques and to provide guidelines for selecting a suitable rendering technique for a given application.

\vspace{-0.5em}
\subsection*{Contribution}
We provide a qualitative and quantitative comparison of techniques for rendering 3D line sets with transparency. For our evaluation, we have systematically selected a set of techniques that we believe are representative for the different principal approaches that are available today.
Even though our evaluation study has been performed solely on 3D line sets, the results are also applicable to other application scenarios where transparency is used to reveal otherwise hidden structures.

Through our study, users and practitioners can gain an understanding of the principal implications of using a certain technique and become aware of their major strengths and limitations with respect to quality, memory requirements, and performance. Since we use a range of different sized data sets with vastly different internal structures, our evaluation hints to specific data-dependencies of certain rendering concepts. 
We tried to individually select a transparency setting for each data set that reveals important features in a meaningful way. Thus, we consider our results representative of typical use cases of transparent line rendering. For each technique, we also analyze the pre-process that is required to build the data representations needed for rendering and perform a thorough evaluation of rendering performance.

Moreover, we have modified LL to improve scalability with the number of fragments, and MLAB to make it less dependent on the order of fragments per pixel. For LL, we developed GPU-friendly variants of shell-sort and priority-queues through the min-heap data structure, resulting in a performance increase of a factor of 2-3. Our implementation of MLAB uses a discrete set of depth intervals and can considerably reduce the number of incorrectly merged fragments.


We have made our implementations publicly available~\cite{KernNeuhauser2020}, the test environment using NVIDIA's RTX~\cite{Maack2020}, all data sets~\cite{KernLineSets2020}, and all benchmark results for image quality and performance evaluation. We have also included additional descriptions of how to use the implementations and apply them to other data sets.

\vspace{-0.5em}
\section{Related Work}
Prior work~\cite{Maule2011,Wyman2016} has compared some of the many different rendering  techniques for transparent geometry. These evaluations, however, have mainly focused on the use of techniques for real-time graphics effects in scenes comprised of a few spatially extended and homogeneous transparent objects with rather low depth complexity. Thus, the suitability of techniques for visualization tasks as outlined in our work is difficult to infer from available evaluations. To the best of our knowledge, an evaluation and comparison of techniques for rendering large 3D line sets, including ray-based approaches and scenes with extremely high depth complexity and high-frequency transmittance functions, has not been performed.

\vspace{-0.5em}
\subsection{Object-order techniques}
Several approaches have been proposed to blend the fragments falling into the same pixel in correct visibility order without having to resort to an explicit sorting of geometry. 
Everitt et al.~\cite{Everitt2001} presented depth peeling, which renders for each pixel in the $i$-th rendering pass the $i$-th closest surface point using a second depth buffer test against the values from the previous pass.
In early work by Carpenter~\cite{Carpenter1984}, the A-buffer was introduced as a data structure that stores the unordered set of fragments falling into each pixel. These fragments are then sorted explicitly based on the stored depth information.
Yang et al.~\cite{Yang2010} used per-pixel linked lists to store a variable number of fragments per pixel on the GPU, after which they are sorted to blend the fragments in correct order.
Contrary to the linked lists, the $k$-Buffer~\cite{Bavoil2007} stores only the $k$ nearest fragments, and merges fragments heuristically if more than $k$ fall into the same pixel.

In scenarios where the $k$-Buffer is not applicable, fragments have to be blended heuristically.
Adaptive Transparency~\cite{Salvi2011} operates on $k$ fragments and aims to store an approximation of the transmittance function per pixel. Alpha blending is then performed in a second pass using this approximation.
Maule et al.~\cite{Maule2013} proposed Hybrid Transparency, which aggregates fragments using a $k$-Buffer and merges them heuristically with respect to depth and opacity. Even though this approach is order-independent, it is not able to cope with scenes containing many layers.
Multi-Layer Alpha Blending (MLAB)~\cite{Salvi2014} is a single-pass technique that uses a fixed number of per-pixel transmittance layers to approximate the transmittance along a view ray. When all layers are occupied and the current fragment creates a new layer, the two most appropriate adjacent layers are merged in turn.
Stochastic Transparency~\cite{Enderton2010} uses weights to blend or discard fragments based on opacity.
Weighted-Blended Order-Independent Transparency~\cite{McGuire2013} proposed to use weights based on occlusion and distance to the camera to merge fragments.

Recently, M\"unstermann et al.~\cite{Muenstermann2018} introduced Moment-Based Order-Independent Transparency (MBOIT). Rojo et al.~\cite{Rojo2019} demonstrated the embedding of importance-based transparency control into MBOIT.
MBOIT approximates the transmittance function pixel-wise by power moments or trigonometric moments, and applies logarithmic scaling to the absorbance to enforce order-independency and facilitate additive compositing.
Moment-Based transparency builds upon Fourier opacity mapping~\cite{Jansen2010}, which represents transmittance as a low-frequency distribution dependant on depth, and approximates these distributions using trigonometric moments, i.e., Fourier coefficients.

Another category of techniques render transparent layers using multiple samples per pixel, for example,
Stochastic Layered Alpha Blending~\cite{Wyman2016} and Phenomenological Transparency~\cite{McGuire2017}. 
The latter technique also incorporates physical processes to create realistic effects of translucent phenomena.
However, these techniques significantly increase the number of generated fragments, which is problematic in scenarios where the depth complexity is extremely high.
As such, we do not consider them in our study.

We do also not consider particle-based~\cite{Sakamoto2012} and voxel-based~\cite{Crassin2009, Laine2010} rendering techniques for transparent geometry. Especially when used to render space-filling line sets, these techniques significantly increase the number of rendered primitives or the resolution of the used voxel grid and require substantial modifications to render geometric shapes with fine geometric details and sharp outlines.

\vspace{-0.5em}
\subsection{Image-order techniques}
Image-order techniques for rendering line primitives make use of ray tracing.
Advances in hardware and software technology have shown the potential of ray tracing as an alternative to rasterization, especially for high-resolution models with many inherent occlusions.
Developments in this field include advanced space partitioning and traversal schemes~\cite{Wald:2006:RTAS,Woop:2006:BKD,Wald:2017:OCR:3024362.3025410}, and optimized GPU implementations
\cite{Aila2009hpg,Laine2010,Parker2010}, to name just a few. Wald et al.~\cite{conf/scivis/WaldKJUPP15} proposed the use of ray tracing in combination with a tree-based search structure for particle locations to efficiently find those particles a ray has to be intersected with. Kanzler et al.~\cite{Kanzler2018} built upon voxel ray tracing and proposed a GPU rendering technique for large 3D line sets with transparency.
They use an approximate voxel model for 3D lines, using quantization of line-voxel intersection points to a discrete set of locations on voxel faces. Ray tracing is then performed using the regular grid as an acceleration structure.

Ray tracing of line sets can be performed on analytic or polygonal tube models by using common acceleration structures like kD-trees or bounding volume hierarchies to accelerated ray-object intersections.
CPU and GPU ray tracing frameworks like OSPRay~\cite{OSP} and OptiX~\cite{Parker2010} can be used for this purpose. 
OSPRay builds on Intel's Embree ray tracing kernels~\cite{Wald2014} and has built-in support for rendering fixed-radius opaque streamlines and B\'ezier curve primitives.
Han et al.~\cite{han_ray_2019} further extended OSPRay with a module for rendering Generalized Tube Primitives, supporting varying radii, bifurcations, and transparency. 
NVIDIA's RTX ray tracing through the OptiX interface~\cite{NVIDIA2018} uses RT cores on current GPUs to perform hardware-accelerated ray-primitive intersection tests.
OptiX also provides an interface to implement custom shaders, which, in our current scenario, can also be used to analytically intersect rays with tubes.


\section{Line Rendering}
\label{sec:techniques}
We classify line rendering algorithms into two major groups: object-order and image-order.
Object-order techniques use rasterization of geometric primitives to let the GPU compute the fragments falling into each pixel in an arbitrary order.
Although the order is first given by the order in which rendering calls are issued for each primitive, this order is not given when processing each fragment in the fragment shader stage.
For transparency, these techniques use either fragment merge heuristics or 2-pass approaches to ensure (correct) transmittance and visibility.
In contrast, image-order techniques use ray tracing to find the surface points seen through the pixels. 
The correct visibility order of the points along a ray is established by using a space-partitioning scheme to traverse a ray in front-to-back order through space.
\subsection{Object-Order}
Object-order techniques can be classified into accurate and approximate techniques. Accurate techniques guarantee the exact visibility order of rendered fragments. Approximate techniques violate this order by blending a fragment's color over a color that already contains the color of a fragment that is closer to the camera. Although approximate techniques typically have bounded memory and rendering constraints, accurate approaches come with either unbounded rasterization load or unbounded memory requirements.

\subsubsection*{\textbf{Depth Peeling}}
\label{sec:dp}
Depth Peeling (DP)~\cite{Everitt2001} generates pixel-accurate renderings of transparent geometry by rendering the scene multiple times and using the depth buffer to achieve ordered blending, each transparent layer at a time.
DP utilizes the depth buffer hardware test to successively obtain the next closest layer and performs standard front-to-back blending into the current framebuffer.

\begin{figure}[h!]
	\centering
	\begin{overpic}[width=0.32\linewidth,clip,trim= 40mm 80mm 320mm 0mm]
		{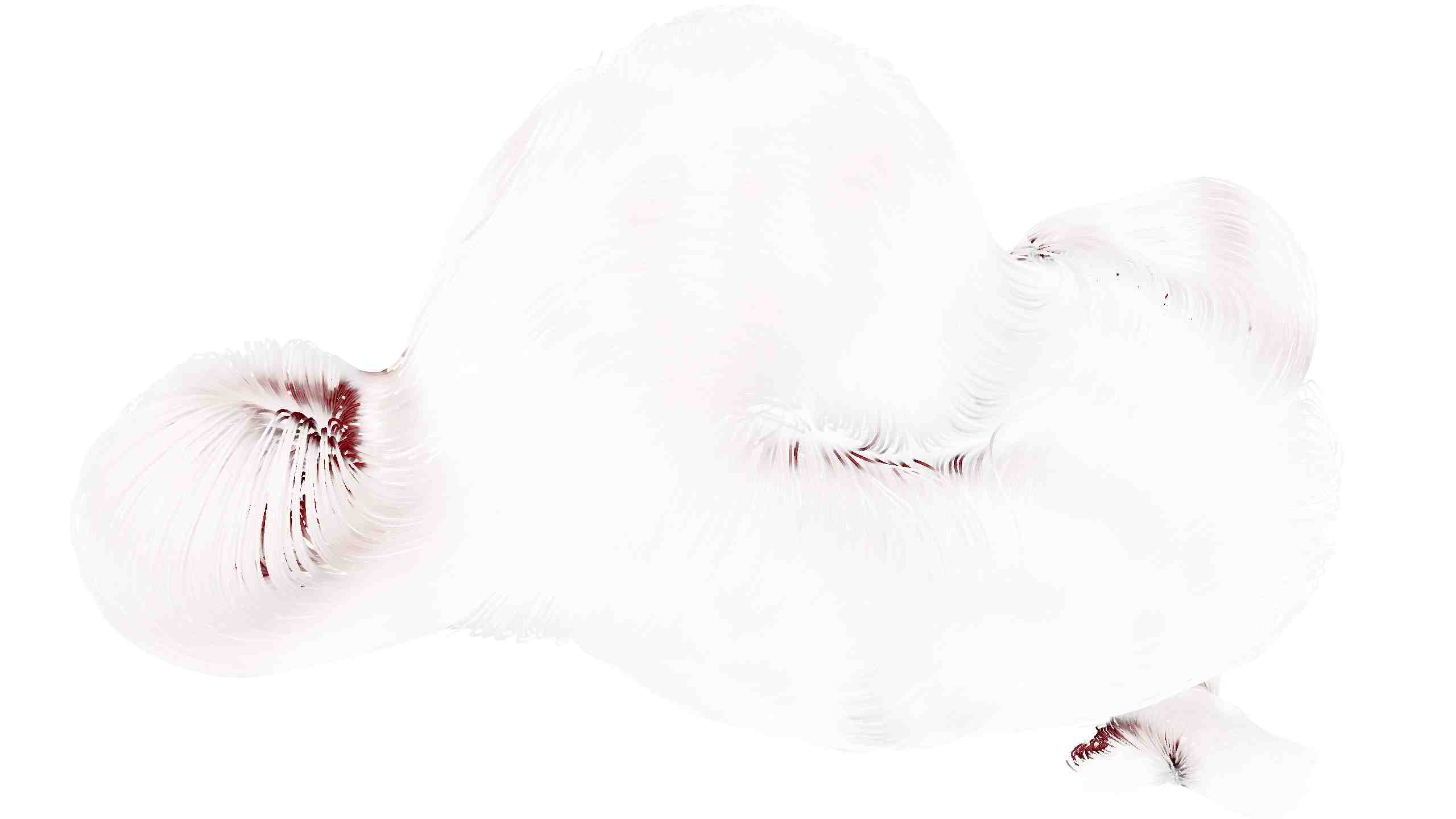}
	\end{overpic}
	\begin{overpic}[width=0.32\linewidth,clip,trim= 40mm 80mm 320mm 0mm]
		{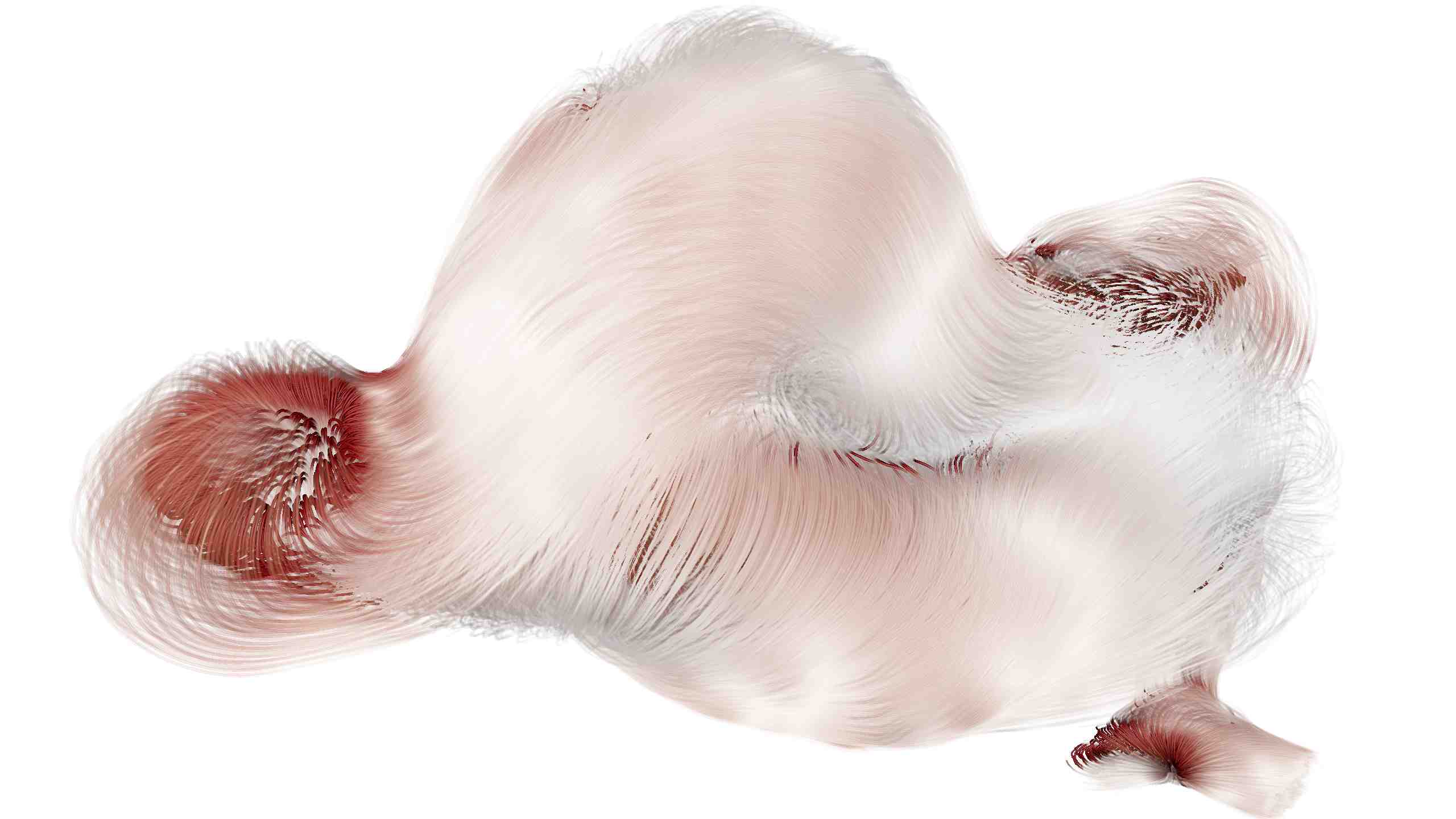}
	\end{overpic}
	\begin{overpic}[width=0.32\linewidth,clip,trim= 40mm 80mm 320mm 0mm]
		{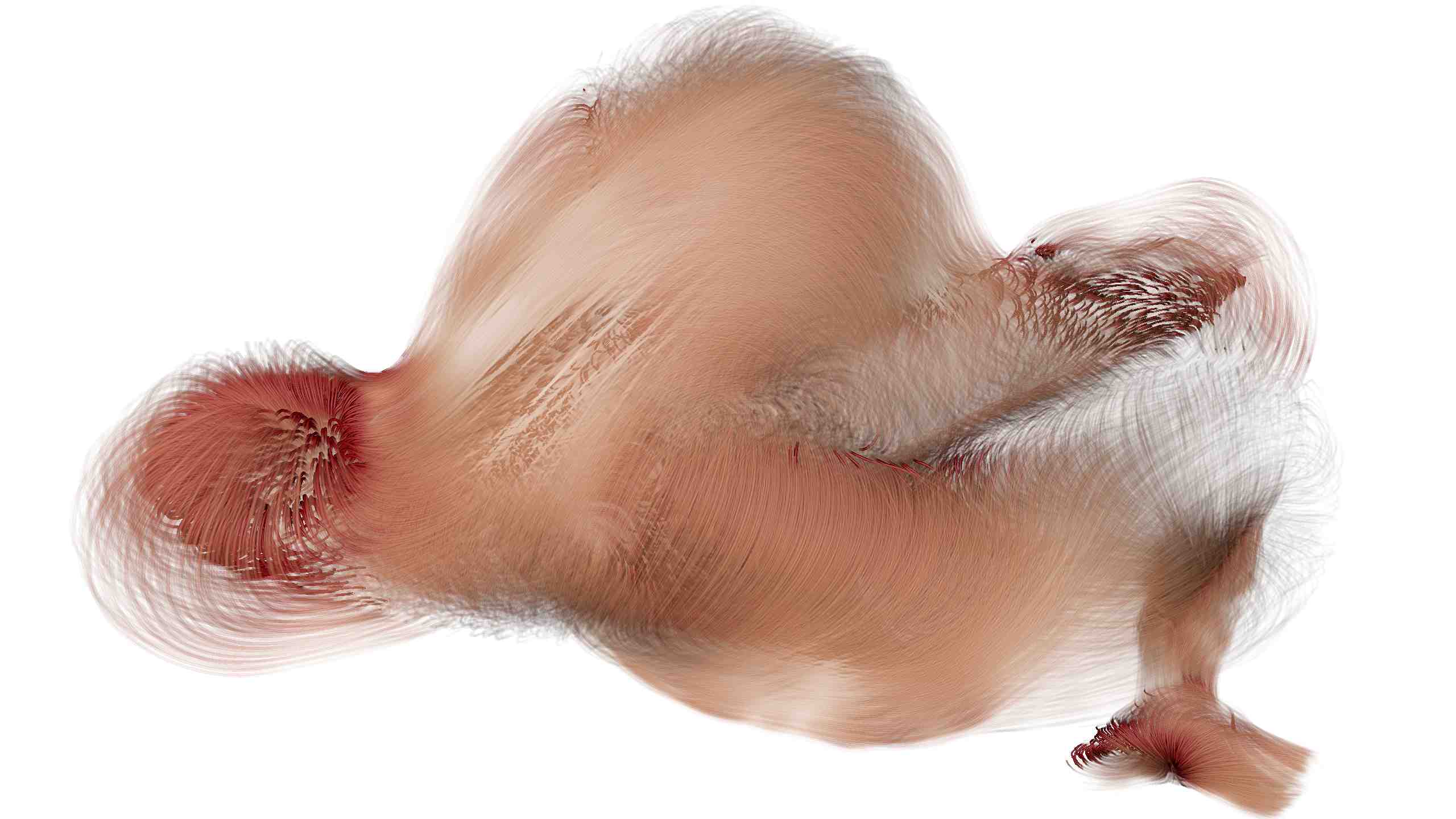}
	\end{overpic}
	\caption{Intermediate results of depth peeling for layers 1, 10, and 50 of a semi-transparent line set. Note that even at layer 50 notable differences appear in the final result.}
	\label{fig:depth_peeling}
	\vspace{-1.5em}
\end{figure}

A unique property of DP is that it does not require any additional memory besides a second depth buffer. On the other hand, DP needs to ``peel'' layers, i.e., render the scene as many times as the depth complexity of the scene. In our application scenario, where the scenes are comprised of many thousands of thin and often space-filling objects, a huge number of rendering passes typically has to be performed. As indicated in Fig.~\ref{fig:depth_peeling}, terminating the blend passes after a fixed number of times is dangerous, because deep layers with high opacity can contribute significantly to the final pixel color.

Due to the high rendering cost of DP, its performance is typically about 1-2 orders of magnitude below that of all other alternatives we consider. Therefore, we decided to use DP solely for generating ground truth images of transparent lines, against which the results of other techniques are compared.

\subsubsection*{\textbf{Per-Pixel Linked Lists}}
\label{sec:ll}
While DP has bounded memory constraints and unbounded rasterization load, Per-Pixel Linked Lists (LL)~\cite{Yang2010} come with bounded rasterization load yet unbounded memory requirements.
LL renders the scene only once, and stores all generated fragments in linked lists over all pixels.
Then, for every pixel, a pixel shader is invoked which traverses the list and stores all fragments belonging to that pixel in a GPU buffer resource. The fragments are then sorted wrt. their depth.
Finally, the fragments are blended in sorted order to produce the final pixel color.
Besides the global fragment buffer, LL requires a head buffer that stores, for every pixel, the offsets to the first fragment in the linked list, and an atomic counter that tracks the number of inserted fragments to enable concurrent gathering of new fragments into the fragment buffer. LL assumes that the GPU buffer resource is large enough to store all rendered fragments; otherwise, it fails to correctly render the scene.
To reduce the memory requirements, one commonly stores fragment colors in 32-bit unsigned integers, with 8 bits per color and $\alpha$ channel.

Even if the available GPU memory is large enough, which is often the case on high-end GPUs, sorting the many hundreds or even thousands of fragments per pixel can become a performance bottleneck. Although simple sorting algorithms such as bubble sort or insertion sort are suitable for small numbers of fragments, they do not scale well to large numbers of fragments.
To address this limitation, we have incorporated alternative sorting algorithms that achieve better scalability and can be implemented on top of GPU-friendly data structures, i.e., GPU versions of shell-sort and priority-queue through the min-heap data structure. 

Shell-sort~\cite{Shell1959} is an in-place sorting algorithm based on insertion-sort.
It is specifically designed to achieve improved sorting performance for large arrays by exchanging elements that are far apart from each other.
Shell-sort subdivides the array into $k$ subarrays by sorting each $k$-th element of the array via insertion sort.
$k$ in this context defines the offset between elements and is hence called the gap.
The gap is iteratively decremented in $l$ iterations using a pre-defined sequence $(k_1,k_2,...,k_l)$ of $l$ numbers.
The $a_i$-th element is then sorted in the $i$-th iteration.
Note that $k_l$ is always 1 since every element needs to be sorted in the last iteration.
In our work, we used $l = 4$ and a gap sequence of $(24,9,4,1)$ (based on Table 1 in~\cite{Ciura2001}) to efficiently order elements with an average depth complexity of 124 elements per node.

Priority-queues are implemented with the min-heap data structure.
A min-heap is a full binary tree where each node contains a key defining the priority (or order) of the element.
For each parent node, the key of its children is either equal or smaller than its own key.
Heap data structures are commonly implemented as binary trees.
In our case, the depth value of each fragment represents the key.
That is, after each insert operation, the root node is the currently closest element in the min-heap.

Upon insertion of all fragments in the heap, the next closest fragment is iteratively obtained by removing the root node from the heap until the heap is empty.
Root removal is implemented by setting the element with the least priority as the new root and sinking it down until it is correctly sorted. This process takes $O(\log n)$ time for a heap with $n$ elements.
Since the root has to be obtained $n$ times, the total time complexity is $O(n \cdot (\log n))$, which is faster than the depth complexity of the previously mentioned sorting algorithms.

\begin{figure}[h!]
    \vspace{-0em}
	\centering
    \begin{overpic}[width=0.85\linewidth,clip]
		{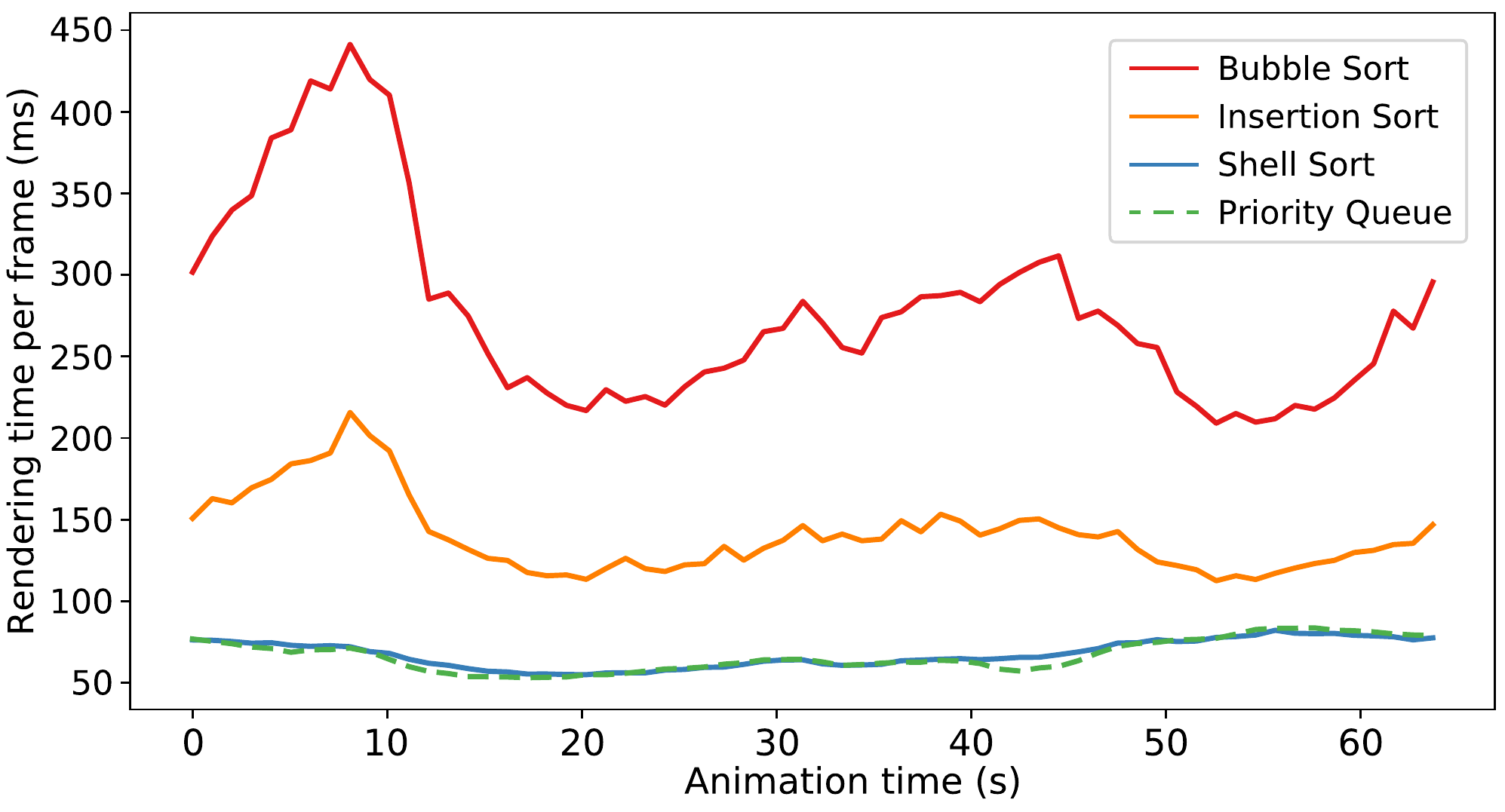}
	\end{overpic}
	\caption{Rendering times for a flight around the ANEURYSM data set using LL and different sorting algorithms.
	}
	\label{fig:linked_lists}
 	\vspace{-1.0em}
\end{figure}

Both sorting algorithms have been embedded into LL to improve its performance. Fig.~\ref{fig:linked_lists} shows performance graphs for renderings of the aneurysm data set from many different views. As can be seen, shell-sort and priority-queues significantly improve the sorting performance by a factor of 2 to 3 on average and keep the sorting time almost constant over all frames. Due to the slightly better performance of priority-queues for other data sets, we decided to use this version of LL in our evaluations.%

\subsubsection*{\textbf{Multi-Layer Alpha Blending}}
\label{sec:mlab}
Multi-Layer Alpha Blending (MLAB)~\cite{Salvi2014} is a single-pass technique. It belongs to the class of transparency rendering techniques that are bounded in both memory consumption and rendering load. This class of techniques does not perform exact visibility sorting of fragments but strives to approximate the transmittance and color. MLAB does this by heuristically merging fragments into a small number of layers, a so-called blending array, that are finally merged into the pixel color.

The blending array consists of $k$ buffers, into which incoming fragments are merged. Each of the $k$ buffers stores the blended colors and transmittances of the fragments merged into it, as well as a depth value. Each fragment is placed into a corresponding buffer based on its depth.
When all layers are occupied and the current fragment creates a new layer, the two most appropriate adjacent layers are merged in turn.

We chose MLAB because of its simplicity, performance, and low memory consumption.
On the other hand, it is clear that with only $k$ layers---eight by default in our current implementation---MLAB is not always able to accurately reconstruct the colors and transmittances for scenes with high depth complexity. In particular, the quality of MLAB is highly dependent on the order in which fragments are generated, as the outcome of the heuristic merge operation depends highly on this order. Some of the inaccuracies shown in Fig.~\ref{fig:teaser} are due to this dependency.

Additionally, MLAB is not frame-to-frame-coherent if the order in which fragments are rendered to the intermediate buffers is not guaranteed over time, resulting in flickering artifacts during animation. We prevent this error by explicitly enabling order-preserving pixel synchronization (see~\cite{PixelSync}) so that fragments are processed in the order primitives were issued.
Note that with pixel synchronization enabled, we did not experience any loss or increase in performance.

\subsubsection*{\textbf{Multi Layer Alpha Blending with Depth Bucketing}}
\label{sec:mlabdb}
To make MLAB less dependent on the order in which fragments are generated, we propose a variant that considers a discrete set of depth intervals.
We call this approach MLAB with depth bucketing (MLABDB).
The general idea underlying MLABDB is to discretize the scene into $k$ disjoint buckets that perform MLAB independently for the corresponding depth interval.
Each fragment is thus assigned to a bucket by means of its depth value and merged heuristically into the local corresponding color and transmittance buffer.
Since the buckets are already sorted wrt. depth, blending can finally be performed by blending the buckets' values in front-to-back order.

\begin{figure}[t]
	\centering
	\includegraphics[width=0.95\linewidth]{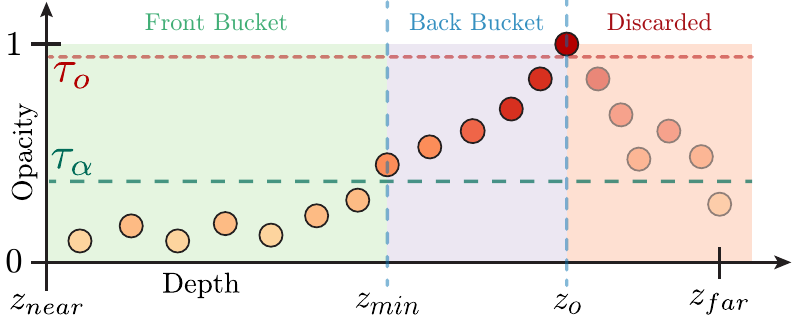}
	\caption{Fragments with opacity, ordered by depth. MLABDB searches for the first fragment with $\alpha \geq \tau_\alpha$ to obtain depth boundary $z_{min}$ of the front bucket. Back bucket bounds are $z_{min}$ and $z_o$, with $z_o$ the depth of the first opaque fragment. Fragments with $z > z_o$ are discarded.}
	\label{fig:mlabdb}
	\vspace{-1em}
\end{figure}

However, only discretizing the scene into buckets of equal intervals produced images with less quality and visible artifacts.
MLAB itself is not order-independent and yields different results per pixel depending on the order of fragments for each bucket and pixel, resulting in artifacts.
To avoid these artifacts, in MLABDB we segment the scene into two buckets and set the boundaries of the buckets heuristically with respect to opacity (compare Fig.~\ref{fig:mlabdb}).

\begin{figure}[h]
	\centering
	\begin{overpic}[width=0.49\linewidth,clip,trim= 80mm 40mm 160mm 0mm]
		{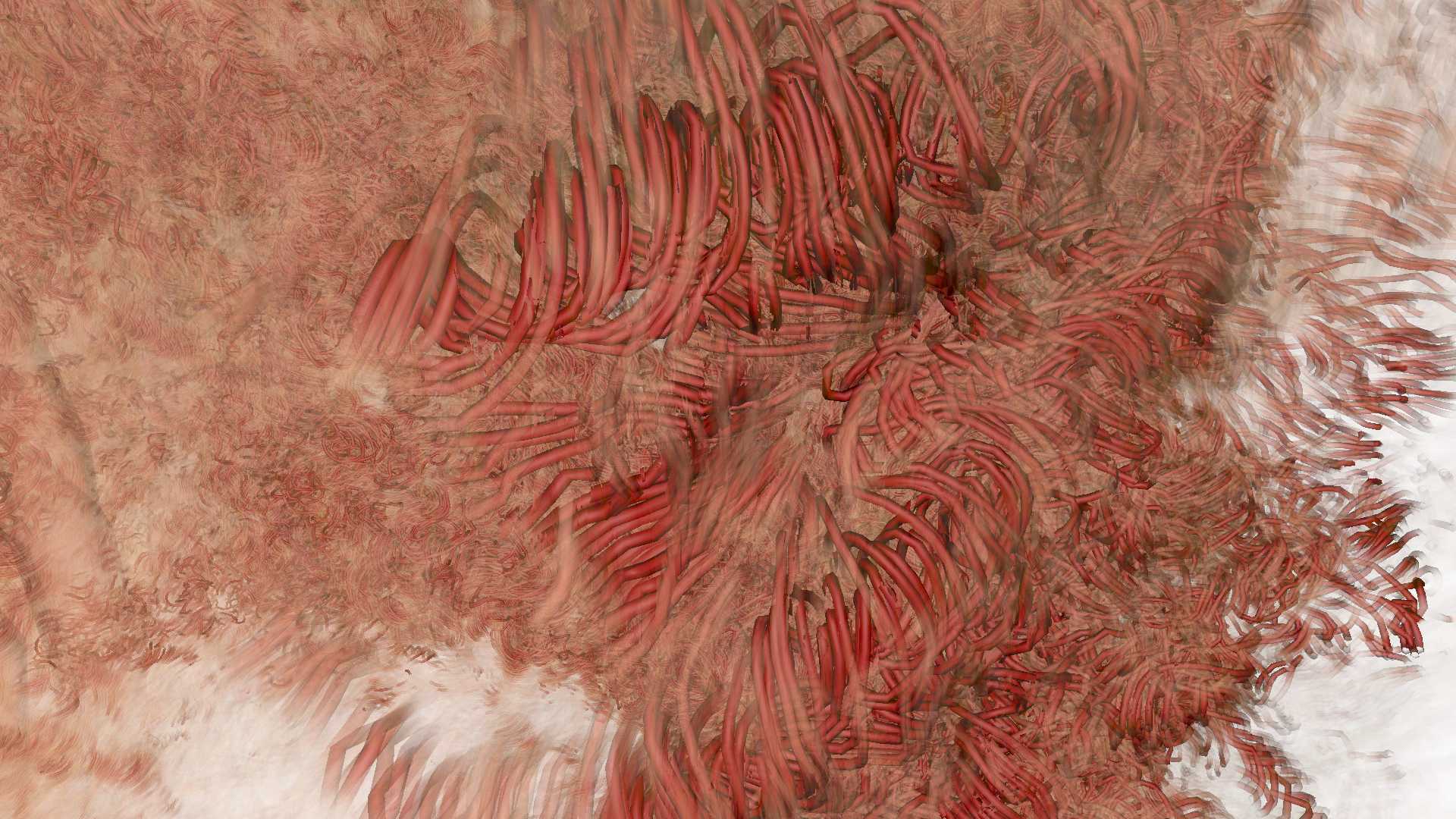}
	\end{overpic}
	\begin{overpic}[width=0.49\linewidth,clip,trim= 80mm 40mm 160mm 0mm]
		{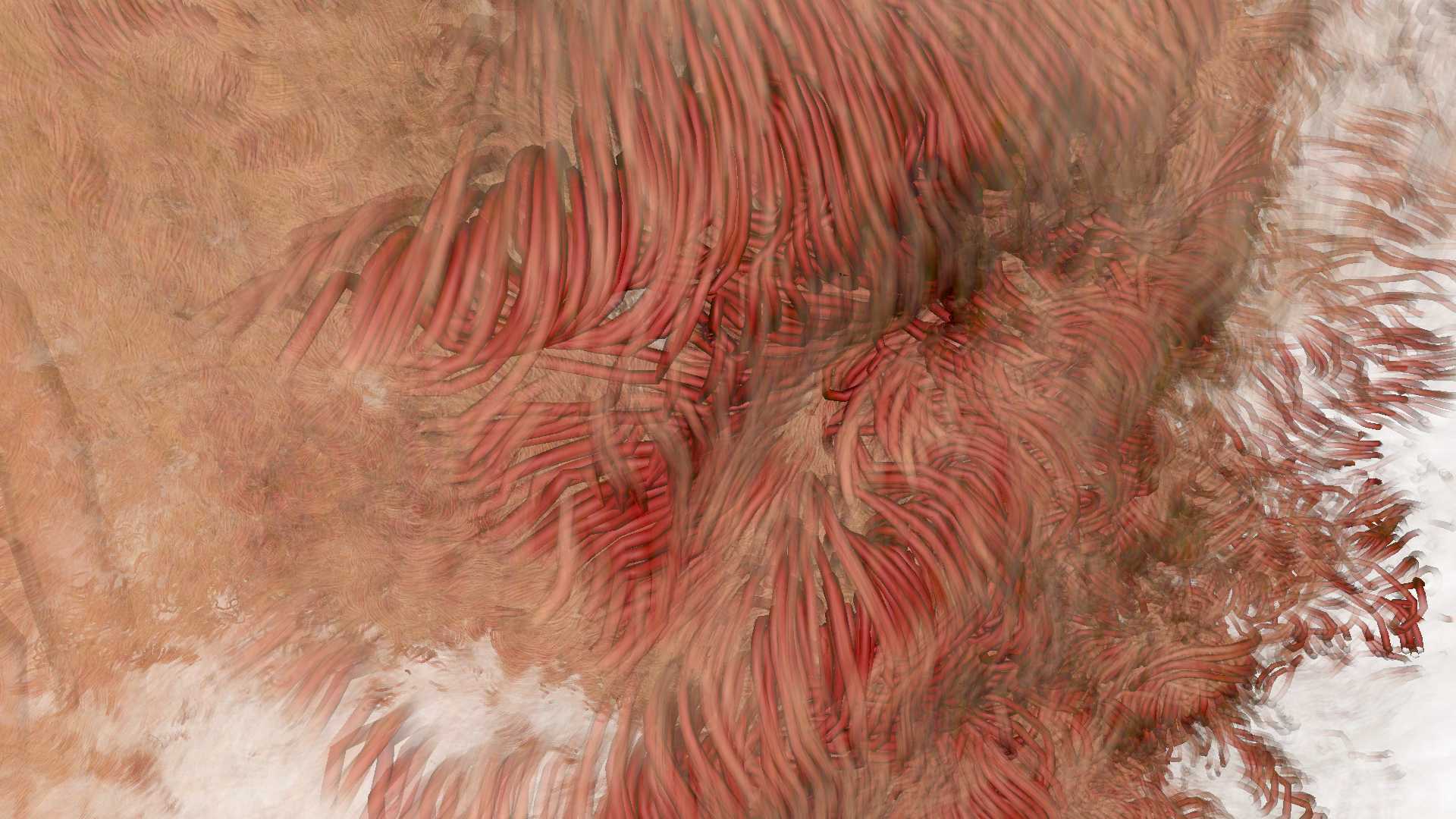}
	\end{overpic}
	\caption{Difference between MLAB (left) and our modified variant MLABDB (right). MLAB reveals interior lines erroneously due to wrong blending order. MLABDB renders correctly in this scenario.}
	\label{fig:mlab_error}
	\vspace{-1.5em}
\end{figure}

MLABDB requires two rendering passes to obtain the final color.
In the first pass, the boundaries of the two buckets are determined. For each gathered fragment, the first fragment with opacity $\alpha$ greater or equal a user-defined threshold $\tau_\alpha$ is maintained. The depth value $z_{min}$ of this fragment represents the upper depth boundary for the first bucket, called front bucket.
Since fragments behind opaque lines should not be merged the first opaque fragment with $\alpha \geq \tau_o$ is preserved. Its depth values $z_o$ and $z_{min}$ define the upper and lower boundary of the second bucket, called the back bucket.
In the second pass, all incoming fragments are assigned to the corresponding bucket by using their depths, and heuristic merges are performed independently for each bucket.
Fragments with a depth value greater than $z_o$ are discarded.
For the front bucket, we found that $n=1$ or $n = 2$ layers were suitable to gather the fragments and avoid order-dependent problems of MLAB, under the assumptions that all fragments of low opacity contribute equally to the image.
The back bucket uses a blending array with four or five layers. As demonstrated in Fig.~\ref{fig:teaser}c and further evaluated in Sec.~\ref{sec:evaluation}, MLABDB can, in many cases, considerably improve the quality of MLAB (compare Fig.~\ref{fig:mlab_error}). On the other hand, since it requires more operations, it is slightly less efficient than MLAB. Note that thresholds need to be set carefully, as visual artifacts can occur as seen in Fig.~\ref{fig:datasets_errors}(a) and discussed in Sec.~\ref{sec:quality}.

\subsubsection*{\textbf{Moment-Based Order Independent Transparency}}
\label{sec:mboit}
Moment-Based Order Independent Transparency (MBOIT) ~\cite{Muenstermann2018} is another variant of transparency rendering techniques with bounded memory and rendering constraints.
It builds upon either power moments or trigonometric moments to approximate the transmittance function per pixel in a stochastic way.
Power moments are used in statistics to reconstruct or approximate functions such as the mean and standard deviation of arbitrarily sampled random distributions.
In addition, MBOIT operates on the logarithm of the transmittances per fragment to enable order-independent additive blending.
The transmittance of the $n$-th fragment at depth $z_f$ and opacity $\alpha_f$ is given by
\begin{equation}
T(z_f) = \prod_{l = 0}^{n - 1}  (1 - \alpha_l), {z_l < z_f}
\end{equation}
The absorbance can then be defined in the logarithmic domain as
\begin{equation}
A(z_f) = -\ln T(z_f) = \sum_{l = 0}^{n - 1}  -ln(1 - \alpha_l)
\end{equation}
The absorbance can be interpreted as a cumulative distribution function of the transmittance at each layer, given that for all $l$ translucent fragments it holds that $z_l < z_f$. The distribution can be described as
\begin{equation}
Z := \sum_{l = 0}^{n - 1} -ln(1 - \alpha_l) \cdot \delta_{z_l},
\end{equation}
where $\delta_{z_l}$ is the Dirac-$\delta$ function.
Using a power moments generating function $b$: $[-1, 1] \rightarrow \mathbb{R}^{m+1}, b(z) = (1,z,z^2,z^3,\cdots,b^m)^T$, for $m$ power moments the transmittance is given by
\begin{equation}
b := E_z(b) = \sum_{l = 0}^{n - 1} -ln(1 - \alpha_l) \cdot b(z_l).
\end{equation}

MBOIT requires two passes to compute the final color.
In the first pass, the $m$ power moments are computed that are required to reconstruct the transmittance function. The first pass requires storing $m$ floating point values per pixel, we use four in our experiments.
In the second rendering pass, the transmittance of each fragment is reconstructed using the pre-computed power moments via
\begin{equation}
T(z_f, b) = \exp(-A(z_f, b)).
\end{equation}
The real absorbance of the fragment is estimated by computing its lower and upper bounds and interpolating in-between these bounds with an interpolating factor $\beta = 0.1$. This factor was determined by testing multiple values and settling for the one giving the best results.
As the quality of reconstruction further degrades with large depth value ranges, the depth values are transformed to logarithmic scale~\cite{Muenstermann2018}.
The final color can then be computed using the total absorbance, stored in the first power moment $b_0$, and the reconstructed transmittance $T(z_f, b)$ (cf. eq.~2 in \cite{Muenstermann2018}).

M\"unstermann et al.~\cite{Muenstermann2018} pointed out that this is problematic for scenes with intersecting geometry and large depth ranges, and this turns out to be especially problematic in the situation where many changes in the transparencies along one single view ray occur (see discussion in Sec.~\ref{sec:quality}).
\subsection{Image-Order}
\label{sec:image}
Image-order techniques guarantee exact visibility order of the surface points along the view rays. They utilize a search structure to efficiently find the objects that need to be tested. Therefore, they often come with increased, yet per-frame constant memory requirements. On the other hand, they have unbounded rendering constraints, since the number of ray-object intersection tests depends on the view direction.

\subsubsection*{\textbf{Voxel-Based Ray-Casting}}
\label{sec:vrc}
\begin{figure}[h]
	\centering
	\includegraphics[width=0.9\linewidth]{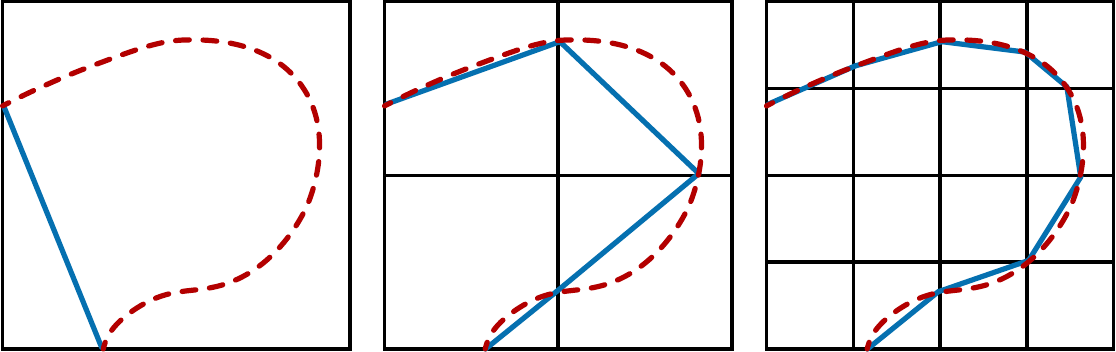}
	\caption{The original curve (red dotted line) is discretized into a voxel grid of 1, 4, and 16 voxels, respectively, from left to right. Per voxel lines are clipped against the voxel faces and linearly connected. The blue curve represents the approximated original curve at the given grid resolution.}
	\label{fig:voxel_raycasting}
	\vspace{0em}
\end{figure}

VRC is an image-order line rendering technique.
It builds upon the voxelization of a line set into a regular voxel grid and performs ray-casting in this grid with analytical ray-tube intersections to correctly blend all intersection points.
For discretizing the lines into the voxel grid (i.e., curve voxelization), each line is subdivided into a set of line segments by clipping the line at the voxel boundaries (see Fig.~\ref{fig:voxel_raycasting}). To obtain a compact representation of these segments, their endpoints are quantized based on a uniform subdivision of the voxel faces (i.e., line discretization).

For every pixel, a ray is cast through the voxel grid, going from voxel face to voxel face using a digital differential analyzer algorithm. Whenever a voxel is hit, it is determined how many lines are stored in that voxel. If a voxel is empty, it is skipped; otherwise, the ray is intersected against the tubes corresponding to each line. 
If multiple intersections with the tubes are found, they are computed and then sorted in place in the correct visibility order. Since tubes can overlap into adjacent voxels, neighboring voxels also need to be taken into account for intersection testing.

A potential weakness of VRC is the approximation quality. Curve voxelization and line quantization introduce an approximation error, which increases with decreasing grid resolution and coarser discretization of voxel faces. Conversely, higher grid resolutions and finer discretizations yield better approximations, but can significantly increase the memory required to store the voxel representation. 

%
\subsubsection*{\textbf{OSPRay CPU Ray Tracing}}
\label{sec:osp}
OSPRay is used to evaluate the performance of CPU ray-tracing for transparent line rendering.
Within OSPRay, there are three options for representing line primitives.
OSPRay's built-in streamlines represent the lines as a combination of analytic cylinder and sphere primitives, suitable for rendering opaque streamlines with a constant radius.
For smoother higher-order curves or transparency, OSPRay can also use Embree's built-in B\'ezier curve primitive directly.
Finally, the Generalized Tube Primitive module~\cite{han_ray_2019} extends OSPRay's original streamline approach to support varying radii, bifurcations, and correct transparency.
The Generalized Tubes module represents the streamlines as a combination of spheres, cylinders and cone stumps, and employs a constructive solid geometry intersection test to ensure correct transparency.
Although this CSG-based intersection comes at some cost, it is required to avoid showing interior surfaces from intersections with the constituent primitives.
To render the primitives, we use OSPRay's built-in scientific visualization renderer, which supports illumination effects such as shadows and ambient occlusion.

It is of course also possible to tessellate the tube primitives and render them in OSPRay as a triangle mesh. In our testing, we found that when using a very low-quality tessellation, the triangle mesh outperformed the Generalized Tubes for transparent geometry, due to the removal of the CSG traversal. However, even with a low-quality tessellation, the memory consumed by the triangle mesh is of concern, moreover, tessellating to a level of detail that matches the quality of the Generalized Tubes or B\'ezier curves will require significantly more primitives, impacting performance. For small- to medium- sized data sets, triangulation may be a reasonable approach.

\subsubsection*{\textbf{RTX Ray Tracing}}
\label{sec:rc}
RTX is used to evaluate the performance of GPU ray-tracing for line rendering.
On the RTX architecture, dedicated hardware, the RT cores, are used to accelerate the traversal of bounding volume hierarchies---utilizing axis-aligned bounding boxes---and the execution of ray-triangle intersection tests.

As the maximum recursion depth on current RTX hardware (32) is too low for data sets with high depth complexity, we opted for an iterative approach. We also note that a recursive approach is likely to be far more expensive than an iterative one.
Our first approach used any hit shaders to accumulate fragments along the rays. However, this can provide only an approximate result, as any hit shaders are not guaranteed to be run in a strict front-to-back order. Thus, we did not pursue this approach further.

\begin{figure}[!ht]
	\includegraphics[width=\linewidth]{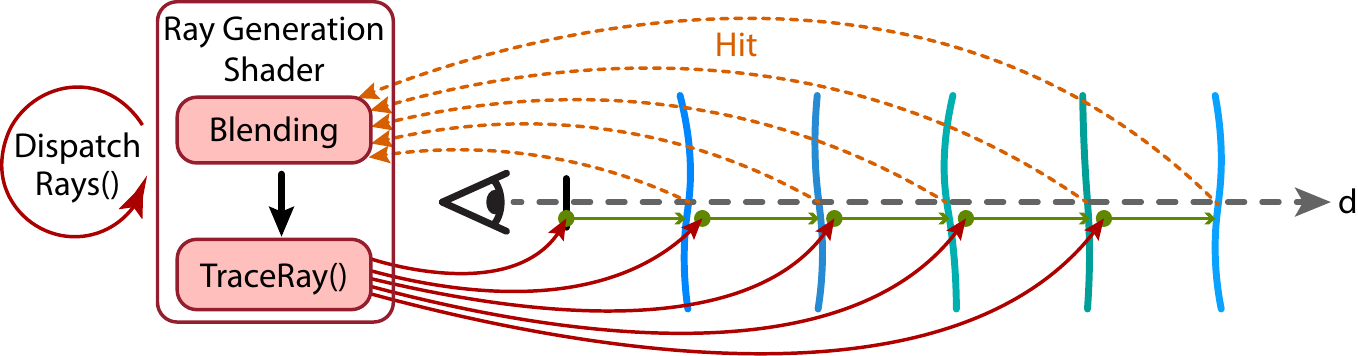}
	\caption{Illustration of iterative ray-casting using the RTX framework. Blue-colored paths represent line primitives of the data set. At each frame, the ray generation shader is called once  and is responsible for iteratively blending over all fragments and issuing new rays (TraceRay()) at intersection point $t_i$ (green arrows). During ray traversal, the intersection point with primitive closest to the viewer is computed. On each hit, color and opacity of line is obtained and sent back to the ray generation shader (orange arrows).
	}
	\label{fig:rtx_raycasting}
 	\vspace{-1em}
\end{figure}

Instead, in our experiments we utilize a closest hit shader in combination with a loop in the ray generation shader that blends the fragments returned by it in front-to-back order (cf. Fig.~\ref{fig:rtx_raycasting}).
Intersection sorting is thus done entirely by the acceleration structure traversal unit.

The closest hit shader also returns its hit distance along the ray, so that the ray generation shader can start the next ray right after the last hit using a very small offset to avoid intersecting the same primitive again.
The loop is terminated by either a sentinel value returned from the miss shader, run when no primitive is hit, or a zero transmittance value.
Although iterative next-hit traversals could, in theory, fail to find all intersection hits (see Wald et al.~\cite{Wald2018}), we have not experienced this problem in our experiments.

The RTX framework can also trace against custom geometry using an intersection shader, which we
utilize to perform analytic intersection tests against the tube representations for each line. 
A ray is first intersected with an infinite tube, and the intersection points are then clipped against the two planes delimiting the tube segment. 
To correctly interpolate the vertex attributes in the closest hit shader, both planes are intersected with a line parallel to the tube and through the clipped point.
The position of the clipped point on this line segment is then mapped to \([0;1]\) and used as interpolation factor in the closest hit shader.

Interestingly, the analytic ray intersection tests are about a factor of two slower than ray-triangle intersection tests, even though the lower primitive count leads to a significantly smaller memory footprint. The high performance for triangles is likely attributable to the hardware acceleration of triangle intersection testing on the RTX hardware, and do not consider analytical tests in the remainder of the evaluation for RTX.
\section{Evaluation}
\label{sec:evaluation}
We evaluate and compare all selected line rendering techniques regarding memory consumption, performance, and quality.
All GPU techniques were run on a standard desktop PC with Intel Xeon processor, 32 GB RAM, and an NVIDIA Geforce TITAN RTX with 24 GB VRAM.
CPU ray tracing was performed on a system with 2 Intel Xeon E5-2640 CPUs at 2.4 GHz and 3.4 GHz boost frequency with 40 CPU threads in total.
We used the Vulkan SDK 1.1.129 with the extension VK\_NV\_ray\_tracing to implement RTX ray tracing and conducted the performance tests using the NVIDIA driver 441.87.
Both CPU and GPU architectures come at roughly the same price, making the comparison fair in terms of financial investment.
Furthermore, all performance measurements (using data sets that fit into memory) were carried out on an NVIDIA Geforce RTX 2060 SUPER with 8 GB VRAM and a single Intel i7-5930K CPU with 12 threads. The performance scale-down compared to the measurements in Sec.~\ref{sec:performance} was roughly a factor of 1.6 -- 3 and 2, respectively.
All images were rendered at a viewport resolution of 1920$\times$1080 for performance and 1280$\times$720 for image quality. When statistics are given for flights around a data set, the camera parameters were set so that most of the viewport is covered by that data set and the entire viewport is covered in zoom-in scenarios. Ground truth images are generated via DP, yet we do not consider DP any further due to the limitations discussed in the introduction.

\subsection{Data Sets}
\label{sec:datasets}

\begin{figure*}[!ht]
	\centering
	\includegraphics[width=0.245\textwidth]{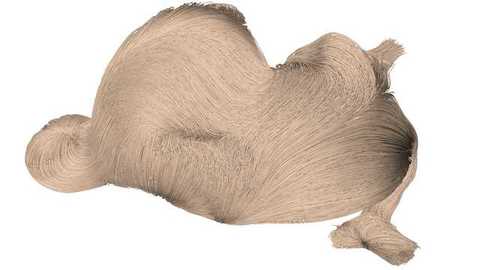}
	\includegraphics[width=0.245\textwidth]{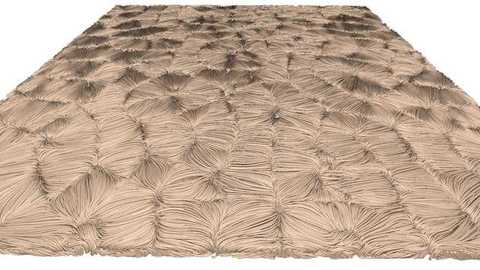}
	\includegraphics[width=0.245\textwidth]{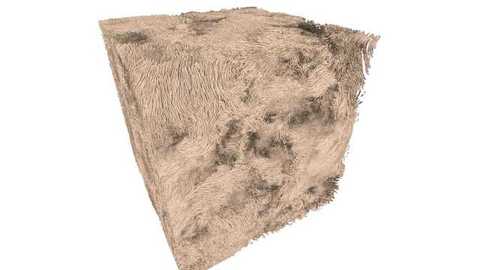}
	\includegraphics[width=0.245\textwidth]{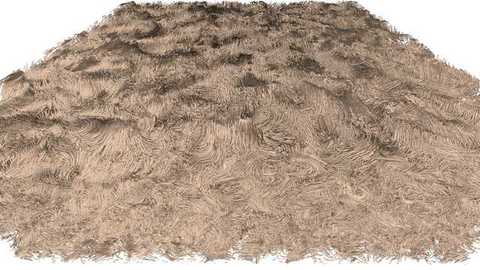}
	\includegraphics[width=0.245\textwidth]{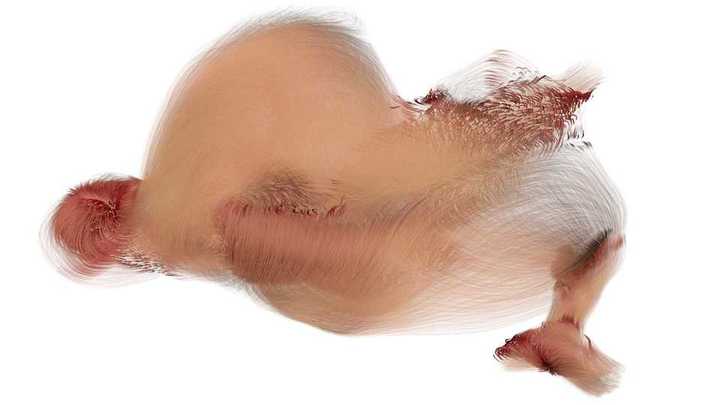}
	\includegraphics[width=0.245\textwidth]{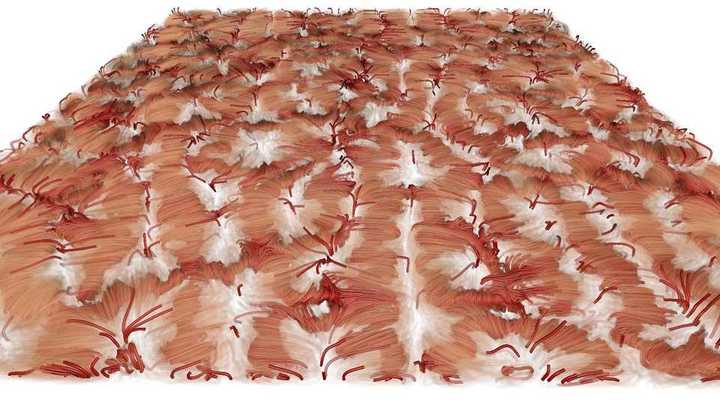}
	\includegraphics[width=0.245\textwidth]{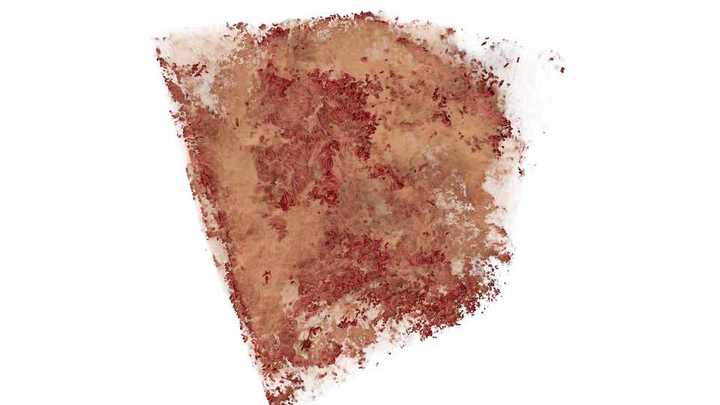}
	\includegraphics[width=0.245\textwidth]{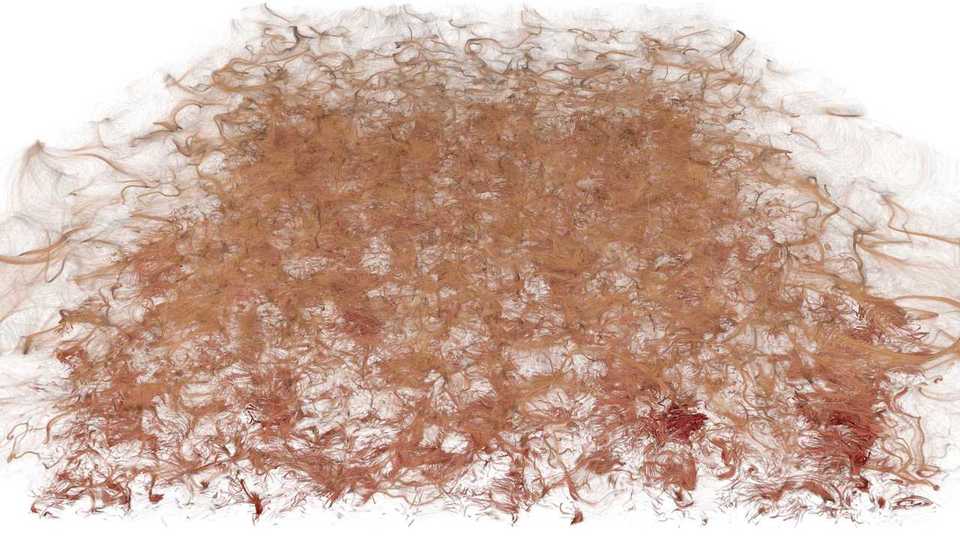}
	\caption{Data sets used in our experiments. Top: Opaque line rendering. Bottom: Transparent line rendering. From left to right: ANEURYSM, CONVECTION, TURBULENCE, and CLOUDS. Transparency greatly aids the ability to visualize important features in the data.}
	\label{fig:datasets}
	\vspace{-1em}
\end{figure*}
Our experiments were performed on data sets with vastly different numbers of lines and line density. For each data set, we selected meaningful transparency assignments, e.g., by mapping physical parameters along the lines or geometric line curvature to transparency. 
The following data sets were used:
\begin{itemize}
	\item \textbf{ANEURYSM:} 9,200 randomly seeded streamlines in the interior of an aneurysm~\cite{Byrne2014}, and advected up to the vascular wall. Vorticity in the flow field from which the lines were extracted is mapped to line transparency.
	\item \textbf{CONVECTION:} 100,000 short streamlines that were uniformly seeded in a Rayleigh-Bernard convection between a heated bottom wall and a cold top wall~\cite{Pandey2018}. Transparency is assigned according to line curvature. 
	\item \textbf{TURBULENCE:} 80,000 long streamlines generated in a forced turbulence field of resolution $1024^3$~\cite{Aluie2013}. Transparency is proportional to the maximum $\lambda_2$ vortex measure along a line.
	\item \textbf{CLOUDS:} 400,000 seeded streamlines in a cloud-resolving boundary layer simulation (UCLA-LES, see ~\cite{Stevens2013}) using a large eddy simulation (LES). Streamline integration was conducted on a voxel grid with resolution 384$\times$384$\times$130.
	The magnitude of vorticity along each streamline is mapped to transparency.
\end{itemize}

Fig.~\ref{fig:datasets} shows all data sets, rendered via opaque and transparent lines. Most data sets are very dense, yet by mapping the selected parameters linearly to transparency, important interior structures can be revealed. Table~\ref{tab:preprocess} gives further information on the number of line segments that need to be rendered as tubes, the memory that is required by the initial line representation, and the memory requirements of the internal data representation used by each technique, as explained in the next subsection.

\subsection{Data Preparation and Model Representation}
All rasterization-based line rendering techniques and RTX render the tubes as triangle meshes. Therefore, each connected sequence of lines first needs to be converted into a set of triangulated tubes that are stitched together to form a closed mesh. 
This pre-process is performed on the GPU.
The meshes are stored in a triangle and shared vertex representation with normal buffer, where each index and vector component is represented by a 32-bit value. In an additional attribute buffer, 16-bit per-vertex attribute values are stored. During rendering, these values are mapped to transparency and color. To construct the tubes, for every vertex shared by two lines, the average of the lines' tangent vectors is computed. At the first and last vertex, respectively, the average is just the tangent of the first and last line segment. Three vertices are generated and placed uniformly on a circle around the vertex in the plane orthogonal to the average tangent and containing this vertex. The three vertices from the line start and end points are then connected to form a closed set of tubes. We use the same circle radius for all tubes. The vectors from the initial vertex to the new ones are used as per-vertex normals. From our experiments, we found that three vertices along a circle radius are sufficient to achieve good results from each view direction.

The resulting buffers are used directly as input for LL, MLAB, MBOIT, and RTX. It is worth noting that all rasterization-based techniques can also generate the polygon models on the fly during rendering in a geometry shader, or use a pixel shader that takes the line information as input and analytically tests for intersections with the corresponding tube. However, since rendering the pre-computed geometry is up to a factor of 2 faster, we do not consider on the fly generation in our evaluation.

RTX requires another pre-process to build an AABB hierarchy from the given polygon model. For triangle geometry, the RTX framework supports only a few position formats, and raw data must be converted if it does not already match. For custom geometry, the API requires conservative estimates of the AABB of each primitive. 
Construction of the AABB hierarchy is performed on the GPU via the API. RTX then generates a tree structure, that is traversed by every ray until reaching the leaf nodes where ray-triangle tests are performed.

Since VRC cannot handle polygon models but tests analytically against the tubes during ray-casting, a voxel-based renderable line representation is first built and uploaded to the GPU. For ANEURYSM and CONVECTION, we used an optimized voxel grid resolution of 113$\times$110$\times$128 (x,y,z-dimension) and 128$\times$8$\times$128, respectively, and a quantization level of 16. For TURBULENCE and CLOUDS, we increased the resolutions to $256^3$ and $512^3$, respectively, and used a quantization level of 32. Grid resolutions were chosen to reduce the probability that lines fall onto each other and become indistinguishable.

OSP constructs a bounding volume hierarchy using Embree~\cite{Wald2014}. From the input line data, we build the Generalized Tubes geometry, which consists of a set of analytic spheres, cylinders, and cone stumps, the union of which forms the tube. These primitives are passed to Embree as a user geometry, over which it will construct a BVH. As with RTX, the user geometry must provide a conservative bounds estimate to the BVH builder. In our evaluation, we found that Embree's B\'ezier curves provided better performance for transparent tubes, and in this case, we switch to use Embree's Round B\'ezier curves, available through OSPRay's ``curves'' geometry type. Embree then builds a BVH over the curve primitives as before. The curve primitive is built into Embree, and additional optimizations to the BVH quality may be applied during the build, that are not available for user geometry such as our Generalized Tubes. During rendering, Embree traverses packets of rays in SIMD through the BVH until reaching a leaf node, where intersection tests are performed with Generalized Tubes or B\'ezier curves.

For all data sets and rendering techniques, Table~\ref{tab:preprocess} lists the number of line segments to be rendered (LS), the memory requirement to render all line models (LM), the memory requirement of the used renderable representations (primitive buffers and acceleration structure) on the GPU or CPU, and the time required to build these representations. 
%
%
Table~\ref{tab:preprocess} indicates that VRC performs better than rasterization-based approaches regarding memory requirement. In general, the polygon model requires much more memory than the voxel grid used by VRC, in some cases more than 10 times. The build times, on the other hand, are about a factor of 4 faster compared to VRC. We attribute this difference in build times to the fact that building the voxel representation requires far more arithmetic and memory scan operations for clipping lines at voxel boundaries, counting how many line segments fall into a voxel, and computing indices into per-voxel memory containers. Rasterization-based approaches, on the other hand, require only simple index arithmetic once the number of lines and the vertices per line is known.
\definecolor{Raster}{HTML}{377EB8}
\definecolor{OSP}{HTML}{A65628}
\definecolor{RTX}{HTML}{984EA3}
\definecolor{VRC}{HTML}{FF7F00}

\begin{table}[!tb]
	\centering%
	\caption{Data statistics and memory requirements. Number of line segments in millions (LS), line model size in MB (LM), size of renderable representation in GB (primitive buffers, acceleration structures), and build times in seconds for {\color{Raster}{rasterization-based techniques}}, {\color{OSP}{OSP}}, {\color{RTX}{RTX}}, and {\color{VRC}{VRC}}.}
	\label{tab:preprocess}
	\fontsize{7pt}{7pt}\selectfont
	\begin{tabularx}{1.0\linewidth}{|X<{\centering}|m{4mm}<{\raggedleft}|m{3mm}<{\raggedleft}|>{\color{Raster}}m{3mm}<{\raggedleft}>{\color{OSP}}m{3mm}<{\raggedleft}>{\color{RTX}}m{3mm}<{\raggedleft}>{\raggedright\color{VRC}}m{3.5mm}|>{\color{Raster}}m{1.5mm}<{\raggedright}>{\color{OSP}}m{0.5mm}<{\raggedleft}>{\color{RTX}}m{0.5mm}<{\raggedleft}>{\color{VRC}}m{3.5mm}<{\raggedright}|}%
		\hline
		  \multicolumn{1}{|>{\centering\arraybackslash}c|}{\textbf{Data Set}} 
		& \multicolumn{1}{>{\centering\arraybackslash}m{4mm}|}{\textbf{LS}} 
		& \multicolumn{1}{>{\centering\arraybackslash}m{3mm}|}{\textbf{LM}} 
		& \multicolumn{4}{>{\centering\arraybackslash}c|}{\textbf{Render. Repr. (GB)}}
		& \multicolumn{4}{>{\centering\arraybackslash}c|}{\textbf{Build Times (s)}}\\ \hline

		ANEURYSM    & 2.3  & 34 & 0.39&0.38&2.04&0.05 & 0.8&0.2&0.6&3.6  \\ 
		CONVECTION & 9.9  & 151 & 1.67&1.62&5.27&0.04 & 3.2&0.3&2.8&11.1  \\ 
		TURBULENCE  & 17.5 & 267 & 3.01&3.01&9.38&0.50 & 7.7&1.1&5.1&23.6  \\ 
		CLOUDS      & 39.6 & 610 & 4.63&6.12&14.5&0.89 & 13.5&4.5&9.5&42.1 \\ \hline%
	\end{tabularx}%
	\vspace{-1em}
\end{table}

We find that RTX consumes significantly more memory than the other alternatives, partly because of the larger number of triangles that is finally stored in the BVH acceleration structure.
In addition to that, space-partitioning schemes for dense line sets become increasingly inefficient and run into the problem of either clipping lines at the boundaries, thus duplicating vertex information, or using overlaps, which significantly increases the number of regions to be tested.
OSP achieves the best performance for building the acceleration structures since it considers only the initial line segments and comes with a highly optimized multi-threaded BVH build routine provided by Embree.


\subsection{Per-Frame Memory Requirements}
\begin{figure}[!h]	
	\centering
	\begin{overpic}[width=0.525\linewidth,clip,trim= 0mm 00mm 02mm 03mm]
		{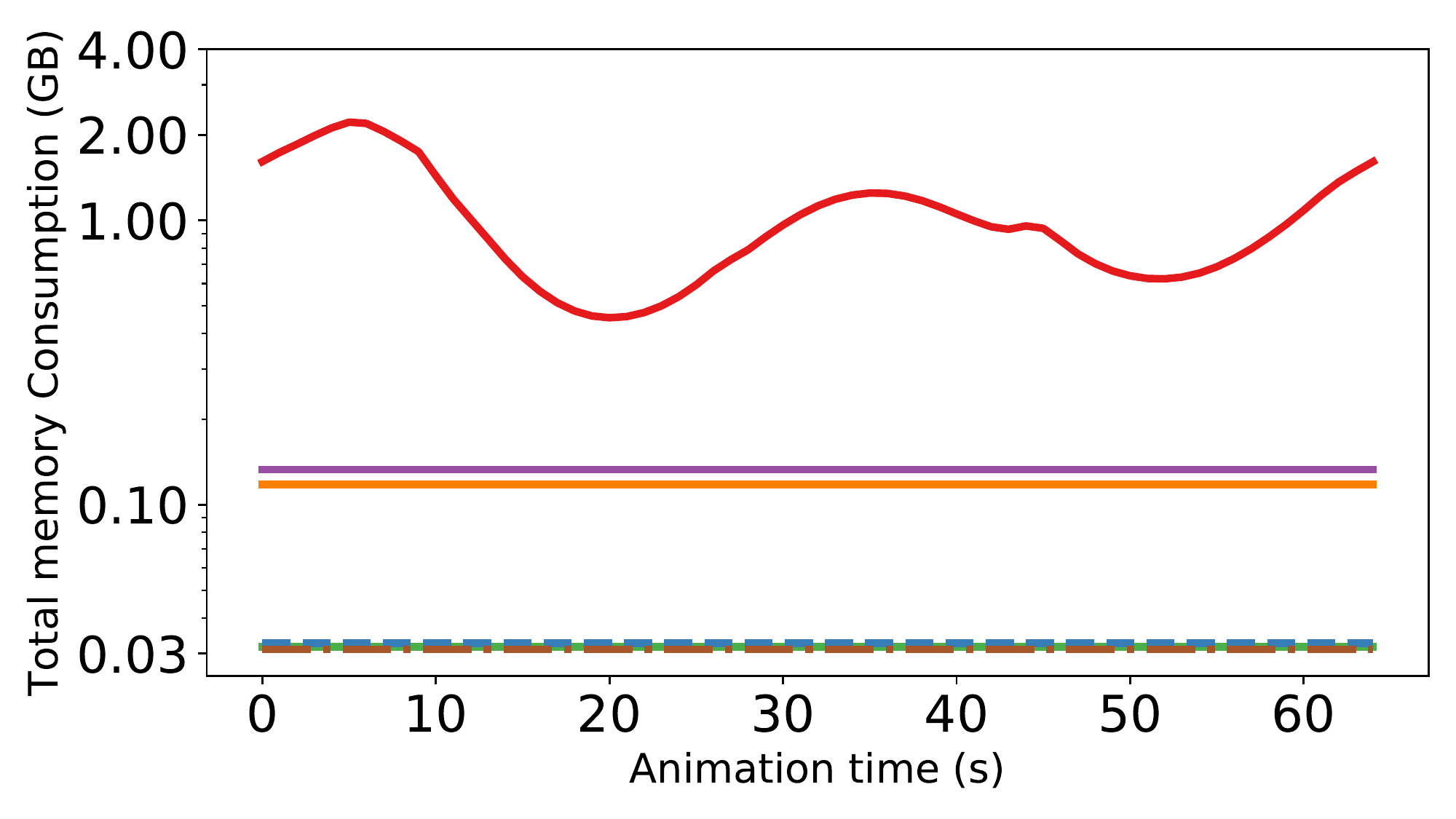}
		\put(40,28){\textbf{\scriptsize\textcolor{gray}{ANEURYSM}}}
	\end{overpic}
	\begin{overpic}[width=0.454\linewidth,clip,trim= 27mm 00mm 02mm 03mm]
		{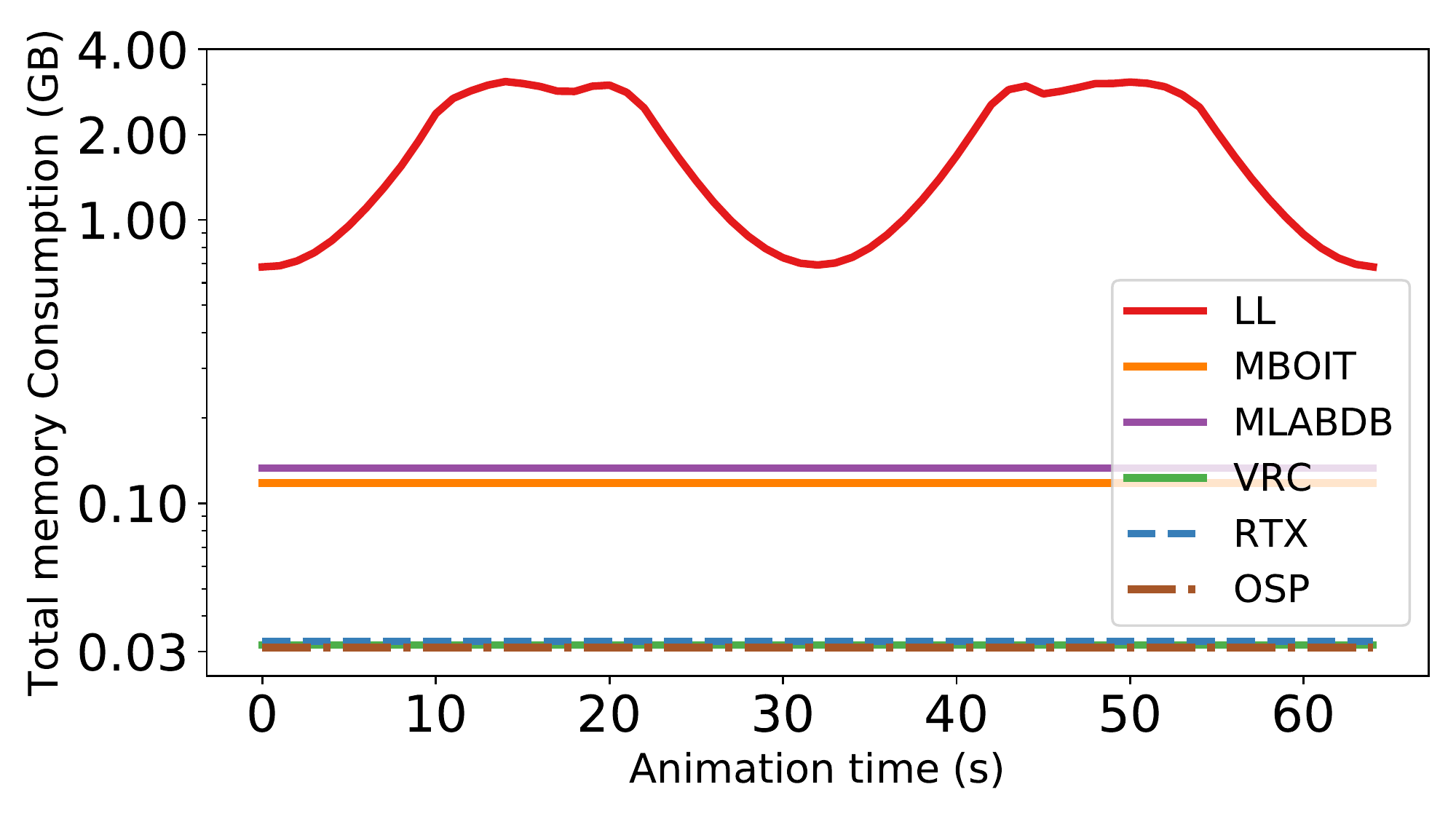}
		\put(26,32){\scriptsize\textbf{\textcolor{gray}{TURBULENCE}}}
	\end{overpic}
	
	\caption{Memory consumption (in addition to internal model representation) of techniques for different views of ANEURYSM and TURBULENCE.
	}
	\label{fig:memory_aneurysm}
	\vspace{-1.0em}
\end{figure}
To analyze the additional memory consumption of each technique during rendering, we render the models along pre-recorded paths around them, including two zoom-ins.
For ANEURYSM and TURBULENCE, the graphs in Fig.~\ref{fig:memory_aneurysm} show the memory requirements per frame (not including the internal model representation) by each technique.

OSP, RTX, and VRC do not require any additional memory beyond the internal renderable model representation. MLABDB and MBOIT require an additional per-pixel buffer with $k$=5 or $k$=4 layers, respectively. MBOIT stores four 32-bit floating point power moments in these layers. The memory consumption of this buffer is negligible compared to that of the renderable representation. LL, on the other hand, needs to keep a buffer with as many entries as the maximum number of fragments that can be generated for any of the possible views. Due to performance issues, this buffer is usually allocated once in a pre-process, using a prescribed maximum number of fragments. The memory consumption of LL can exceed the available GPU memory, especially when rendering at higher viewport resolutions.

In general, if the data set is too large to fit into available GPU memory (4 -- 8 GB VRAM on recent commodity graphics cards) rendering the entire line set per frame is not possible anymore.
Here, object-order techniques can simply split up line sets into chunks and render those separately at each frame.
VRC can split the voxel-based representation into chunks and proceed the same way.
RTX can generate chunks of the data and creates an AABB hierarchy for each chunk individually.
This requires the traversal of several AABBs at the same time which can lead to a drop in performance.
For LL, memory requirements can be reduced by using screen-space partitioning, so that only subsets of fragment lists have to be stored per pixel at once.
%
\vspace{-0.5em}
\subsection{Rendering Performance}
\label{sec:performance}

\begin{figure*}[!ht]	
	\centering
	\begin{overpic}[width=0.245\textwidth,clip,trim= 0mm 10mm 2mm 2mm]
		{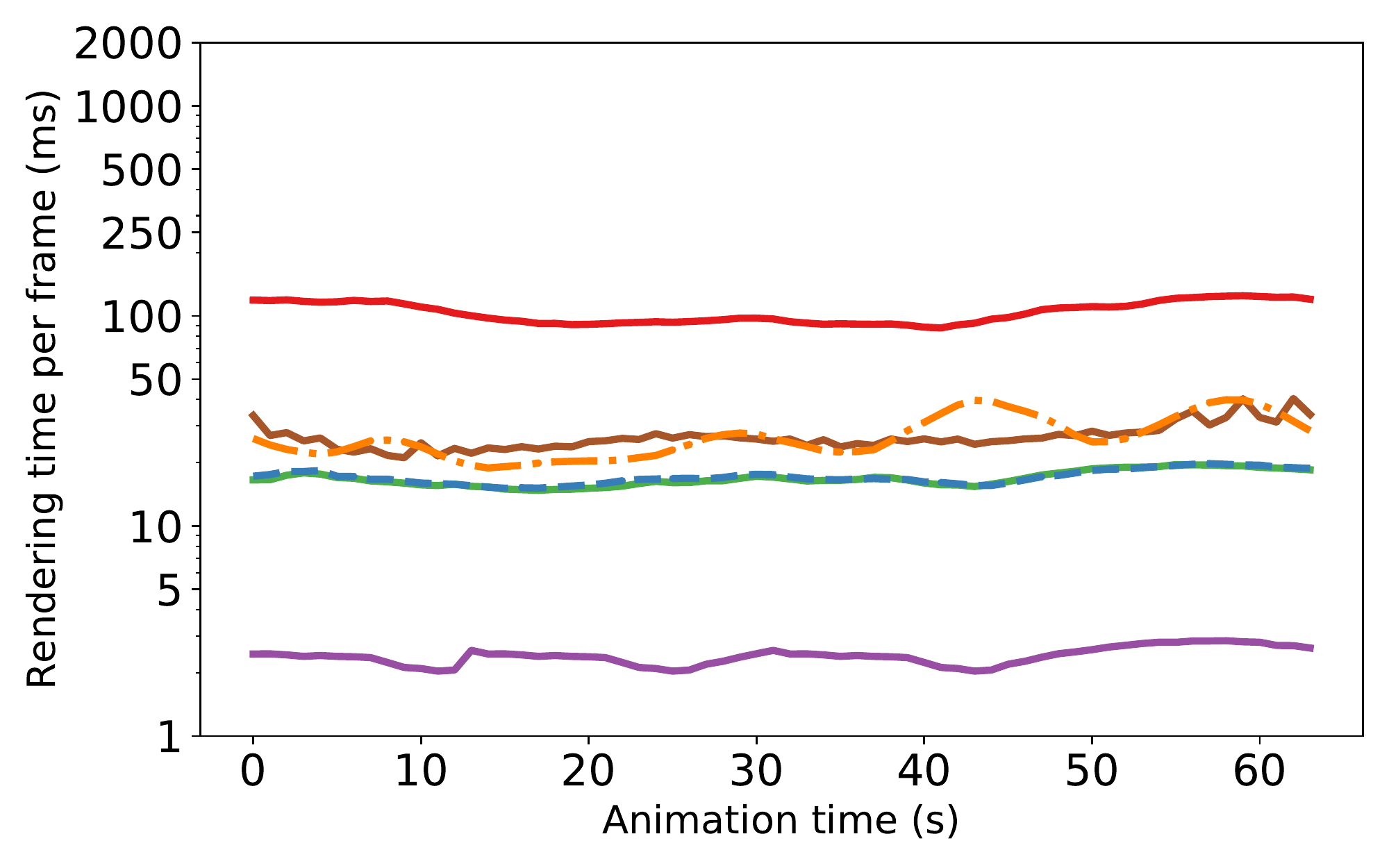}
	\end{overpic}%
	\begin{overpic}[width=0.245\textwidth,,clip,trim= 0mm 10mm 2mm 2mm]
		{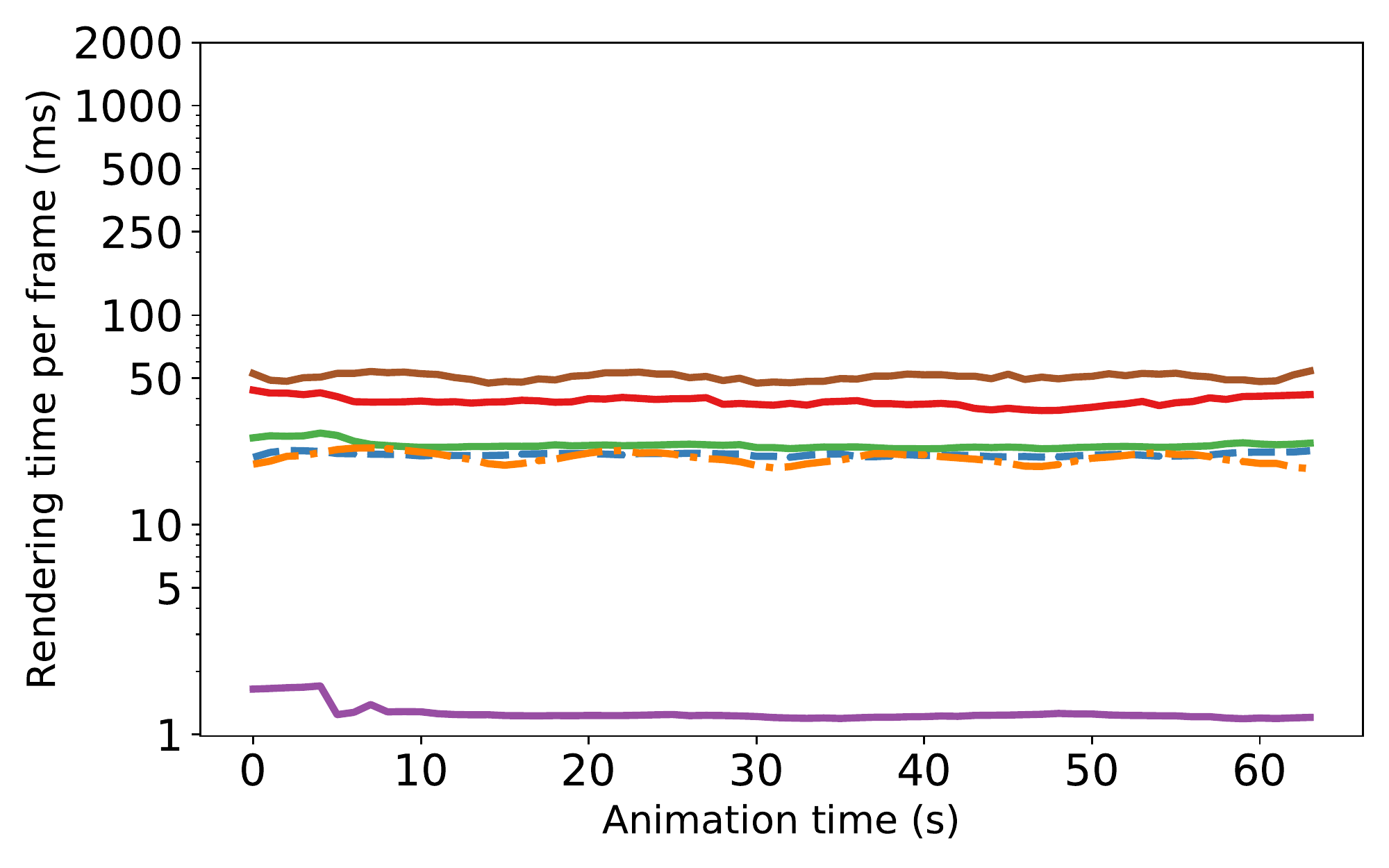}
	\end{overpic}%
	\begin{overpic}[width=0.245\textwidth,clip,trim= 0mm 10mm 2mm 2mm]
		{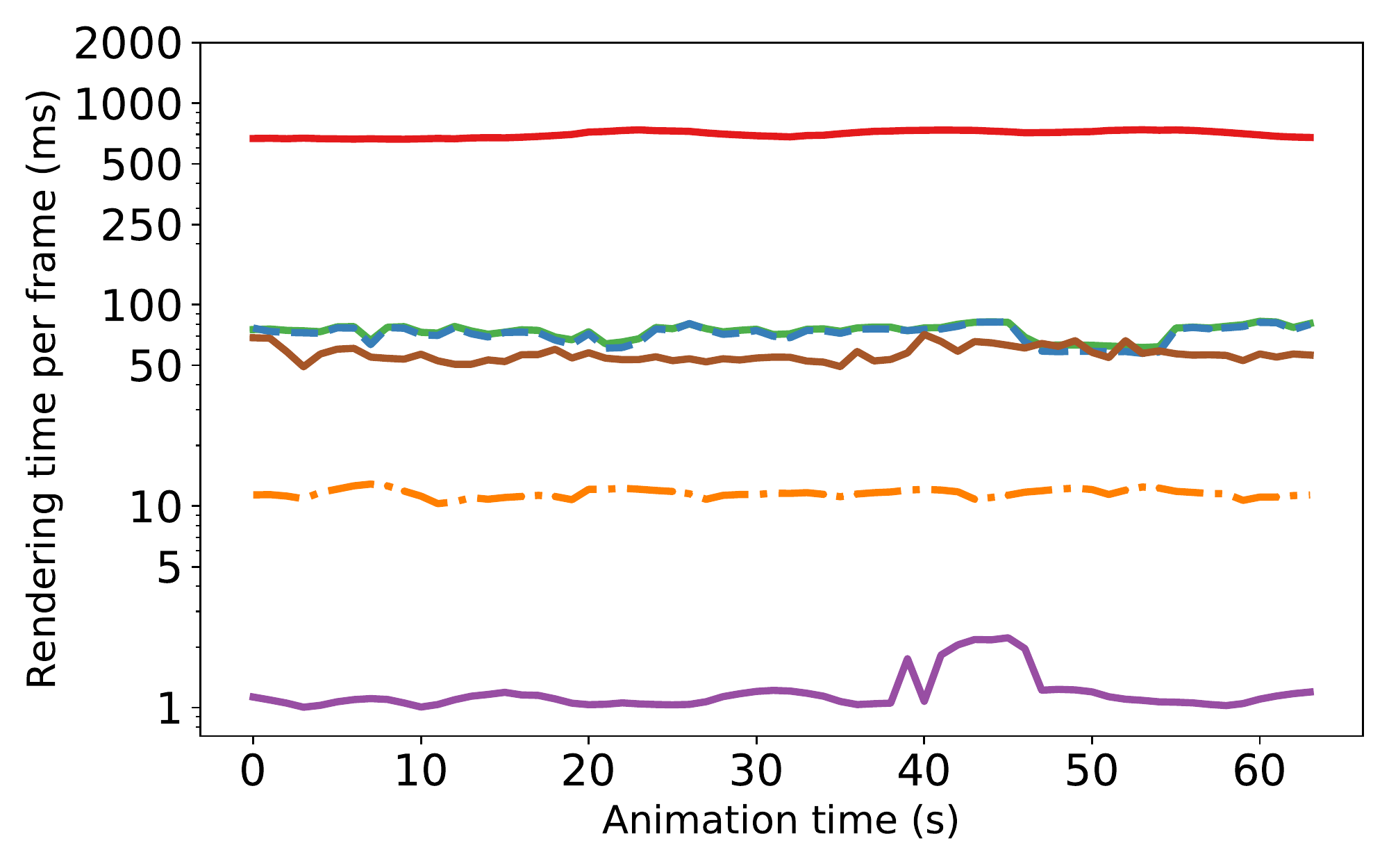}
	\end{overpic}%
	\begin{overpic}[width=0.245\textwidth,clip,trim= 0mm 10mm 2mm 2mm]
		{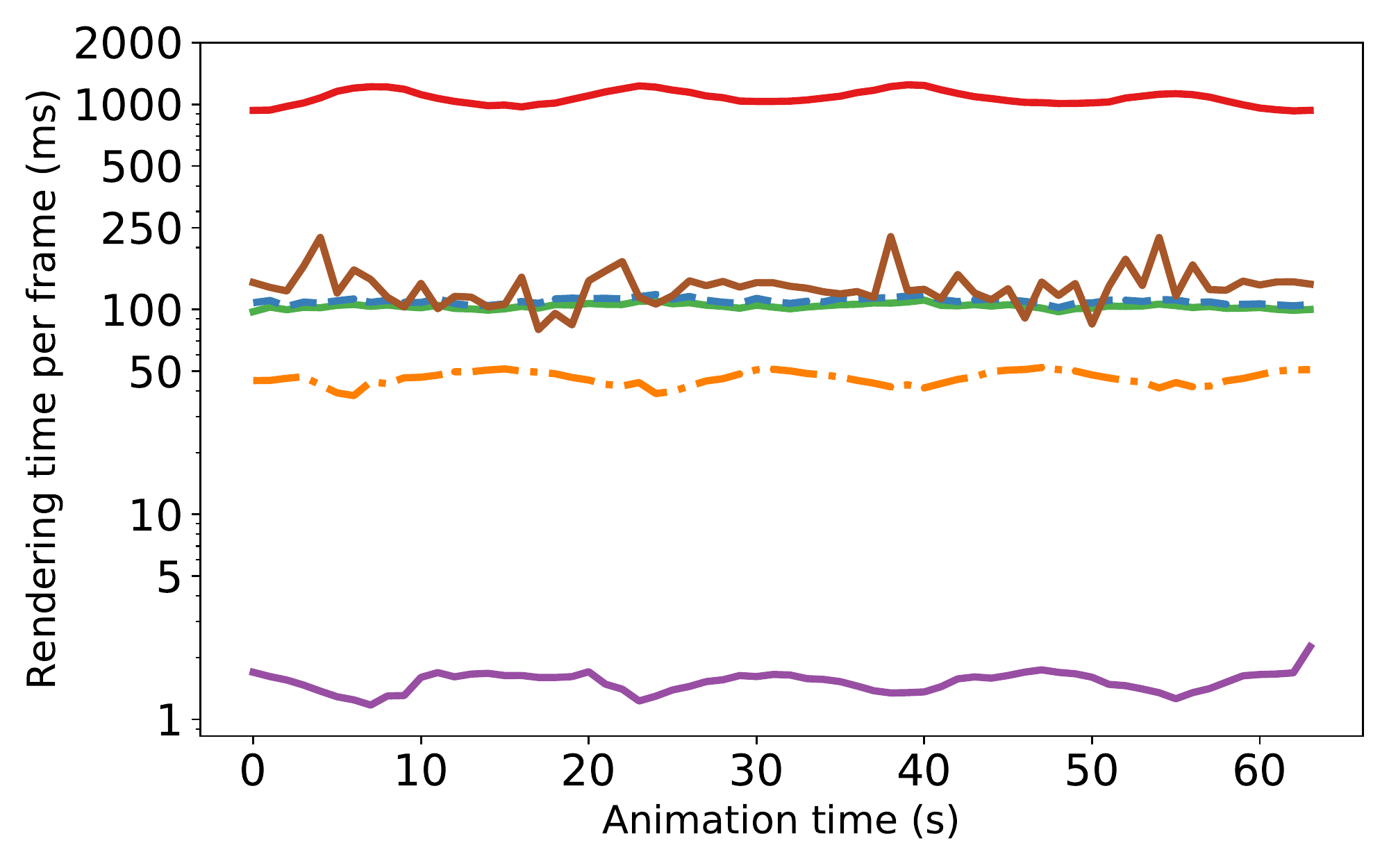}
	\end{overpic}

	\begin{overpic}[width=0.245\textwidth,clip,trim= 00mm 10mm 2mm 2mm]
		{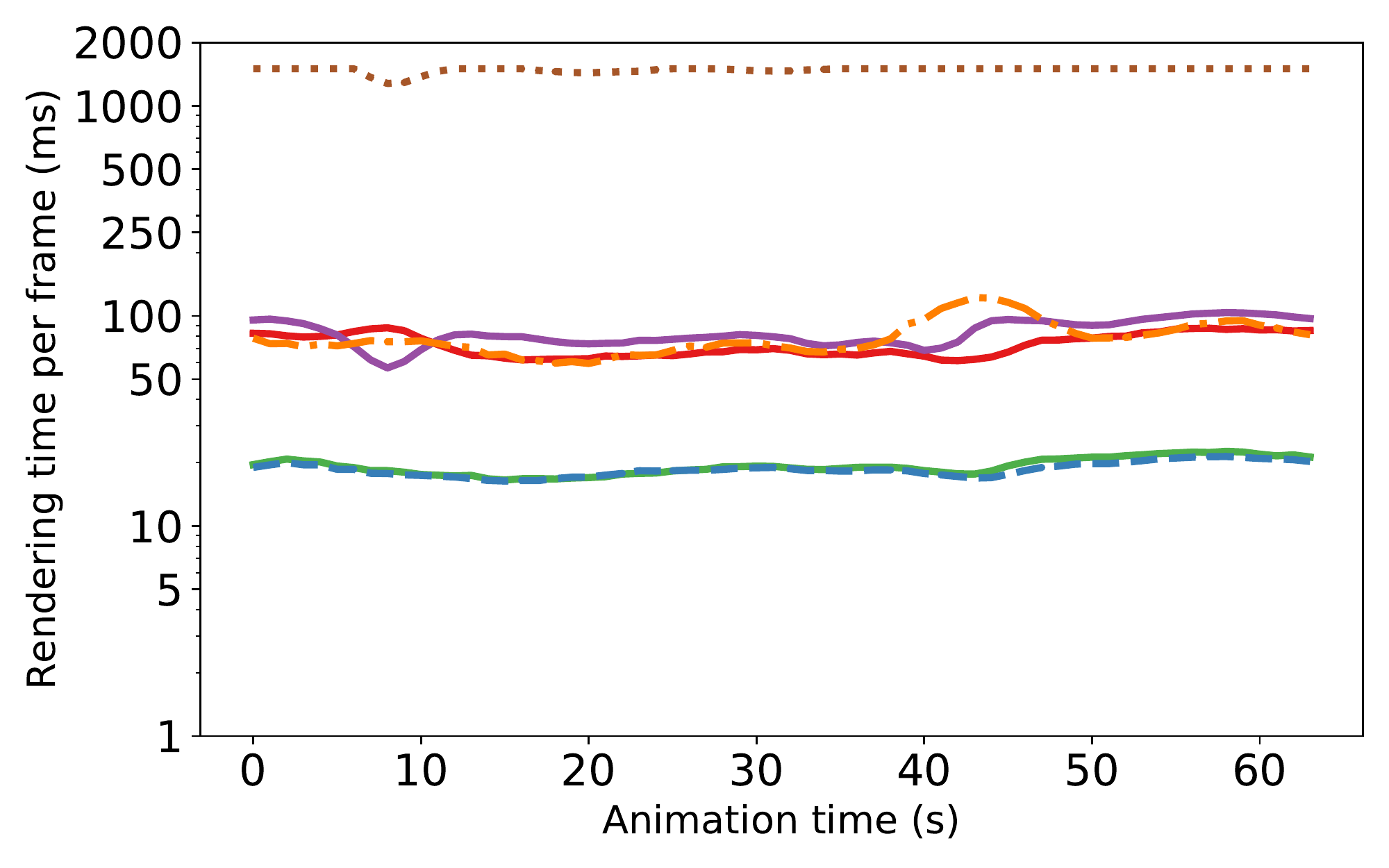}
	\end{overpic}%
	\begin{overpic}[width=0.245\textwidth,clip,trim= 00mm 10mm 2mm 2mm]
		{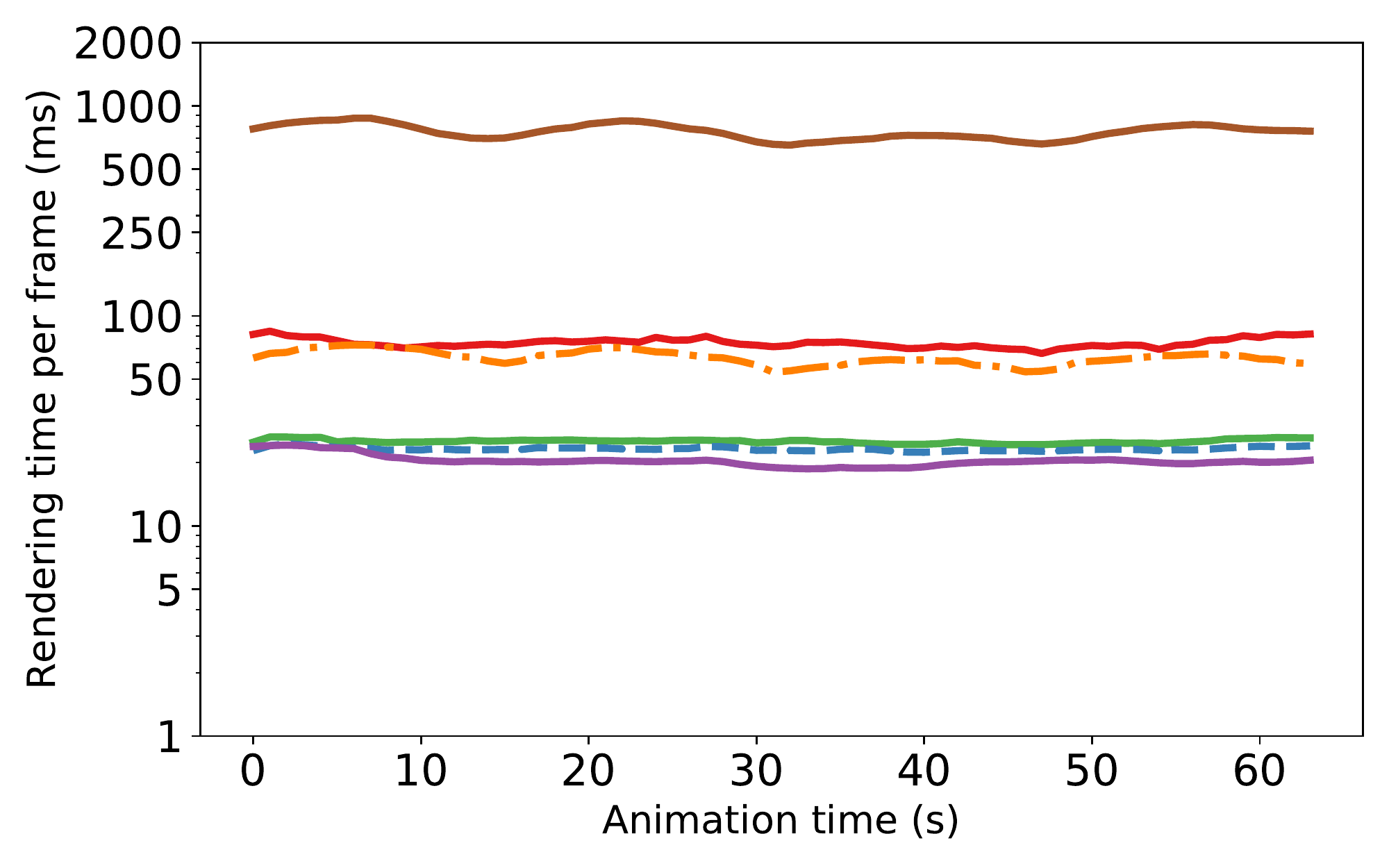}
	\end{overpic}%
	\begin{overpic}[width=0.245\textwidth,clip,trim= 00mm 10mm 2mm 2mm]
		{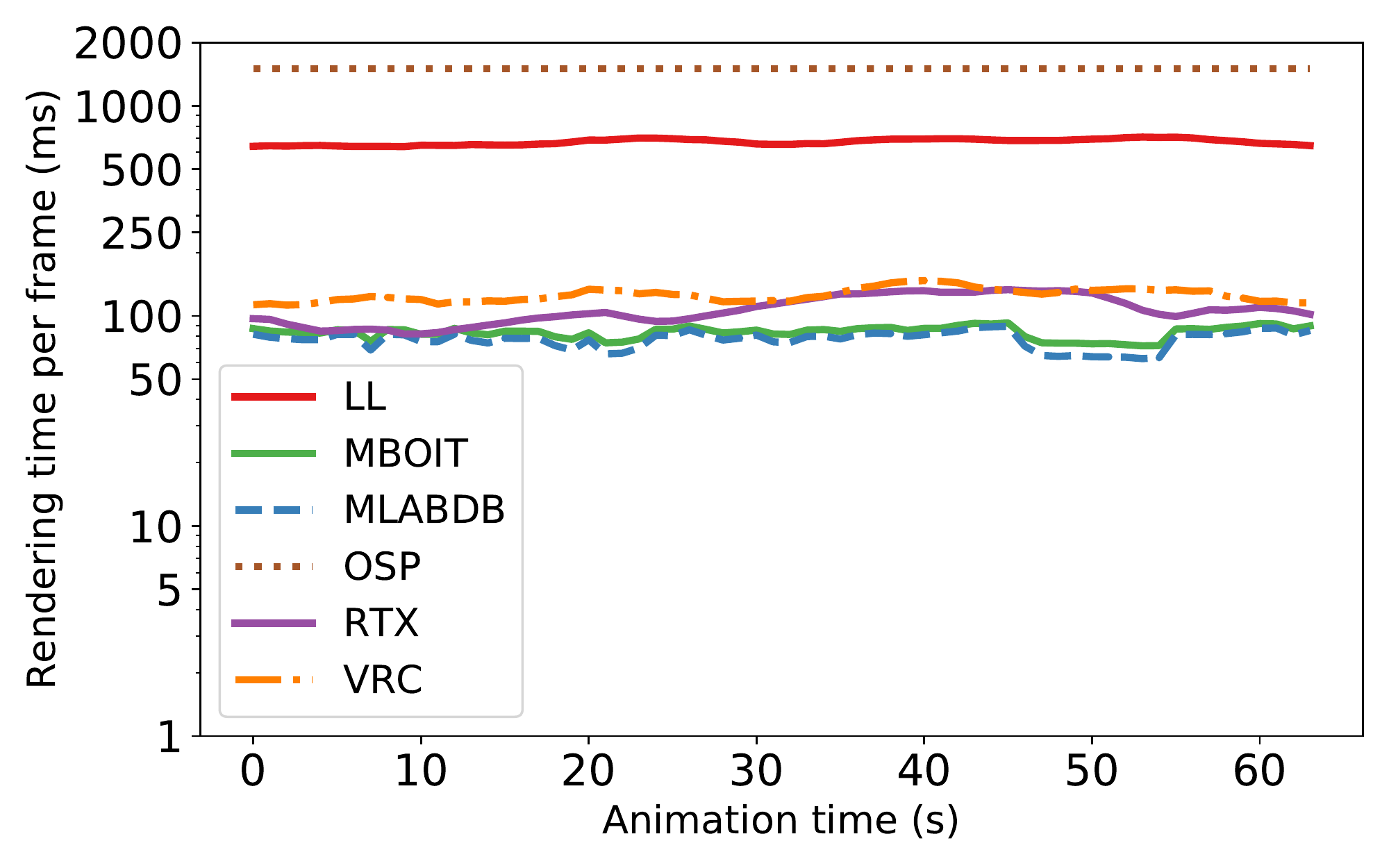}
	\end{overpic}%
	\begin{overpic}[width=0.245\textwidth,clip,trim= 00mm 10mm 2mm 2mm]
		{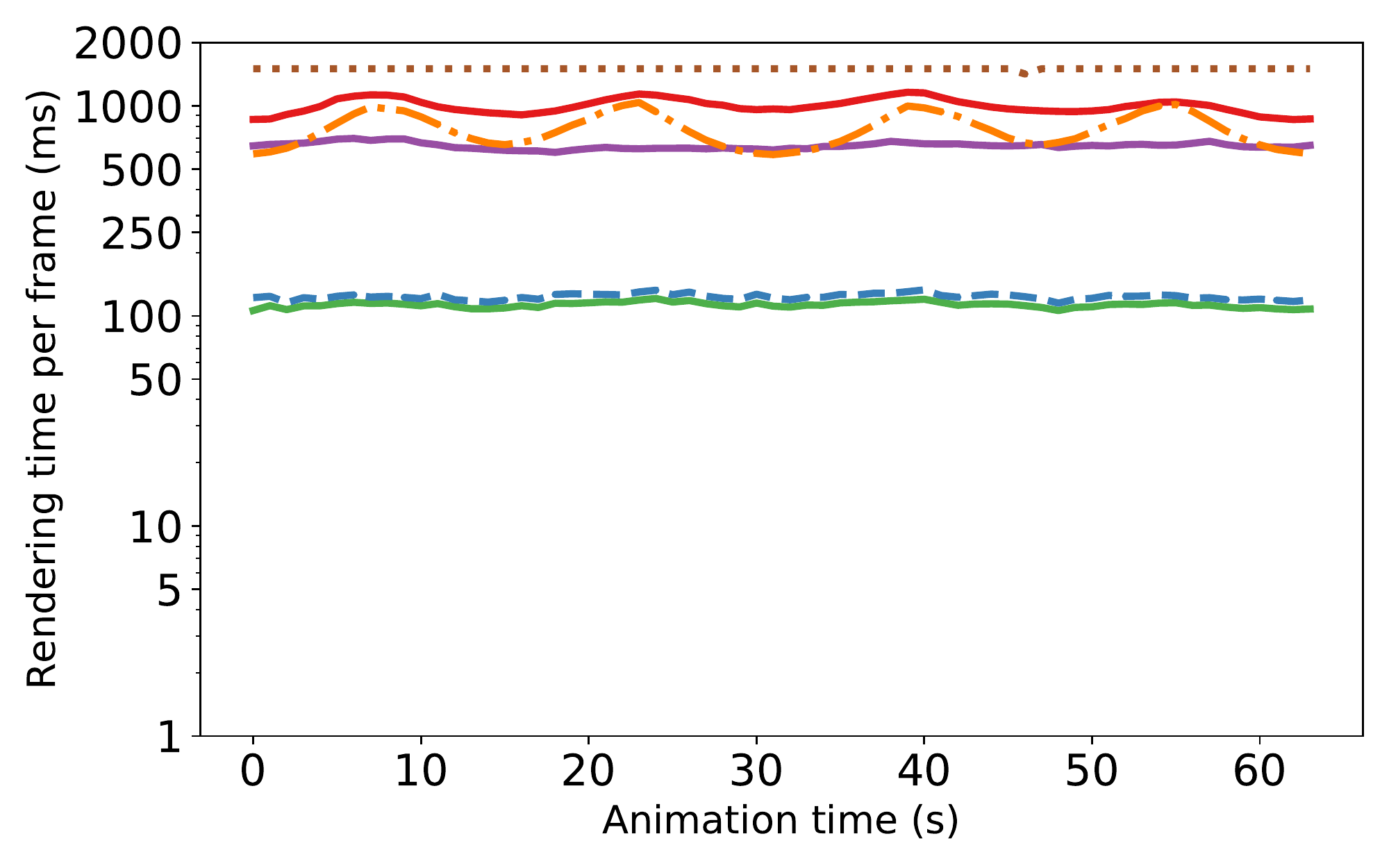}
	\end{overpic}

	\begin{overpic}[width=0.245\textwidth,clip,trim= 00mm 0mm 2mm 2mm]
		{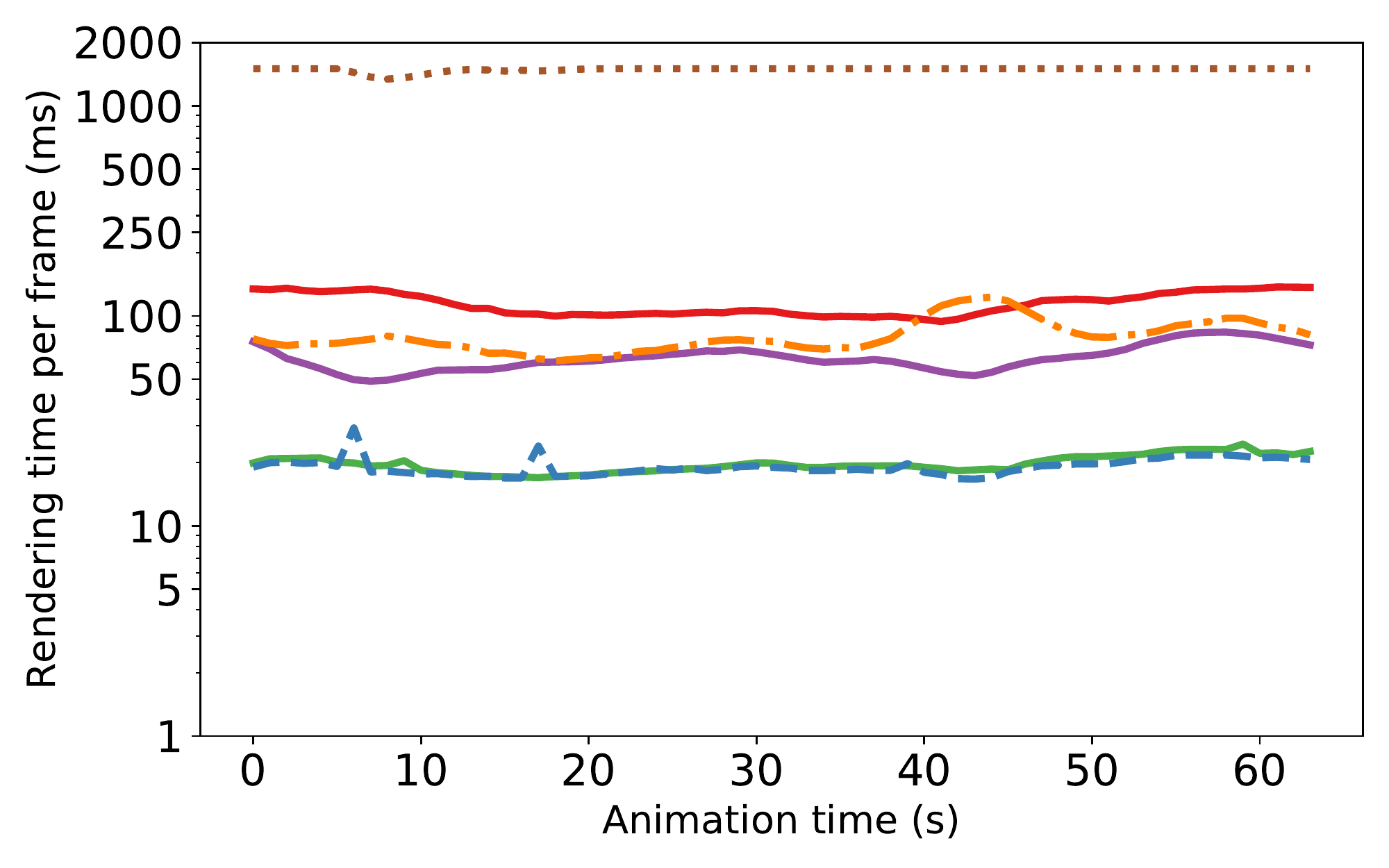}
		\put(38,15){\scriptsize\textbf{\textcolor{gray}{ANEURYSM}}}
	\end{overpic}%
	\begin{overpic}[width=0.245\textwidth,clip,trim= 00mm 0mm 2mm 2mm]
		{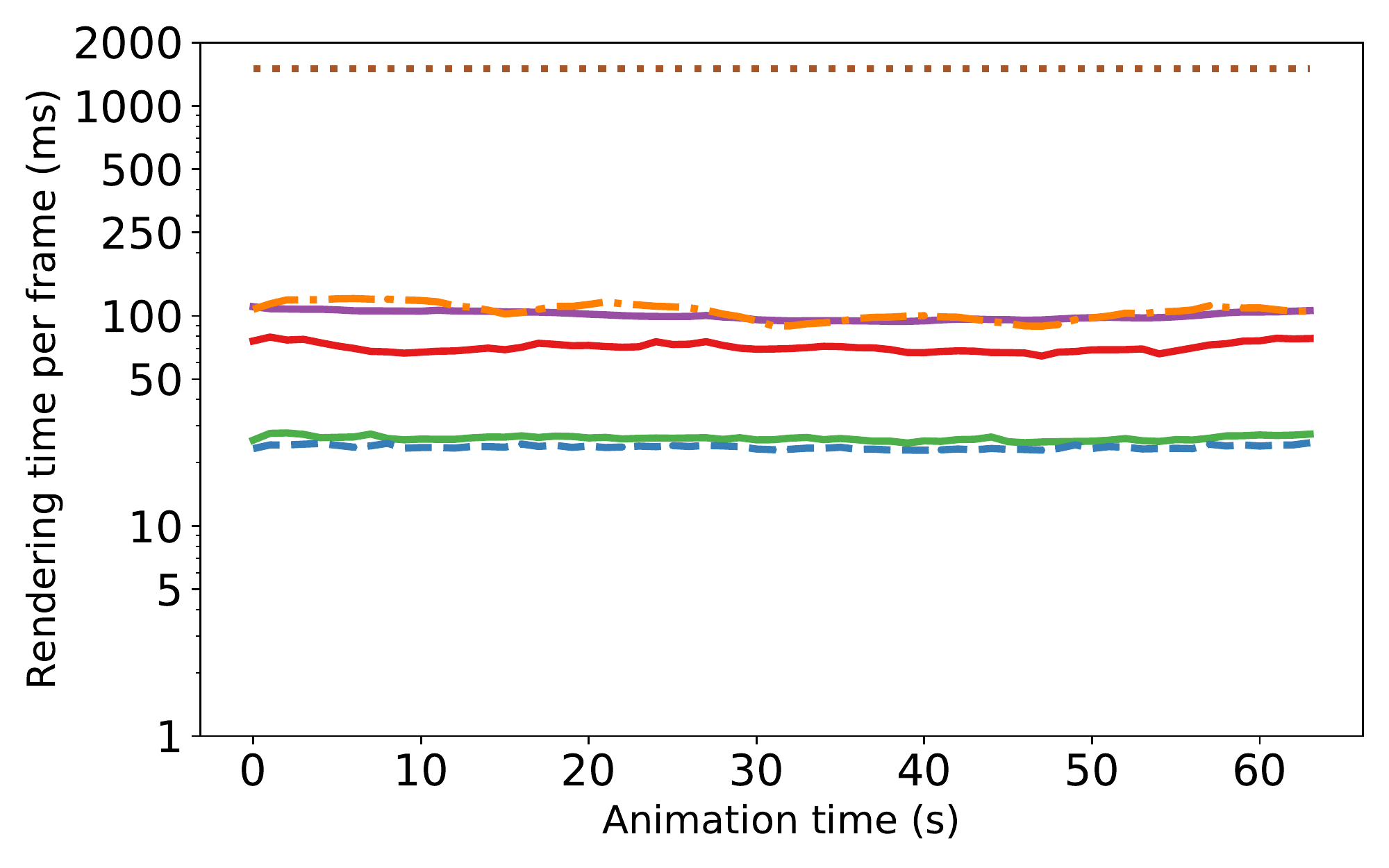}
		\put(34,15){\scriptsize\textbf{\textcolor{gray}{CONVECTION}}}
	\end{overpic}%
	\begin{overpic}[width=0.245\textwidth,clip,trim= 00mm 0mm 2mm 2mm]
		{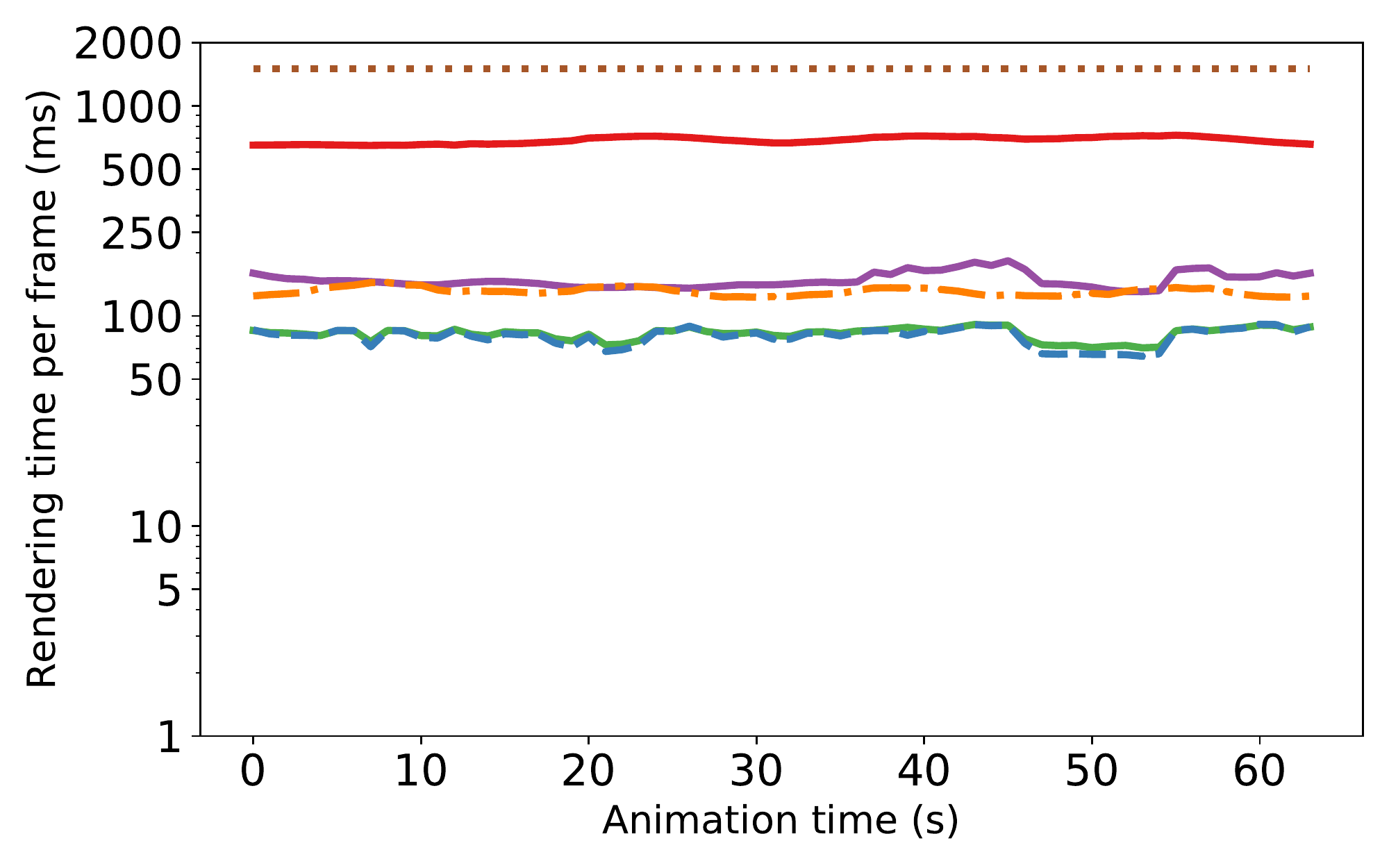}
		\put(34,15){\scriptsize\textbf{\textcolor{gray}{TURBULENCE}}}
	\end{overpic}%
	\begin{overpic}[width=0.245\textwidth,clip,trim= 00mm 0mm 2mm 2mm]
		{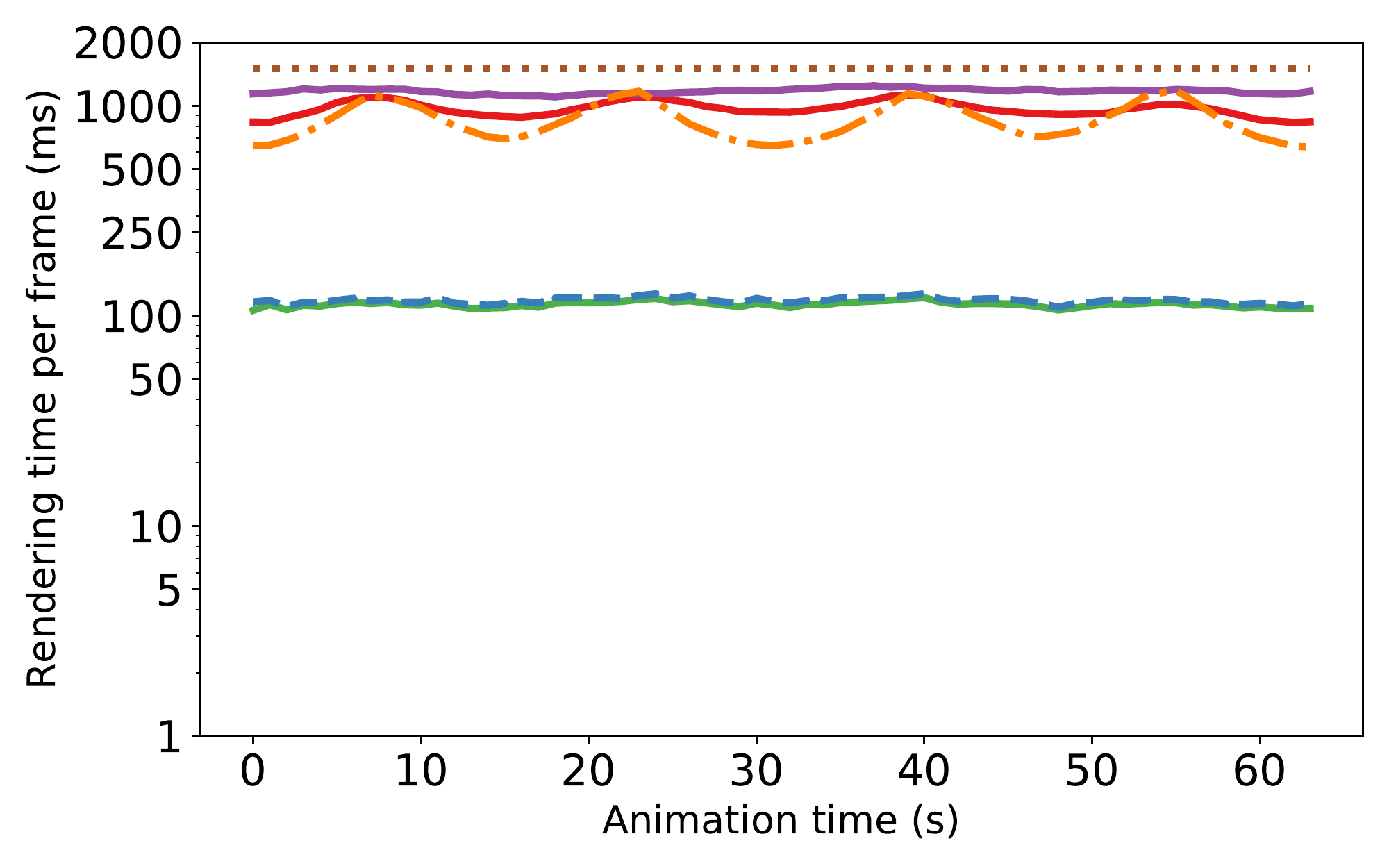}
		\put(42,15){\scriptsize\textbf{\textcolor{gray}{CLOUDS}}}
	\end{overpic}

	\caption{Rendering performance of all techniques using opaque lines (first row), transparent lines with opaque features of interest (middle row), and highly transparent lines (last row). The 1st, 2nd, 3rd, and 4th columns are for ANEURYSM, CONVECTION, TURBULENCE, and CLOUDS, respectively. For OSP, the line is dotted if rendering time exceeds 1500ms.}
	\label{fig:performance}
	\vspace{-1em}
\end{figure*}
Each data set was rendered three times along the pre-recorded flight paths, each time with different transparency settings and zoom levels. In this way, the dependencies between performance and the amount of transparency can be analyzed. We investigate the rendering of opaque lines, semi-transparent lines with an assignment of transparency that gives meaningful results, and lines with constant low transparency (e.g. below 0.15). Even though the latter setting, in general, does not produce meaningful results but mostly blurs out directional information, it is used to demonstrate how either technique behaves in this worst-case scenario. Performance measures are given in Fig.~\ref{fig:performance}. 

The rendering times of MLABDB and MBOIT are almost constant for different transparency settings since both techniques always render the entire data set and cannot exploit early-out strategies to skip fragments that are occluded by opaque ones. Rendering performance is mainly affected by the number of lines to be rendered and limited by the polygon throughput of the GPU.
Over time, depending on the orientation of lines, different amounts of fragments are generated, which results in slightly varying rendering times along the flight paths.  
In contrast, the performance of LL is highly dependent on the number of generated fragments. Although LL renders the entire line set once, similar to MLABDB and MBOIT, the generated fragments need to be stored and sorted. 
Thus, MLABDB and MBOIT render significantly faster than LL, especially for larger data sets. Over time, LL shows performance variations similar to those of MLABDB and MBOIT. 

Since image-order techniques can effectively employ early-ray termination, their performance depends strongly on the transparency setting. When highly opaque lines are rendered (see first column in Fig.~\ref{fig:performance}), the performance of RTX, OSP, and VRC is similar or even faster than that of the rasterization-based techniques. The performance of OSP is usually below that of RTX and VRC, as the CPU does not provide any hardware acceleration for ray tracing.

For opaque lines, CONVECTION seems to be an outlier regarding the relative performance differences between OSP and the other techniques.
OSP uses Embree's BVH builder, yet since a user geometry is used, Embree is not able to split the geometry to reduce the amount of overlap between BVH nodes. 
Densely overlapping data sets with long line segments will result in a poorer quality BVH, with more overlap between the nodes.
However, RTX uses a triangulated representation and can still perform these spatial splits.
For the CONVECTION, since the rolls are laid out in a flat sheet, OSP ends up traversing and intersecting a large amount of the BVH and most of the primitives in aggregate over the image. 
In the remaining data sets, the higher amount of occlusion helps reduce the amount of traversal needed. Especially on TURBULENCE and CLOUDS, even though there are a large number of lines, OSP only sees the box exterior for opaque lines and traverses very little of the data. 
Data sets with transparent lines do not benefit from the higher level of occlusion, and a large amount of the data must be traversed.

With increasing transparency, OSP falls behind the other techniques -- rendering required more than 2000ms and 6000-10000ms to complete for TURBULENCE and CLOUDS, respectively -- as the renderer must now traverse much further into the data.
Consequently, more tree-traversal operations and intersections tests need to be computed.
Although the same holds for RTX, its rendering times can compete with the performance of approximate techniques for line sets up to 100K, potentially due to the hardware acceleration provided for ray traversal and triangle intersection.
The worst-case scenario for OSP and RTX is the highly transparent case, where the majority of view rays have to be traversed until they leave the domain, due to the level of transparency. Thus, the performance drops significantly for large data sets.

It is interesting to note that LL, in many scenarios, can render very efficiently due to the GPU's capability to sort the fragments for many pixels in parallel. The more transparent the fragments are, and the less effectively image-order techniques can exploit early-ray termination, the better the relative performance of LL becomes.

The evaluation shows that, in particular for larger data sets, VRC renders slower than the approximate techniques. This performance difference is mainly due to the traversal of the voxel grid, which is not supported by an acceleration structure to enable empty space skipping, and sorting of multiple ray-tube intersections in the same voxel.  
The relative performance of VRC, on the other hand, is not much affected by increasing transparency. Although, in this case, many more intersection tests need to be performed, GPU-based voxel traversal can be performed very efficiently and does not impact performance as severely as BVH traversal.
VRC outperforms LL by about a factor of 4 and higher for large line sets such as TURBULENCE (cf. last column in Fig.~\ref{fig:performance}). Again, for ANEURYSM and CONVECTION with many empty regions that need to be traversed on the finest voxel level, the relative performance of VRC compared to the rasterization-based approaches decreases.
For CLOUDS, traversing large voxel grids ($512^3$) greatly reduces VRC's performance and leads to rendering times similar to LL when looking from a diagonal angle into the line set.

\subsection{Image Quality}
\label{sec:quality}
LL, VRC, OSP, and RTX simulate the effect of transparency accurately. VRC introduces errors due to the voxelization and quantization of lines into a regular voxel grid; however, the visibility order of lines in the grid is handled correctly. In the worst-case, multiple lines can fall on top of each other in the grid, resulting in an incorrect blending order. In our experiments, we did not perceive visual artifacts caused by this effect.

Inaccuracies in MLAB are caused by lines that are not rendered in correct visibility order, and that are merged heuristically using a limited number of transmittance layers. For scenes with high depth complexity, and even when low transparency is used, the accumulation of errors leads to visible artifacts. Most prominent are errors caused by incorrect merging of fragments with high opacity, i.e., when two such fragments are merged in the wrong order into the same transmittance layer  (Fig.~\ref{fig:teaser}(c) and Fig.~\ref{fig:mlab_error}). If lines are by chance rendered in the correct visibility order, or few opaque fragments are blended into different transmittance layers, MLAB can nevertheless generate accurate results (see Fig.~\ref{fig:teaser}(a) for an example). The view-dependent nature of MLAB, i.e., errors can suddenly appear or disappear depending on whether the rendering order matches the current visibility order, makes it less time coherent. 

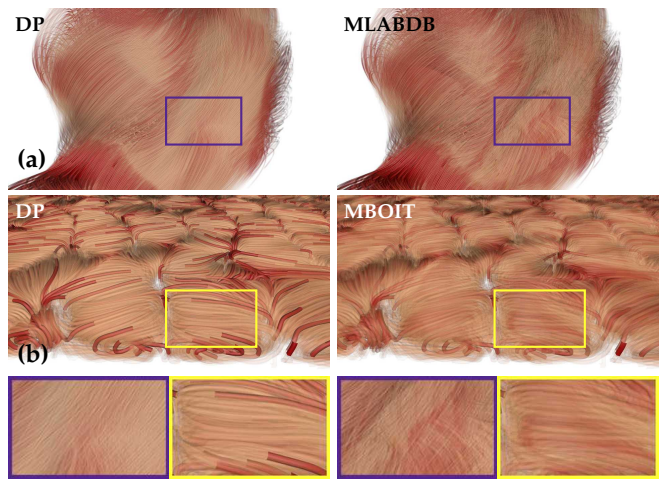
\begin{figure}[h!]
	\centering
	\begin{subfigure}{0.48\linewidth}
		\begin{tikzpicture}
		\node[anchor=south west,inner sep=0] at (0,0) {\includegraphics[width=1\textwidth]{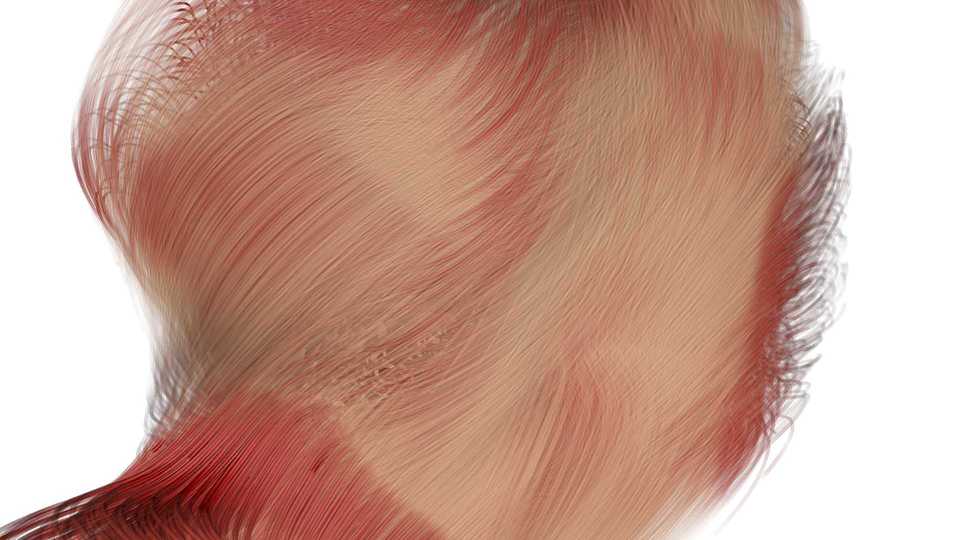}};
		\draw[MarkPurple,thick] (2.1,0.6) rectangle (3.1,1.225);
		\node[scale=0.8, anchor=west] at (0, 2.2) {\textbf{DP}};
		\node[scale=1, anchor=west] at (0, 0.4) {\textbf{(a)}};
		\end{tikzpicture}
		\vspace*{-9pt}
	\end{subfigure}%
	\hspace{1mm}%
	\begin{subfigure}{0.48\linewidth}
		\begin{tikzpicture}
		\node[anchor=south west,inner sep=0] at (0,0) {\includegraphics[width=1\textwidth]{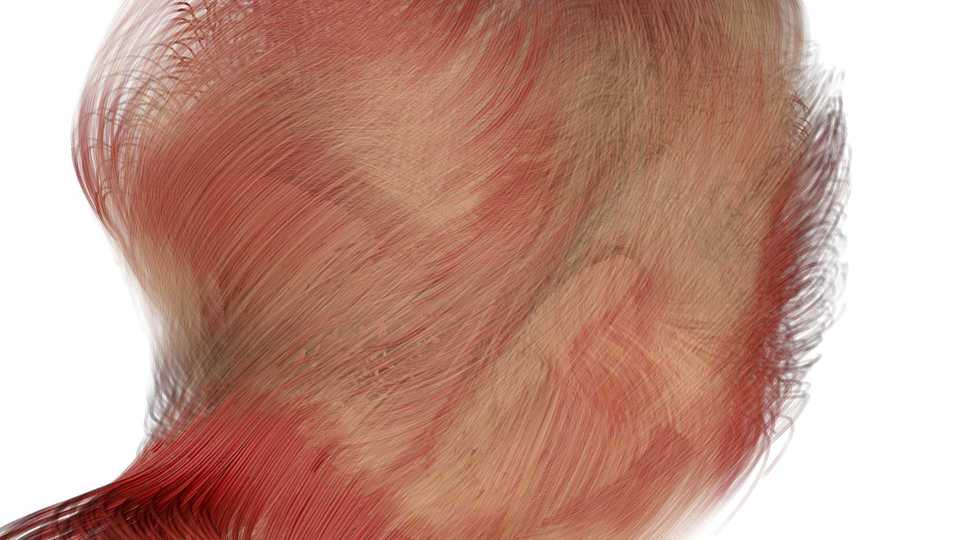}};
		\draw[MarkPurple,thick] (2.1,0.6) rectangle (3.1,1.225);
		\node[scale=0.8, anchor=west] at (0, 2.2) {\textbf{MLABDB}};
		\end{tikzpicture}
		\vspace*{-9pt}
	\end{subfigure}
	
	\begin{subfigure}{0.48\linewidth}
		\begin{tikzpicture}
		\node[anchor=south west,inner sep=0] at (0,0) {\includegraphics[width=1\textwidth]{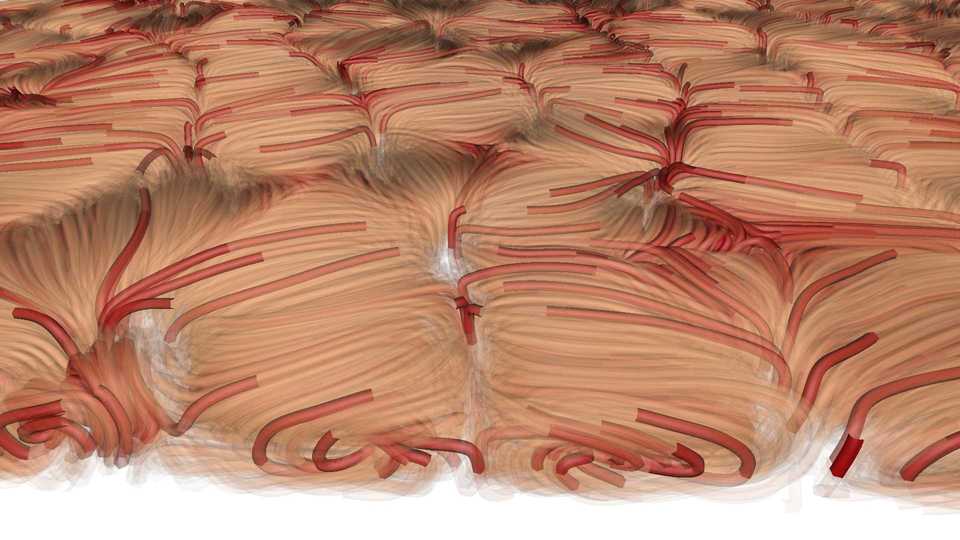}};
		\draw[MarkYellow,thick] (2.1,0.4) rectangle (3.3,1.15);
		\node[scale=0.8, anchor=west, text=white] at (0, 2.2) {\textbf{DP}};
		\node[scale=1, anchor=west] at (0, 0.3) {\textbf{(b)}};
		\end{tikzpicture}
		\includegraphics[width=0.5\linewidth]{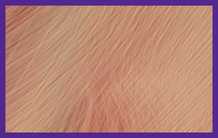}%
		\includegraphics[width=0.5\linewidth]{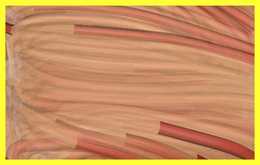}
		\caption*{\textbf{(a)} PSNR = 31.15, SSIM = 0.841}
	\end{subfigure}%
	\hspace{1mm}%
	\begin{subfigure}{0.48\linewidth}
		\begin{tikzpicture}
		\node[anchor=south west,inner sep=0] at (0,0) {\includegraphics[width=1\textwidth]{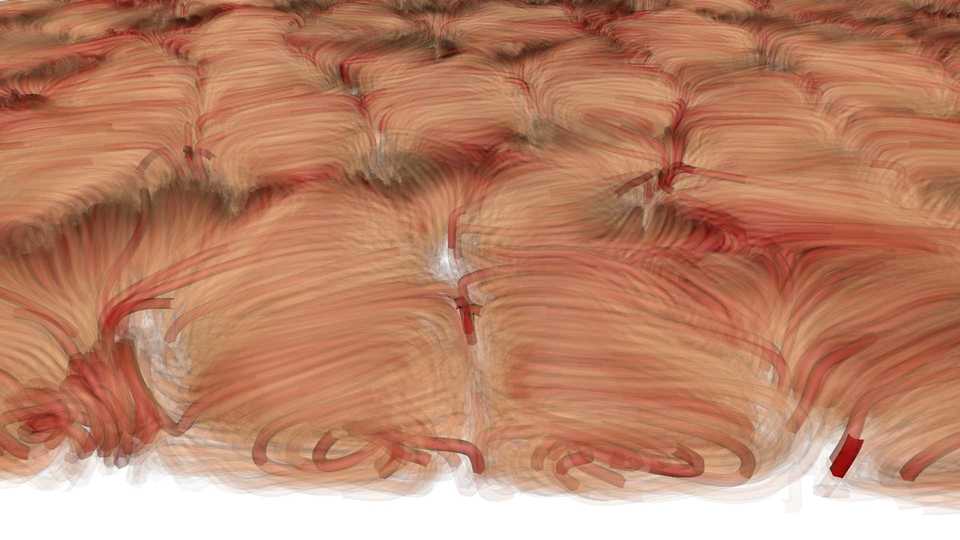}};
		\draw[MarkYellow,thick] (2.1,0.4) rectangle (3.3,1.15);
		\node[scale=0.8, anchor=west, text=white] at (0, 2.2) {\textbf{MBOIT}};
		\end{tikzpicture}
		\includegraphics[width=0.5\linewidth]{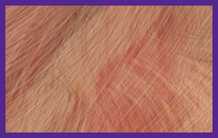}%
		\includegraphics[width=0.5\linewidth]{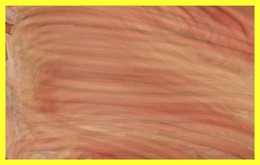}
		\caption*{\textbf{(b)} PSNR = 29.80, SSIM = 0.82}
	\end{subfigure}
	\caption{\textbf{(a)} Ground truth (left) and MLABDB (right).  MLABDB can produce hard visual  ``cuts'' due to depth segmentation. \textbf{(b)} Ground truth (left) and MBOIT (right). MBOIT tends to blur out features and underestimate opacity, erroneously revealing interior lines.}
	\label{fig:datasets_errors}
	\vspace{-1.5em}
\end{figure}

MLABDB can avoid the order-induced artifacts introduced by single MLAB when rendering opaque or nearly opaque lines. However, as mentioned in Sec.~\ref{sec:mlabdb}, thresholding has to be done carefully. In Fig.~\ref{fig:datasets_errors}(a), bucketing leads to hard cuts in color, revealing the depth segmentation of the line set. This artifact occurs primarily when using transfer functions with sudden opacity changes, which conflicts with MLABDB's assumption of many transparent layers occluding opaque ones. Even though these artifacts can be avoided by manually adapting the threshold for the front bucket according to the selected transfer function, this kind of user intervention is not practical in general.  

MBOIT replaces the transmittance function along a line of sight by a low-frequency approximation. Thus, two major types of artifacts can occur: First, as shown in Fig.~\ref{fig:teaser}(b) and Fig.~\ref{fig:datasets_errors}(b), the accumulated opacity of multiple transparent lines is either highly overestimated or underestimated. These over- and under-estimations can lead to misinterpretations of the visualization, as translucent or opaque lines can appear prominent or be missing in the final image.
These errors are due to sudden changes in opacity when mapping the importance of features with step-functions, meaning that MBOIT cannot accurately handle hard transitions in the mapped opacity.
Second, since a low-frequency approximation is used, MBOIT tends to smooth out the transmittance distribution across the pixel image. As shown in Fig.~\ref{fig:teaser}(d) and in Fig.~\ref{fig:datasets_errors}(b), sharp edges between lines with higher opacity are not preserved. This effect, in particular, can hamper a more detailed analysis of the directional structure of important lines, and it tends to smooth out the directional information in important regions.

%
%
\subsection{Quantitative Assessment}
\begin{figure*}[!htb]	
	\centering
	\begin{overpic}[width=0.245\textwidth,clip,trim=0mm 10mm 0mm 10mm]
		{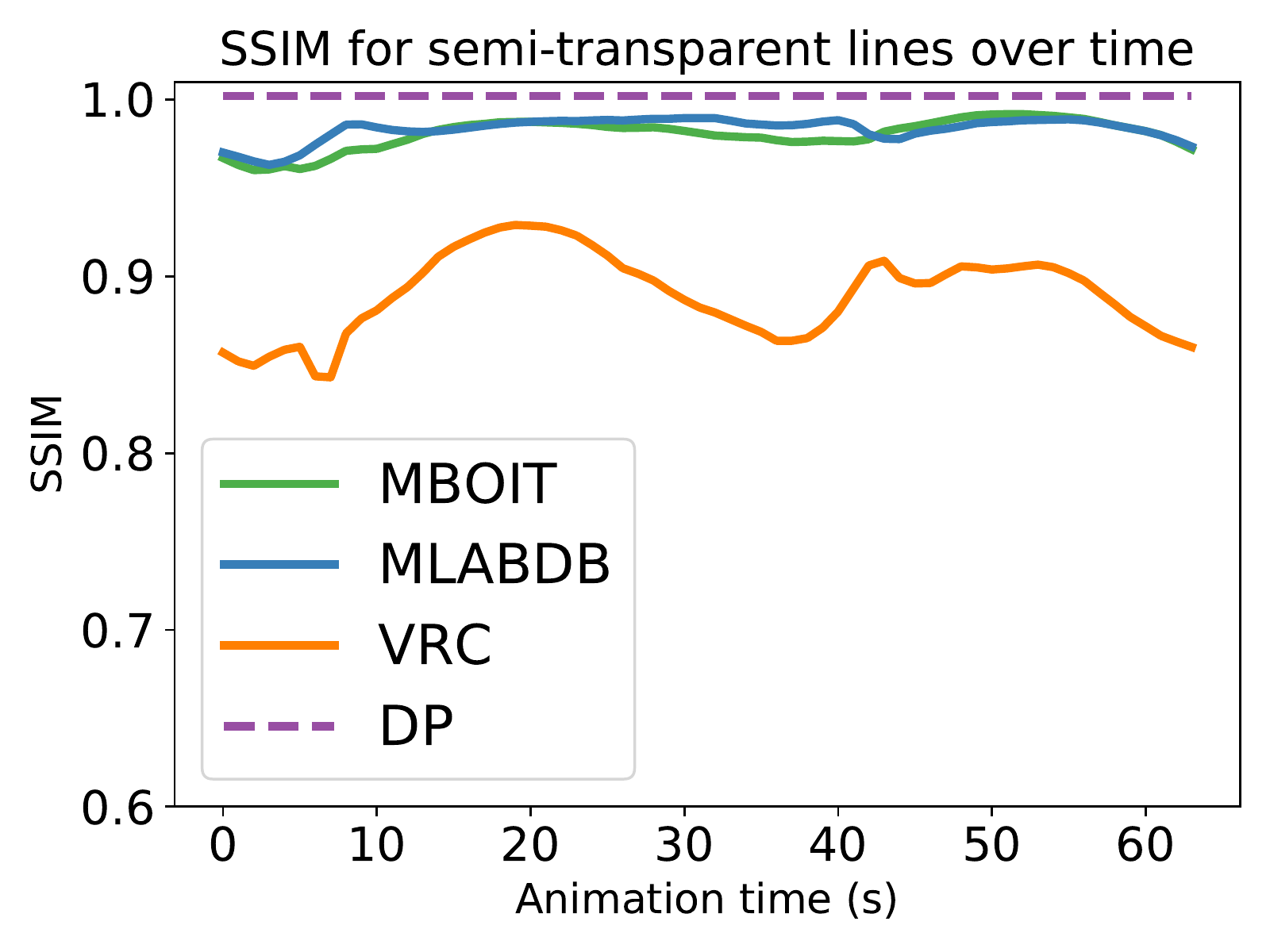}
	\end{overpic}%
	\begin{overpic}[width=0.245\textwidth,clip,trim=0mm 10mm 0mm 10mm]
		{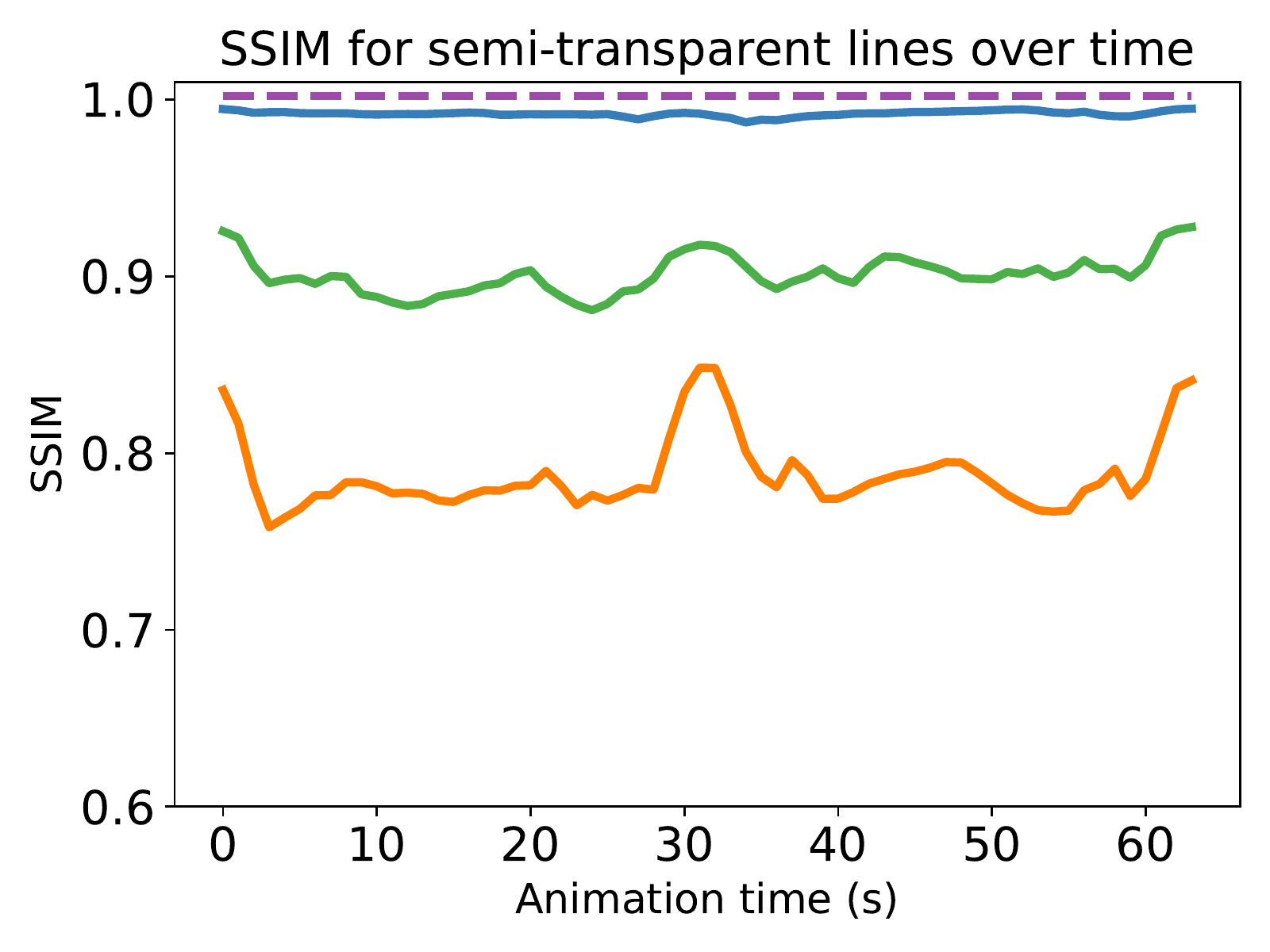}
	\end{overpic}%
	\begin{overpic}[width=0.245\textwidth,clip,trim=0mm 10mm 0mm 10mm]
		{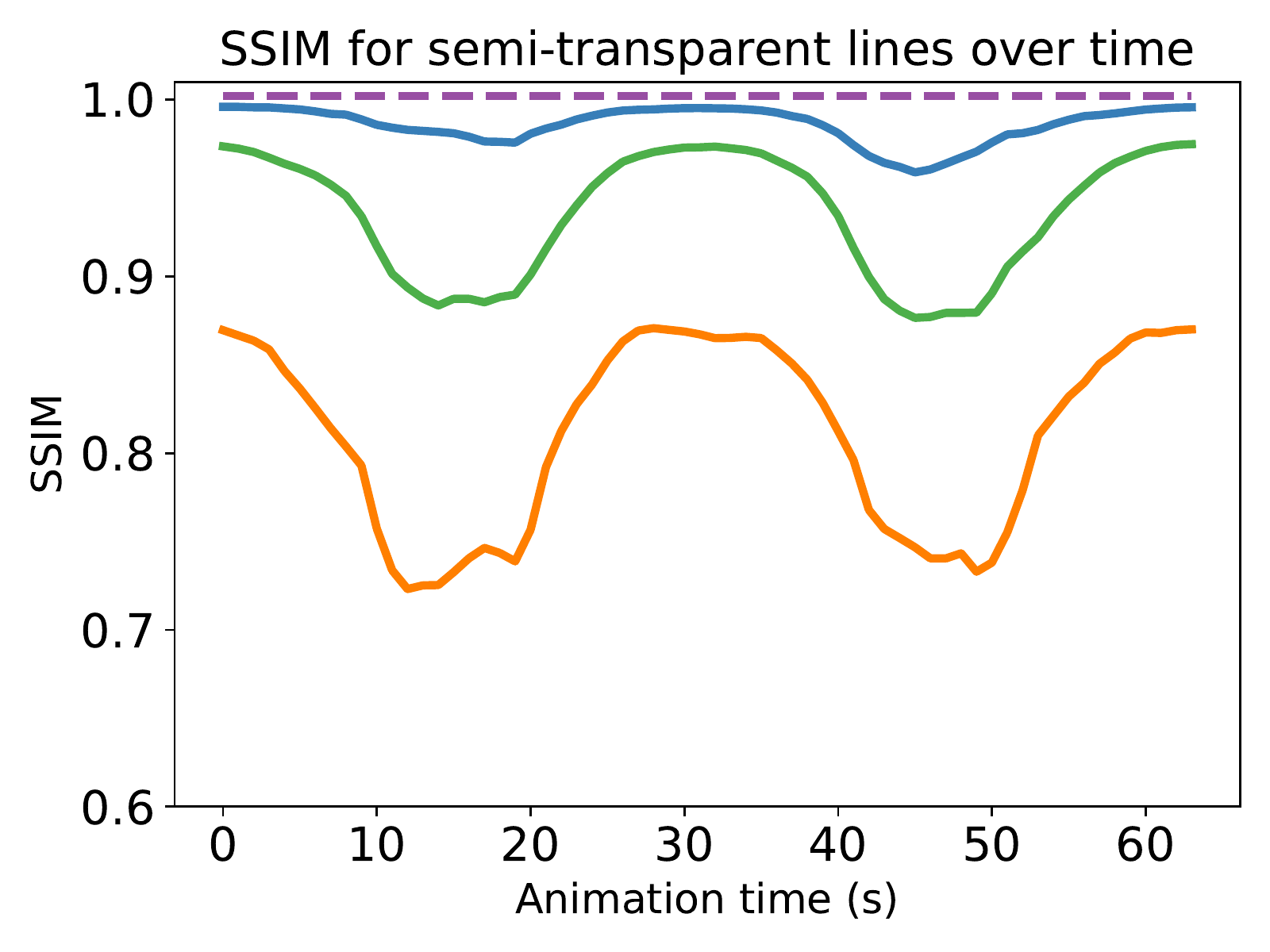}
	\end{overpic}%
	\begin{overpic}[width=0.245\textwidth,clip,trim=0mm 10mm 0mm 10mm]
		{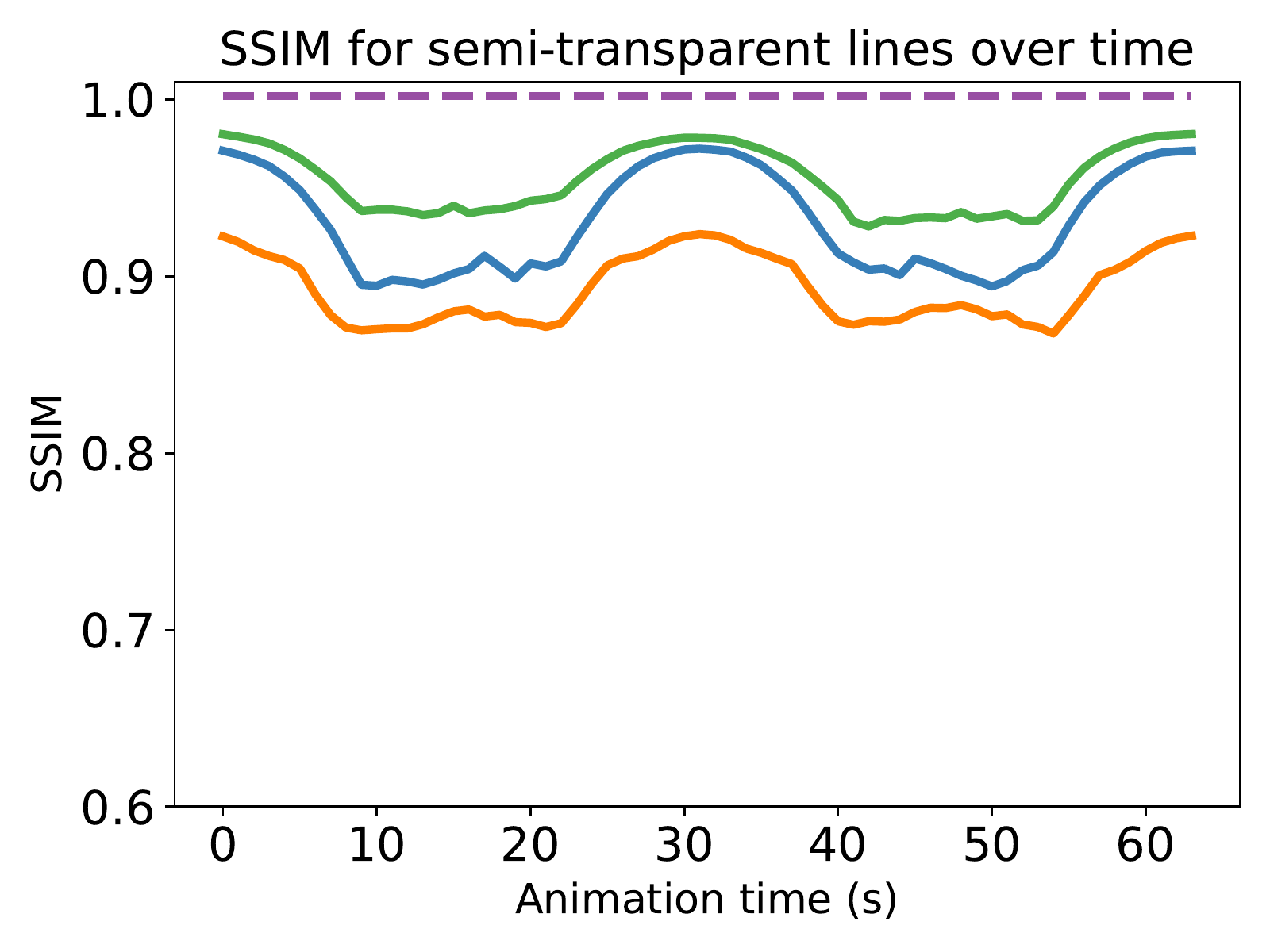}
	\end{overpic}
	\begin{overpic}[width=0.245\textwidth,clip,trim=0mm 0mm 0mm 10mm]
		{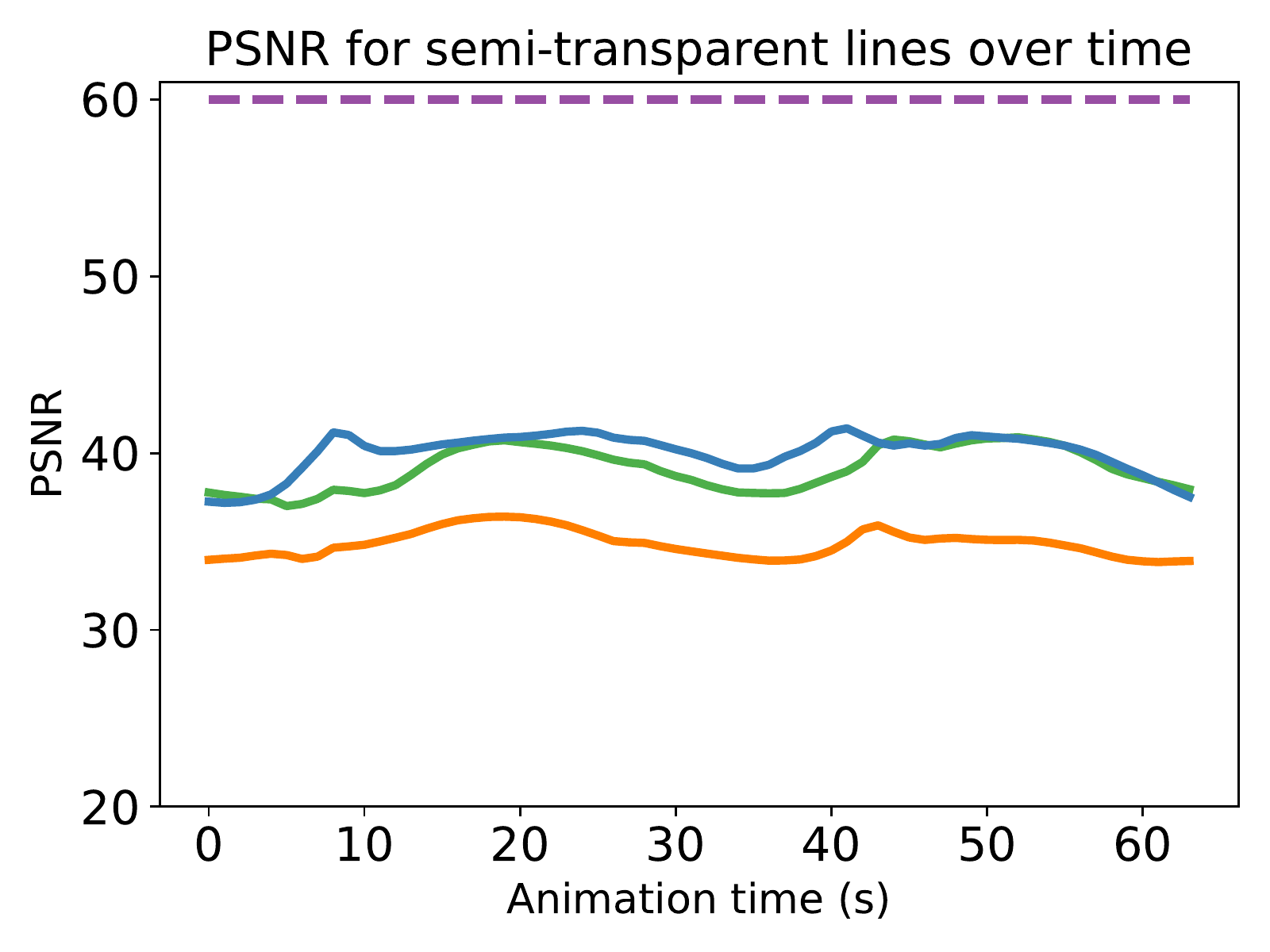}
		\put(38,18){\scriptsize\textbf{\textcolor{gray}{ANEURYSM}}}
	\end{overpic}%
	\begin{overpic}[width=0.245\textwidth,clip,trim=0mm 0mm 0mm 10mm]
		{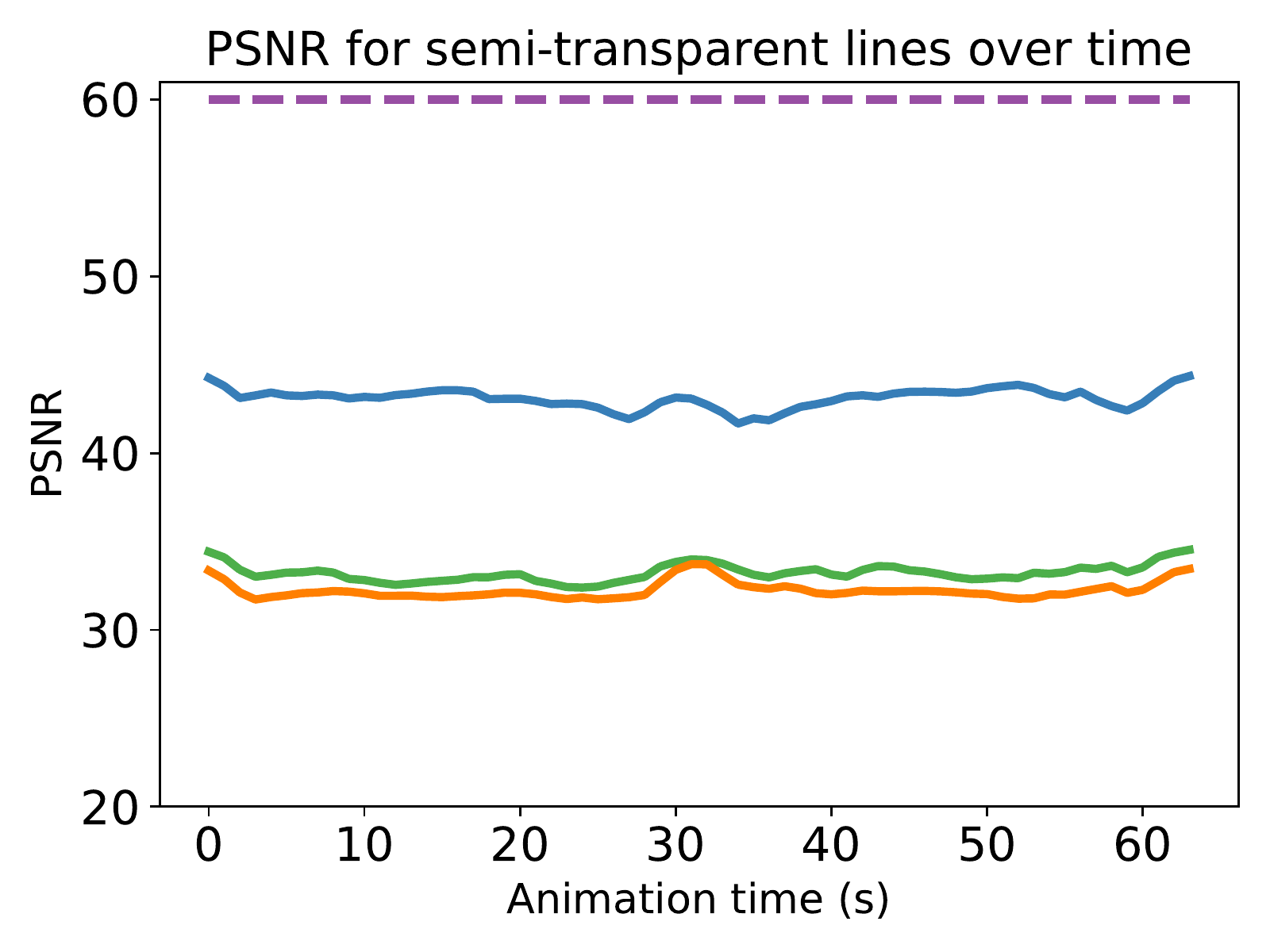}
		\put(34,18){\scriptsize\textbf{\textcolor{gray}{CONVECTION}}}
	\end{overpic}%
	\begin{overpic}[width=0.245\textwidth,clip,trim=0mm 0mm 0mm 10mm]
		{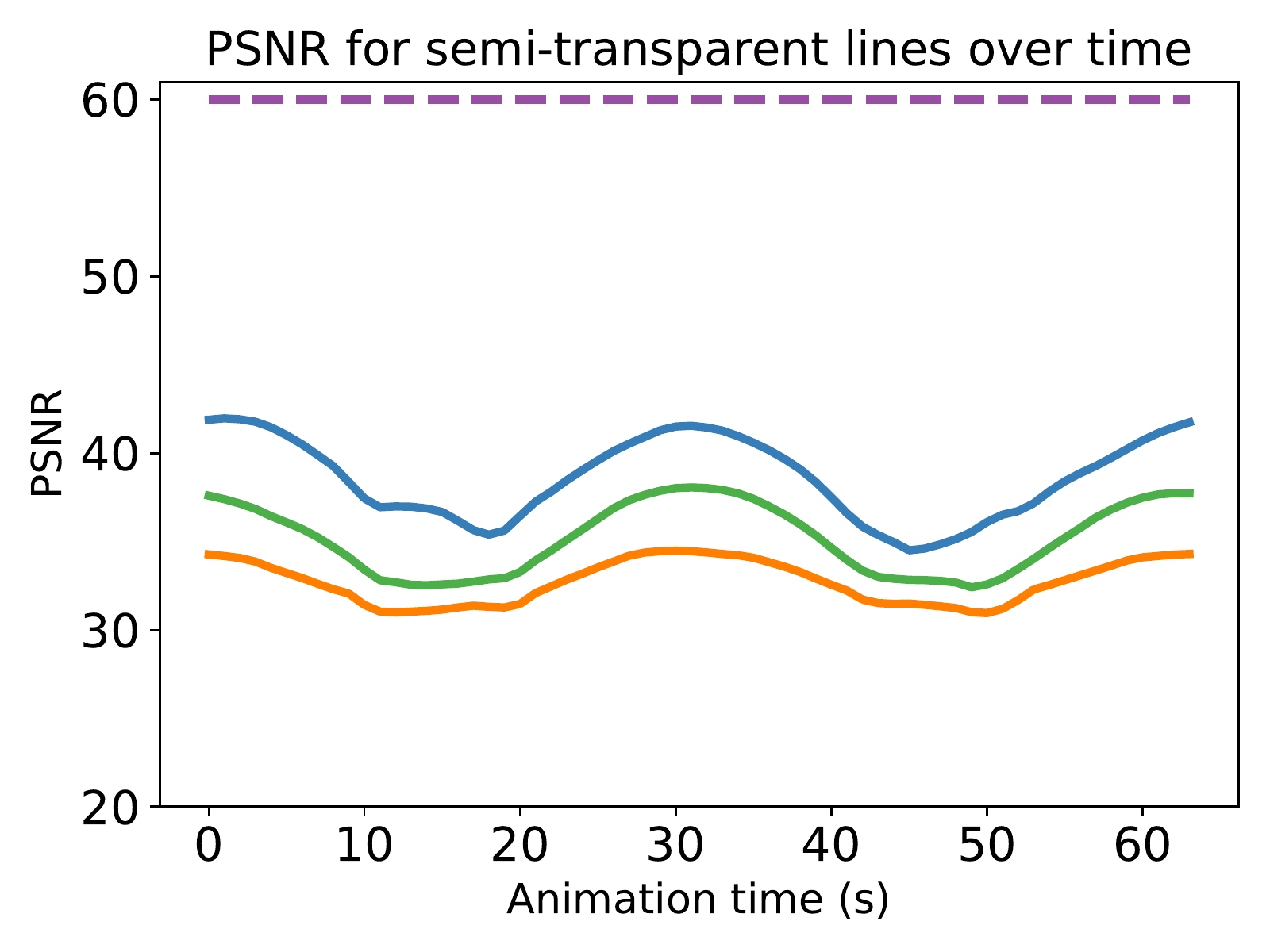}
		\put(36,18){\scriptsize\textbf{\textcolor{gray}{TURBULENCE}}}
	\end{overpic}%
	\begin{overpic}[width=0.245\textwidth,clip,trim=0mm 0mm 0mm 10mm]
		{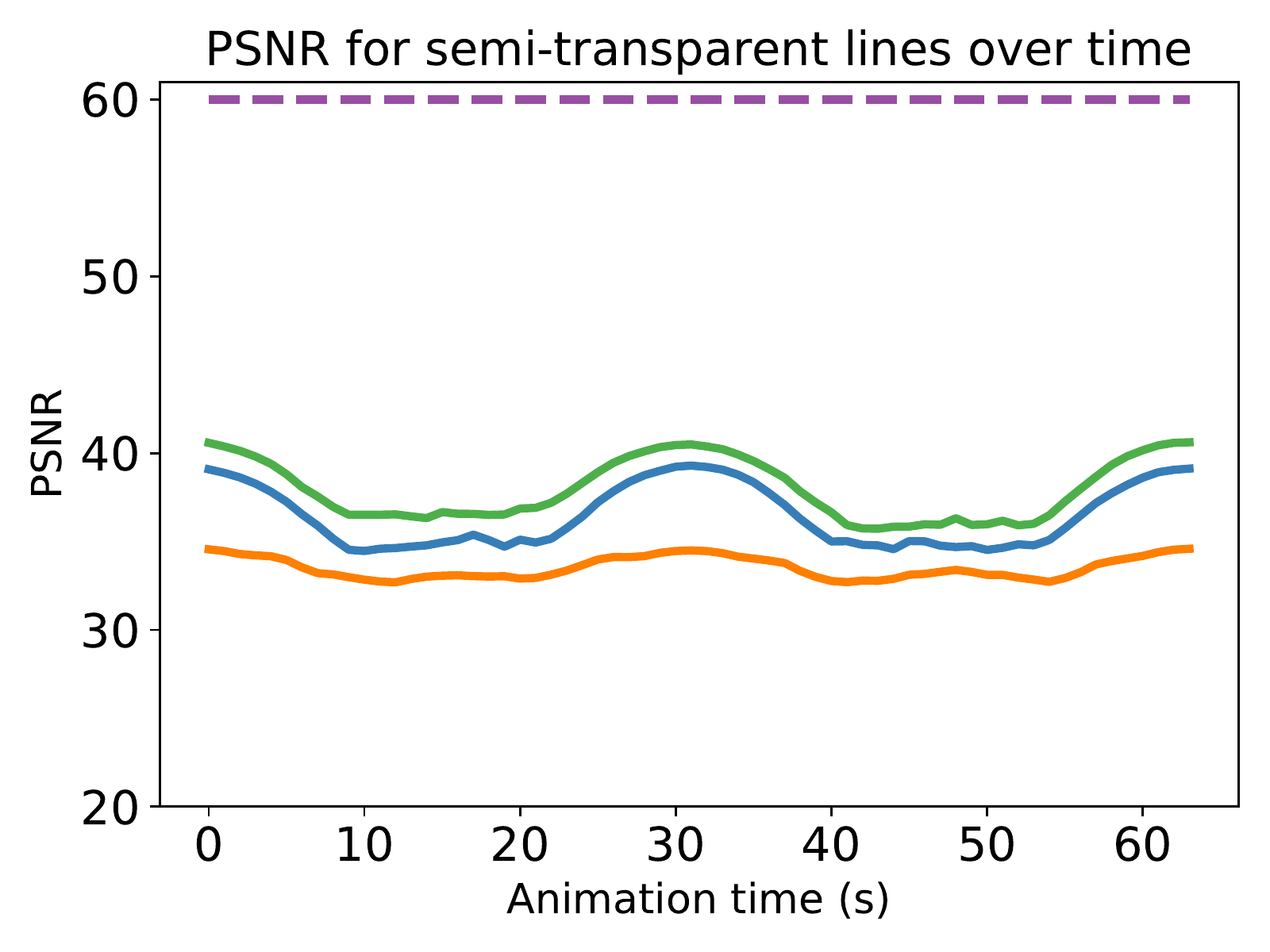}
		\put(42,18){\scriptsize\textbf{\textcolor{gray}{CLOUDS}}}
	\end{overpic}
	\caption{Error metrics of all techniques using transparent lines with opaque features of interest. The 1st, 2nd, 3rd, and 4th columns are for ANEURYSM, CONVECTION, TURBULENCE, and CLOUDS, respectively. The First row shows SSIM for each technique, the second shows PSNR. A higher value is better.}
	\label{fig:quality}
	\vspace{-1em}
\end{figure*}

To further quantify the error that is introduced by the different line rendering techniques, all data sets are rendered along the pre-recorded flight paths using the transparency settings described before.
Lines are illuminated by a headlight and colored via Blinn-Phong shading. 
For each image, the Peak-Signal-To-Noise Ratio (PSNR) and Structural Similarity Index (SSIM)~\cite{Wang2004} are computed between the ground truth rendering using DP, and MLABDB, MBOIT, and VRC, respectively. 
We do not consider LL, RTX, and OSP, since they correctly simulate transparency. 
For each setting, the accuracy measurements are plotted as line graphs over time.
Renderings using semi-transparent settings are shown in Fig.~\ref{fig:quality}, all other settings are shown in Fig.~\ref{fig:quality_appendix}.
An in-detail discussion and all results are provided in Appendix~\ref{app:quantitative}.

\subsection{Visual Quality vs. Per-Pixel Error}
\label{sec:error_cases}
Interestingly, when looking at images where the SSIM and PSNR values show a lower quality of VRC and higher quality of MBOIT and MLABDB, these differences are not reflected in the visual quality of the results. 
Since transparency is handled correctly by VRC, even the close-up views appear similar to the ground truth, and even visible artifacts introduced by the alternative methods do not manifest. 

In this section, we analyze in further detail the relations between visual image quality and per-pixel error. 
For each data set, we analyze two views: one view where all techniques come visually close to the ground truth while producing only a small number of pixel-wise errors (case A), and one that reveals typical artifacts of MBOIT and MLABDB using a most meaningful transparency settings with sharp transitions and alternating high transparency and opacity (case B).

Besides an image-to-image comparison, the analysis is additionally supported by visualizations of the absolute per-pixel differences to the ground truth.
These plots are grayscale images with black regions highlighting large color differences.
Significant differences in images are marked by colored rectangles and supported by close-up views.
Image comparisons of all cases (Fig. A.x) are given in Appendix~\ref{app:quantitative}.

Fig.~\ref{fig:error_aneursym_appendix} depicts a scenario where the viewer is looking through the entire ANEURYSM data set with alternating opaque lines (red-colored) and transparent lines (orange-ocher).
For case A (cf. Fig.~\ref{fig:error_aneursym_appendix}(a), corresponding to frame 5 of the first column in Fig.~\ref{fig:quality}), we observe that all techniques are close to the ground truth image with the best image produced by MLABDB.
MBOIT overestimates the transparency of few transparent fragments occluding opaque lines highlighted by per-pixel error plots (cf. black regions in Fig.~\ref{fig:error_aneursym_appendix}(a) (MBOIT)).
VRC produces high pixel errors in regions close to the viewer since line inaccuracies affect larger areas of pixels.
However, the quality of VRC becomes better with larger distance to the camera, leading to results indistinguishable from the ground truth.
Wrt. case B (cf. Fig.~\ref{fig:error_aneursym_appendix}(b)), the quality of both MBOIT and MLABDB is worse than VRC.
MBOIT struggles to approximate sharp transitions in transparency leading to high over- and underestimations, whereas MLABDB is not able to correctly merge opaque fragments. Bucketing is impossible here, leading to visual artifacts.
However, VRC remains stable and, besides line inaccuracies, is very close to the ground truth.

\begin{figure*}[!ht]	
	\centering
	\begin{subfigure}{0.245\linewidth}
		\setlength{\lineskip}{0pt}
		\begin{tikzpicture}
		\node[anchor=south west,inner sep=0] at (0,0) {\includegraphics[width=1\linewidth]{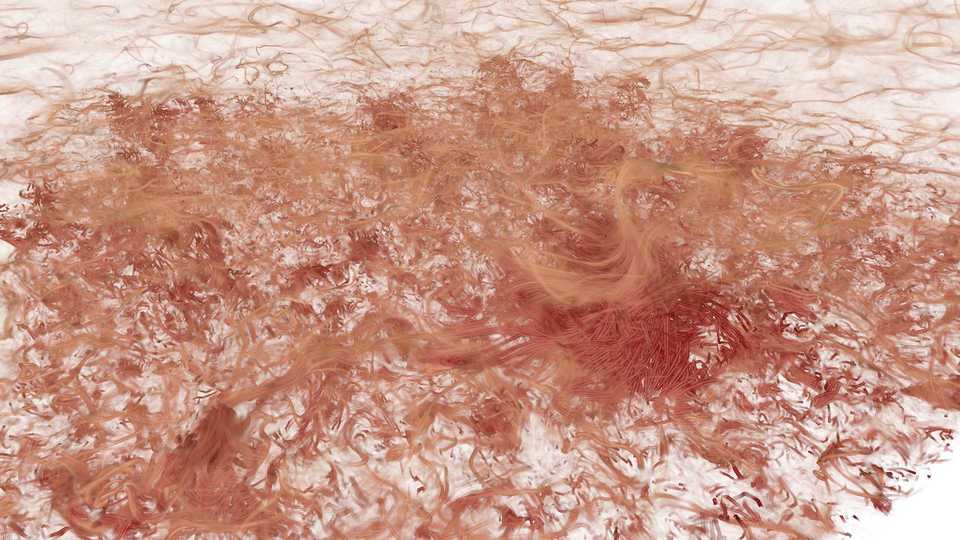}};
		\draw[MarkPurple,thick] (0.5,1.3) rectangle (1.3,1.8);
		\draw[MarkYellow,thick] (2.5,0.8) rectangle (3.3,1.3);
		\node[scale=0.8, anchor=west] at (0, 2.25) {\textbf{DP}};
		\end{tikzpicture}
		\includegraphics[width=0.499\linewidth]{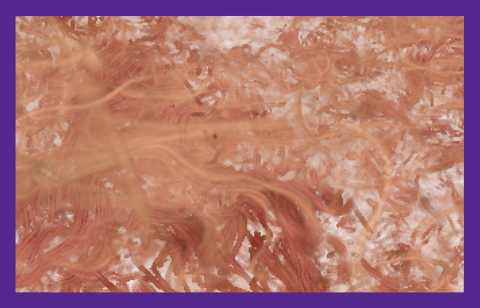}%
		\includegraphics[width=0.499\linewidth]{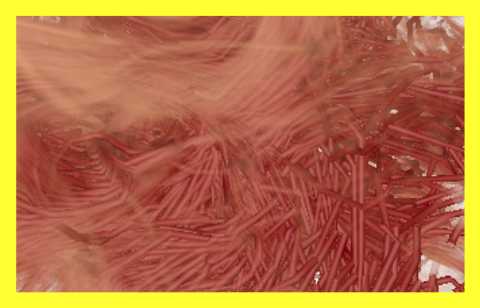}
		\caption*{}
	\end{subfigure}%
	\hspace{1mm}%
	\begin{subfigure}{0.245\linewidth}
		\setlength{\lineskip}{0pt}
		\begin{tikzpicture}
		\node[anchor=south west,inner sep=0] at (0,0) {\includegraphics[width=1\linewidth]{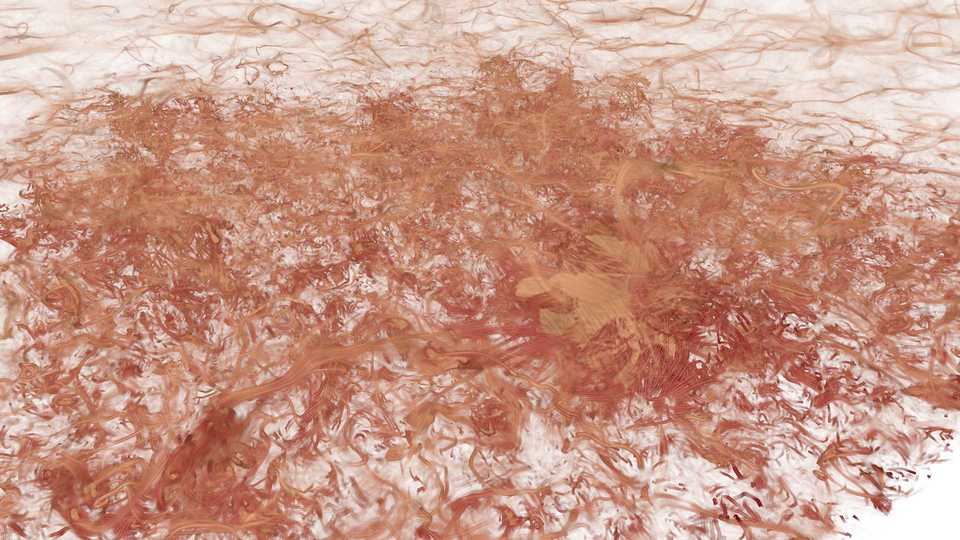}};
		\draw[MarkPurple,thick] (0.5,1.3) rectangle (1.3,1.8);
		\draw[MarkYellow,thick] (2.5,0.8) rectangle (3.3,1.3);
		\node[scale=0.8, anchor=west] at (0, 2.25) {\textbf{MBOIT}};
		\end{tikzpicture}
		\includegraphics[width=0.499\linewidth]{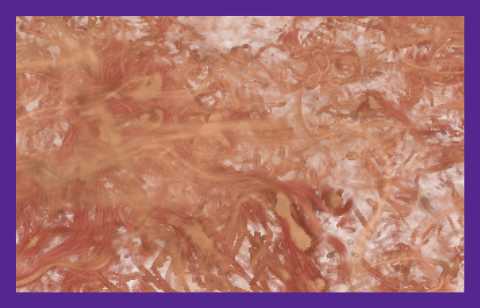}%
		\includegraphics[width=0.499\linewidth]{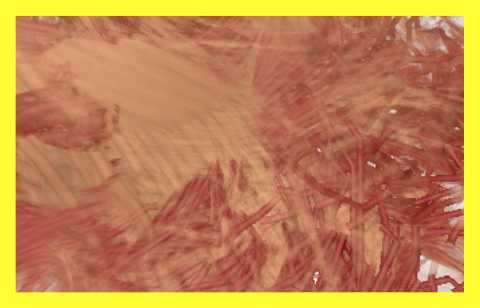}
		\caption*{PSNR = 34.71, SSIM = 0.920}
	\end{subfigure}%
	\hspace{1mm}%
	\begin{subfigure}{0.245\linewidth}
		\setlength{\lineskip}{0pt}
		\begin{tikzpicture}
		\node[anchor=south west,inner sep=0] at (0,0) {\includegraphics[width=1\linewidth]{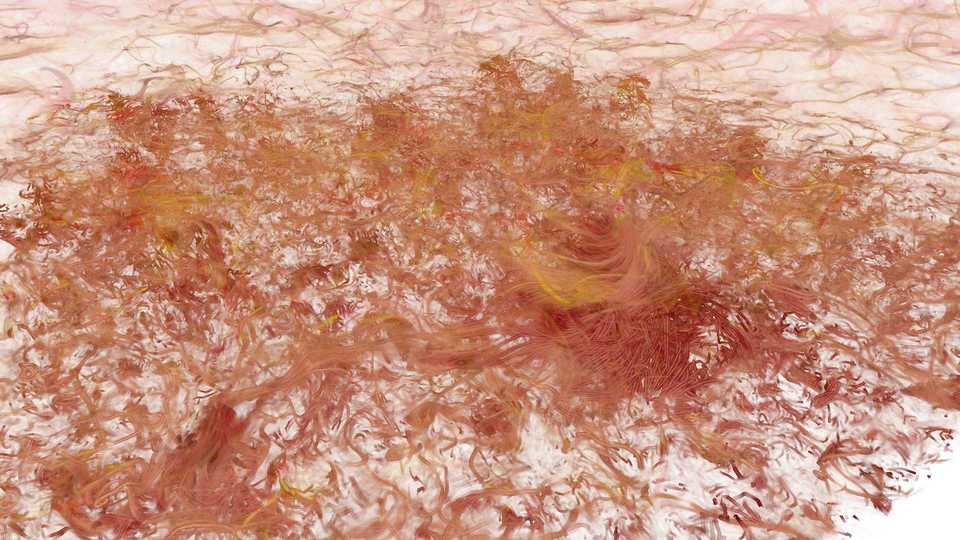}};
		\draw[MarkPurple,thick] (0.5,1.3) rectangle (1.3,1.8);
		\draw[MarkYellow,thick] (2.5,0.8) rectangle (3.3,1.3);
		\node[scale=0.8, anchor=west] at (0, 2.25) {\textbf{MLABDB}};
		\end{tikzpicture}
		\includegraphics[width=0.499\linewidth]{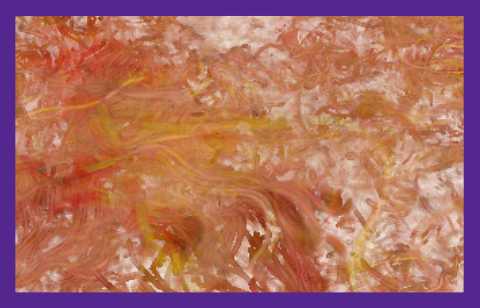}%
		\includegraphics[width=0.499\linewidth]{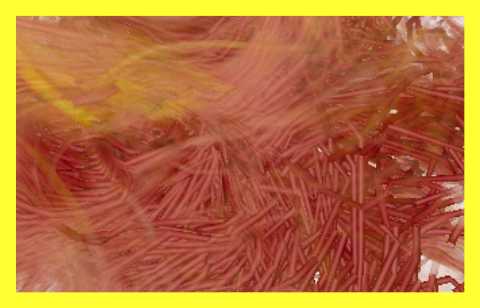}
		\caption*{PSNR = 32.11, SSIM = 0.940}
	\end{subfigure}%
	\hspace{1mm}%
	\begin{subfigure}{0.245\linewidth}
		\setlength{\lineskip}{0pt}
		\begin{tikzpicture}
		\node[anchor=south west,inner sep=0] at (0,0) {\includegraphics[width=1\linewidth]{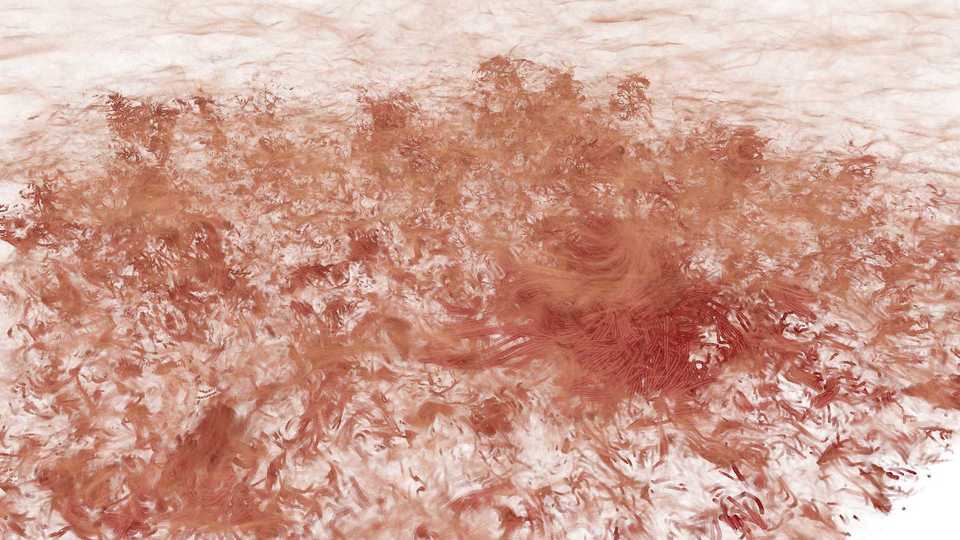}};
		\draw[MarkPurple,thick] (0.5,1.3) rectangle (1.3,1.8);
		\draw[MarkYellow,thick] (2.5,0.8) rectangle (3.3,1.3);
		\node[scale=0.8, anchor=west] at (0, 2.25) {\textbf{VRC}};
		\end{tikzpicture}
		\includegraphics[width=0.499\linewidth]{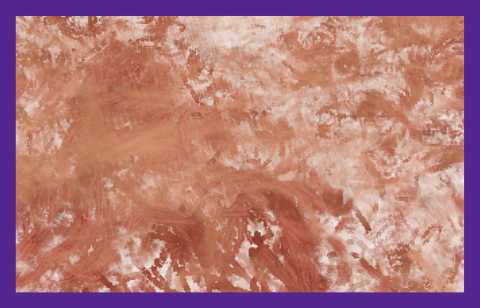}%
		\includegraphics[width=0.499\linewidth]{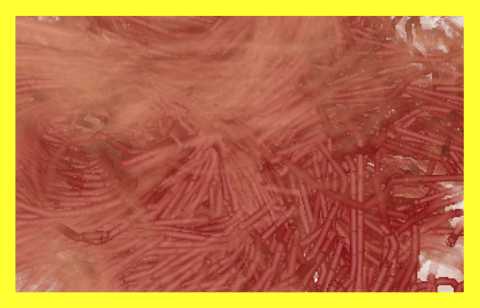}
		\caption*{PSNR = 30.27, SSIM = 0.750}
	\end{subfigure}
	\caption{From left to right: CLOUDS rendered with DP, MBOIT, MLABDB, VRC. 
	Images show renderings using sharp transitions between low and high transparency.
	Blue and yellow rectangles highlight differences between techniques; close-up views are shown below each image.}
	\label{fig:error_clouds}
	\vspace{-1em}
\end{figure*}
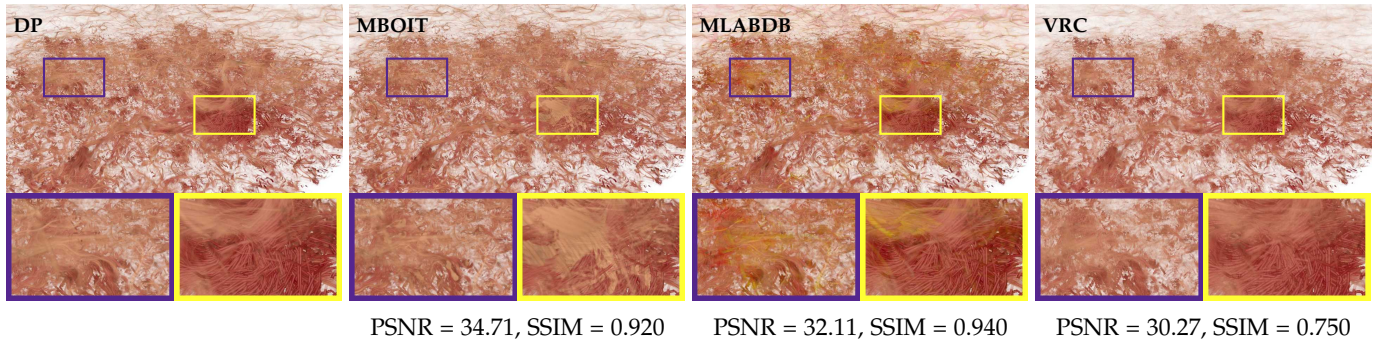

Another example is given in Fig.~\ref{fig:error_convectionrolls_appendix} when looking from a diagonal angle into the entire CONVECTION line set, with the number of transparent lines increasing with the distance to the viewer.
For case A (cf. Fig.~\ref{fig:error_convectionrolls_appendix}(a)), all techniques are visually close to the ground truth with MLABDB working best.
Here, MBOIT exhibits small errors in transmittance approximation, as many opaque lines are occluded by transparent ones.
These errors propagate towards the background as the number of fragments increases with distance, leading to further image quality degradation.
With opaque lines more present in this case, VRC produces a number of wrong line silhouettes due to curve discretization, emphasized in per-pixel error plots along the line edges.
For case B, the quality of both MLABDB and MBOIT decreases with larger distance (cf. Fig.~\ref{fig:error_convectionrolls_appendix}(b)).
In particular, MLABDB produces more per-pixel color inconsistencies, depicted by high noise in error plots, toward the background as more and more fragments are merged incorrectly.
On the other hand, MBOIT has difficulty coping with sharp transitions between transparent and opaque fragments.
Interestingly, the image quality of VRC is independent of the viewer's distance or angle, and line inaccuracies do not accumulate with increasing distance.

Fig.~\ref{fig:error_turbulence_appendix} demonstrates the differences of image quality for zoom-out and close-up views.
Here, case A (cf. Fig.~\ref{fig:error_turbulence_appendix}(a)) represents a zoom-out view that corresponds to frame eight of the third column in Fig.~\ref{fig:quality}.
All techniques are able to properly render TURBULENCE and are visually indistinguishable from the ground truth, although MLABDB shows some weaknesses in rendering transparent regions due to incorrect fragment merges. 
However, these pixel errors hardly affect the overall quality of the image.
Wrt. VRC, line inaccuracies are not present here since lines are highly transparent and line edges are not emphasized. 
Case B (cf. Fig.~\ref{fig:error_turbulence_appendix}(b)) shows the impact of zoom-ins on the quality of all techniques.
MLABDB properly renders opaque lines (red-colored tubes), but incorrectly merges fragments in regions with a large number of highly transparent lines, leading to a wrong colored region (orange instead of ocher) after blending.
MBOIT is able to approximate transmittance in transparent regions but fails to display sharp opaque lines where opaque and transparent lines are close by, leading to blurred outline structures in the final image. 
Per-pixel error plots for VRC reveal some line inaccuracies but demonstrate that, overall, a good image quality is achieved by VRC even for this large data set.

The last example demonstrates the impact of large, dense line data with many layers per pixel ($>10000$ at maximum), using high transparency in Fig.~\ref{fig:error_clouds_appendix}(a) and semi-transparent opacity settings in Fig.~\ref{fig:error_clouds} and Fig.~\ref{fig:error_clouds_appendix}(b).
In the first scenario, all techniques are able to properly render the data set.
However, MLABDB produces a few wrongly colored features due to false fragment merging. MBOIT works properly for this case, and produces few over- or underestimations. VRC, on the other hand, tends to underestimate the actual transmittance as some lines fall into the same cell (disadvantage of VRC explained in Sec.~\ref{sec:vrc}).
Weaknesses of all techniques are even more pronounced in the second scenario; for example, there are more wrongly colored features produced by MLABDB. Due to its line approximation, VRC also struggles to fully reconstruct the actual transmittance. 
MBOIT works best for CLOUDS, as there are only a few overestimates of the actual transmittance.

Since MLABDB and MBOIT do not depend on any primitive type, we also investigated the image quality of our proposed approximate techniques when rendering transparent triangle surfaces and point cloud data. The visualization of these types of data with transparency can, similar to line sets, lead to complex rendering scenarios with many transparent layers and multiple occlusions.
However, for those data types, MLABDB and MBOIT showed similar characteristics as for line sets during rendering (see Appendix~\ref{app:isosurface_points}).

In summary, although all techniques show weaknesses some weaknesses in some cases resulting in pixel-errors, they are able to render transparent data sets with high depth complexity at high image quality. 
Moreover, OSP, RTX, and VRC are temporally stable for all transparency settings and data sets. VRC, despite line inaccuracies, comes very close to the ground truth.
\vspace{-0.5em}
\section{Discussion}
In the following sections, we discuss the major characteristics of all rendering techniques, as given in Tab.~\ref{tab:comparison} and also present the outcome of an informal user study to shed light on the perception of the errors that are introduced by object-order approximate techniques.
\subsection{Object-Order}
Equipped with dedicated GPU-friendly sorting algorithms and data structures, LL shows good rendering performance for all but the largest data sets.
LL was never slower than a factor of 4-5 compared to approximate techniques for small data sets. 
For dense data sets with high depth complexity of more than a thousand layers, however, the required GPU buffers can easily exceed available GPU memory, especially for resolutions above 1080p.

Approximate rasterization-based techniques are very fast, work with rendering constraints, and support rendering on hardware with bounded memory.
Rendering constraints for MLABDB and MBOIT involve, for each pixel, the maximum number of fragments stored or model parameters to approximate the transmittance function. These constraints implicitly keep the memory consumption constant over time.

MLABDB and MBOIT provide good quality in many real-world scenarios, particularly in scenes where features of interest are rendered opaque and remaining lines are mapped to high transparency. 
Introduced artifacts are often subtle and local, yet in some views they can cause artifacts that even give a wrong understanding of the line structures. 
Approximate techniques also fail to maintain time coherence, as their per-pixel rendering outcome is dependent on the transmittance function along each pixel.
These errors are especially frequent if high-frequency transfer functions for transparency are used, which can lead to distracting rendering artifacts.
Although approximation errors can be reduced by, i.e., increasing the number of nodes of MLABDB's blending arrays (in example to more than 15 layers) or the number of power moments, such settings considerably reduce rendering performance and increase memory consumption.
\begin{table*}[!ht]
	\caption{Comparison table of all techniques wrt. type of order, render accuracy, performance, render constraints (RenderC), support of bound memory (BoundM), temporal coherency (Temp), changes in color along each pixel (ChColor), dense (large) data sets (Dense), semi-transparent transparency settings (STrans), and support of global illumination (GI) effects. Symbols {\pmark} and {\mmark} represent high or low performance/quality, respectively (two repeating symbols indicate very high / very low performances). \textbf{/} represents ratings for opaque / semi-transparent (left) and highly transparent lines (right).}
	\label{tab:comparison}
	\fontsize{8pt}{8pt}\selectfont
	\centering%
	\begin{tabularx}{1.0\linewidth}
		{|X<{\centering}|m{12mm}<{\centering}|m{12mm}<{\centering}|m{12mm}<{\centering}|m{12mm}<{\centering}|m{12mm}<{\centering}|m{12mm}<{\centering}|m{12mm}<{\centering}|m{12mm}<{\centering}|m{12mm}<{\centering}|m{12mm}<{\centering}|}%
		\hline
		\multicolumn{1}{|>{\centering\arraybackslash}c|}{\textbf{Technique}} 
		& \multicolumn{1}{>{\centering\arraybackslash}c|}{\textbf{Order}} 
		& \multicolumn{1}{>{\centering\arraybackslash}c|}{\textbf{Accuracy}} 
		& \multicolumn{1}{>{\centering\arraybackslash}c|}{\textbf{Perform.}}
		& \multicolumn{1}{>{\centering\arraybackslash}c|}{\textbf{RenderC}}
		& \multicolumn{1}{>{\centering\arraybackslash}c|}{\textbf{BoundM}}
		& \multicolumn{1}{>{\centering\arraybackslash}c|}{\textbf{Temp}}
		& \multicolumn{1}{>{\centering\arraybackslash}c|}{\textbf{ChColor}}
		& \multicolumn{1}{>{\centering\arraybackslash}c|}{\textbf{Dense}}
		& \multicolumn{1}{>{\centering\arraybackslash}c|}{\textbf{STrans}}
		& \multicolumn{1}{>{\centering\arraybackslash}c|}{\textbf{GI}}%
		\\ \hline
		DP 		& Object 		& Exact 			& \mmark\space\mmark 							& \xmark 			& \cmark 	& \cmark  	& \pmark\space\pmark 	& \mmark\space\mmark & \pmark\space\pmark & \xmark \\ \hline
		LL 		& Object 		& Exact 			& \semimark{\pmark}{\mmark}   & \xmark   		& \xmark		& \cmark 	& \pmark\space\pmark 	& \mmark\space\mmark & \pmark\space\pmark & \xmark \\ \hline
		MLABDB 				& Object 		& Approx. 	& \pmark\space\pmark 						& \cmark		& \cmark  	& \xmark		& \mmark  	& \mmark & \pmark & \xmark \\ \hline
		MBOIT 				& Object 		& Approx. 	& \pmark\space\pmark 						& \cmark		& \cmark 	& \xmark 		& \semimark{\mmark}{\pmark} & \pmark & \mmark & \xmark \\ \hline
		VRC 				& Image 		& Approx. 	& \semimark{\pmark}{\mmark} & \cmark	 	& \cmark 	& \cmark 	& \pmark\space\pmark	& \mmark & \pmark\space\pmark  & \cmark \\ \hline
		OSP 				& Image 		& Exact. 			& \semimark{\pmark}{\mmark\space\mmark} & \xmark & \xmark 		& \cmark 	& \pmark\space\pmark 	& \mmark\space\mmark & \semimark{\pmark}{\mmark} & \cmark \\ \hline
		RTX 				& Image 		& Exact 			& \semimark{\pmark\space\pmark}{\mmark}	& \xmark & \xmark 		& \cmark 	& \pmark\space\pmark	& \semimark{\pmark}{\mmark} & \pmark\space\pmark  & \cmark \\ \hline
	\end{tabularx}%
	\vspace{-0.5em}
\end{table*}

\vspace{-0.5em}
\subsubsection*{Informal User Study}
In a simple user study, users were asked to give their assessment of the quality differences between MLABDB and MBOIT, to further shed light on the perceptional differences between these approximate variants. We recruited 24 participants, comprised of 19 computer science students and 5 computer science PhD students, all having a background in computer graphics. The participants were selected to have no color vision deficiency. The students were exposed to the application of line rendering for the first time. None of them knew the visualized vector fields and line sets beforehand. The study was performed using the desktop PC described above. We showed the users the tool and let them work with a data set not included in the study. Then, we performed two different experiments:
\begin{itemize}
	\item To each user, we showed a sequence of eight tri-sets of renderings: the ground truth image first, and then the same view rendered with MLABDB and MBOIT (showing their typical artifacts) side by side. 
	\item Each user carried out three interactive sessions with two data sets, one minute each. First, the same data set was visualized, starting with MLABDB and then using MBOIT; then the experiment was repeated in reverse order using a different data set.
\end{itemize}

Users were then asked to rate the visual quality of the still images and the animations. 
For both experiments, users could select either MBOIT or MLABDB as the best, or ``undecided''. 
We had three renderings where $70\%$ of the users selected MBOIT, three renderings where $63\%$ selected for MLAB, and two renderings where $65\%$ of the users were undecided.
In addition to the comparison of MBOIT and MLABDB, we asked the users to rate the image quality of still images as ``no difference'' (good) to the ground truth, ``acceptable'' (acc), or ``non-acceptable'' (non-acc). For MBOIT, $41\%$ rated it as good, $46\%$ as acceptable, and $13\%$ as non-acceptable. For MLABDB, $44\%$ rated it as good, $32\%$ as acc, and $24\%$ as non-acc.
To assess the image quality over time for each technique, we asked users in a second experiment to rate the quality of the videos similar to image quality. 
For MBOIT, $62\%$ rated it as good quality, $35\%$ as acceptable quality, and $3\%$ as non-acc. For MLABDB, $45\%$ of users rated it as good, $47\%$ as acc, and $8\%$ as non-acc.

About why they scored the renderings as good, acceptable, or non-acceptable, users mentioned that wrong color outputs and rendering order artifacts (line features falsely hidden) were the most disturbing, as well as the hard and abrupt changes produced by MLABDB or suddenly disappearing features (referred to as ``popping'' or ``flickering'') produced by MBOIT during an animation. Some users argued that sometimes even a wrong impression was suggested by both techniques in still images (see Fig.~\ref{fig:teaser} and Fig.~\ref{fig:datasets_errors}).
In animations, most users did not consider these effects as negative, due to the possibility to interact with the data and, thus, reveal missing information.

To conclude, MLABDB can be recommended for non-dense data or a small number of different colors per pixel while using semi-transparent transparency settings.
MBOIT can be suggested for large and dense data with a high variation of color per pixel and transparency settings with smooth transitions.
Also, both techniques can be applied to triangular meshes or point cloud data  (cf. Fig.~\ref{fig:error_isosurface_appendix} and Fig.~\ref{fig:error_points_appendix}, Appendix~\ref{app:isosurface_points}) since rasterization-based approaches operate on any input geometry.

\subsection{Image-Order}
Image-order techniques should be preferred when rather opaque structures are rendered since they can effectively employ early ray termination.
If transparency is used too aggressively, the time required to traverse the acceleration structures can increase significantly. 
The run-time performance varies strongly depending on the selected view and does not scale well with an increasing number of transparent layers.

In general, RTX performs better than OSP for all transparency settings, but OSP required less memory for acceleration structures and less time than RTX to complete builds.
Surprisingly, our RTX solution was able to achieve real-time rendering performance for all line data sets, including CLOUDS for semi-transparent settings. For opaque and semi-transparent settings, its rendering times were superior to VRC and OSP.
In comparison to OSP, best rendering-times were achieved with VRC and RTX throughout all transparency settings. 
VRC's performance was slightly superior to RTX for large data sets rendered with high transparency. 
Results produced by VRC are hardly distinguishable from the ground truth, especially in scenes where the camera is far from the data set, or the entire data set is seen at once through the current viewport (zoomed-out views).

In terms of memory consumption, VRC is recommended if memory is limited due to rendering constraints, which includes the finite size of the voxel grid, a constant line quantization level, and a fixed number of lines covered per cell.
Although OSP is generally limited by the amount of RAM, RTX requires more than twice as much VRAM as the memory size of the model's renderable representation to build acceleration structures and usually requires 3 times more memory to render the models.
Both OSP and RTX currently support only 32-bit integer values to address primitives on the hardware. As such, large data sets must be chunked beforehand into 4GB or less to be rendered using these methods.



\section{Conclusion and Outlook}
In this work, we have discussed and analyzed different rendering techniques for large 3D line sets with transparency wrt. memory consumption, performance, and quality. 
The major findings of our study are that a) approximate techniques can give acceptable quality at high speed and low memory consumption in many use cases, and b) ray-based approaches offer high quality and often at speeds similar to approximate techniques, besides the most extreme cases with overall low transparency. On the other hand, these techniques can require huge memory resources and considerable pre-processing time.

However, regardless of the technique used, transparent line renderings likely fail to communicate spatial relations when large numbers of transparent lines are rendered. 
In these cases, global illumination effects such as soft shadows and ambient occlusion can help to significantly improve the user's perception~\cite{Kanzler2018}.
Such effects can be integrated in a straightforward way into image-order approaches by tracing secondary rays. 
The integration into object-order approaches is more difficult, and can be achieved only with substantial algorithmic changes, and changes to the data structures used. 
If high-quality rendering for large line sets is desired, we believe that image-order approaches should be favored over object-order approaches.

With the current power of RTX GPU hardware, it will be interesting in the future to combine both transparency rendering and global illumination effects to enhance the visual perception of complex data.
Further user studies have to be conducted to shed light on the question of whether transparency rendering of large, dense line data sets is beneficial to the user's perception, or hampers interpretation of trends in the data, and how this may interact with global illumination effects. In terms of interpretation of dense line sets, it would also be interesting to compare transparency rendering techniques with approaches that heuristically filter line sets and render features-of-interest completely opaque.
\section{Acknowledgements}
We would like to thank Max Bandle and Mathias Kanzler, Technical University of Munich, for their support concerning the efficient implementation of per-pixel fragment sorting and voxel-based line rendering, respectively. 
Thanks to Tobias Günther from ETH Z\"urich for providing access to the cloud simulation data which is provided by DKRZ and MPI-M.
We thank Tony Saad and Josh McConnell at the University of Utah CCMSC for the Uintah
simulation. The Richtmyer-Meskhov data is courtesy of Lawrence Livermore National Laboratory.

This evaluation study has been done within the subproject
''Visualization of coherence and variation in meteorological dynamics``
of the Transregional Collaborative Research Center SFB/TRR 165
''Waves to Weather`` funded by the German Research Foundation (DFG).


%

%
%
%


\ifCLASSOPTIONcaptionsoff
  \newpage
\fi



 \bibliographystyle{IEEEtranS}
\bibliography{IEEEabrv,references/oit_references,references/paper,references/paper0,references/hybrid,references/paper_old,references/osp_bibs}
%
%
%

%

\vfill
\enlargethispage{0in}
\vspace{-3.8em}
\begin{IEEEbiography}
	[{\includegraphics[width=1in,height=1.25in,clip,keepaspectratio]{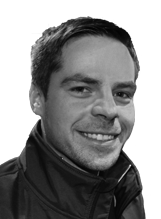}}]{Michael Kern}
	is a PhD candidate at the Computer Graphics and Visualization Group at the
	Technical University of Munich (TUM). He obtained his M.Sc. in computer science from TUM
	in 2016. He did an 6-months internship at the Scientific Computing and Imaging Institute at the University of Utah in 2015 with focus on the visualization of biological data. His general research interests include scientific visualization, ensemble uncertainty analysis, and GPU computing.
\end{IEEEbiography}

\vfill
\begin{IEEEbiography}
	[{\includegraphics[width=1in,height=1.25in,clip,keepaspectratio]{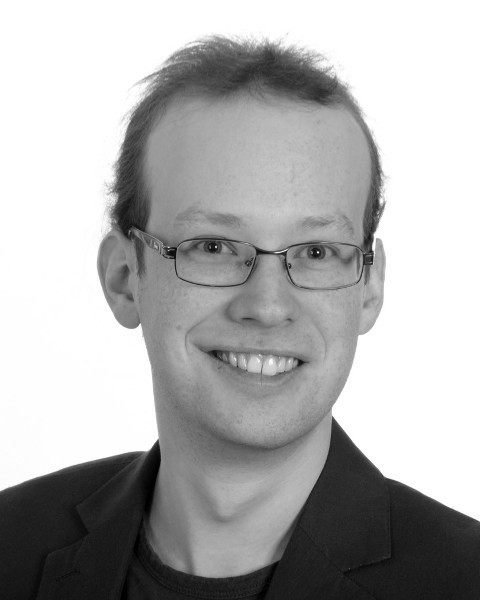}}]{Christoph Neuhauser}
	is a graduate research assistant at the Computer Graphics and Visualization Group at the Technical University of Munich. He received his B.Sc. in computer science from TUM in 2019. Major interests in research comprise scientific visualization and real-time rendering.
\end{IEEEbiography}

\vfill
\begin{IEEEbiography}
	[{\includegraphics[width=1in,height=1.25in,clip,keepaspectratio]{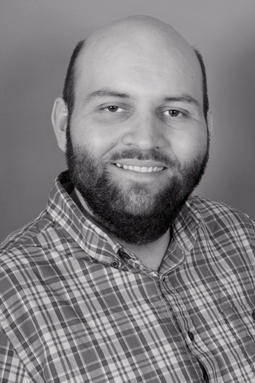}}]{Torben Maack}
	is an undergraduate student at the Technical University of Munich. He received his B.Sc. in computer science at the TUM in 2019. His research interests incorporate areas in real-time rendering and high performance computing.
\end{IEEEbiography}
\vfill
\begin{IEEEbiography}
	[{\includegraphics[width=1in,height=1.25in,clip,keepaspectratio]{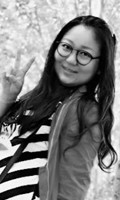}}]{Mengjiao Han}
	is a graduate research assistant at the Scientific Computing and Imaging Institute at the University of Utah. She obtained her B.Sc. in electronic bio-engineering from the Beijing University of Post and Telecommunication in 2010.
	She finally received her M.Sc. in computer science in 2014 at the University of Delaware. Her major interests in research are scientific data visualization for large-scale data, parallel rendering and real-time ray tracing.
\end{IEEEbiography}
\vfill
\begin{IEEEbiography}
	[{\includegraphics[width=1in,height=1.25in,clip,keepaspectratio]{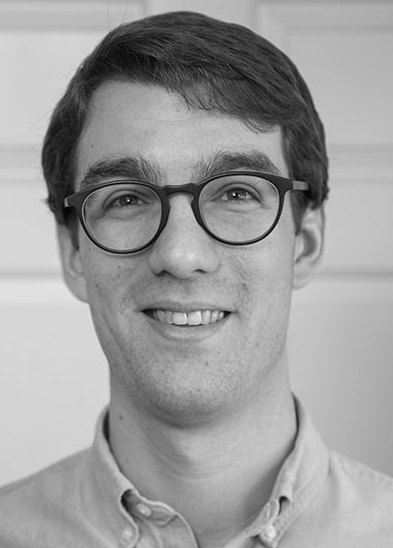}}]{Will Usher}
	is a graduate research assistant at the Scientific Computing and Imaging Institute at the University of Utah. He finished his B.Sc. in physics with a minor in computer science at the University of California, Riverside in 2014. His research interests cover a range of areas in real-time ray tracing, physically based rendering, virtual reality, and high performance GPU computing.
\end{IEEEbiography}
\vfill
\begin{IEEEbiography}
	[{\includegraphics[width=1in,height=1.25in,clip,keepaspectratio]{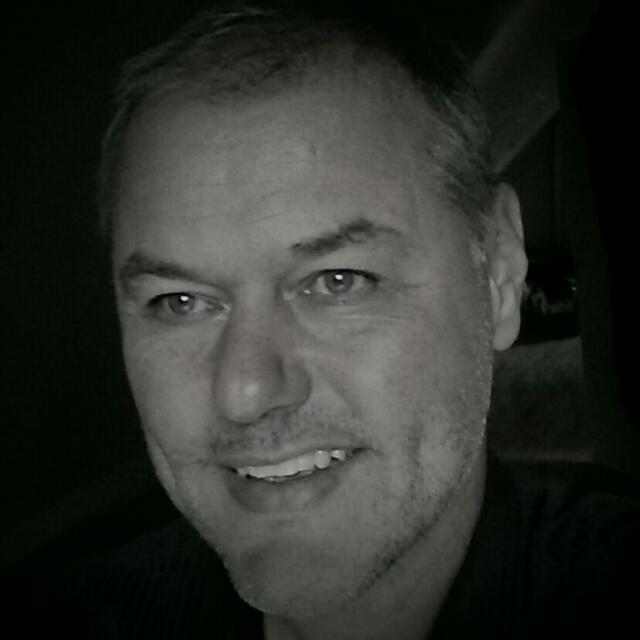}}]{R\"udiger Westermann}
	studied computer science at the Technical University Darmstadt and
	received his Ph.D. in computer science from
	the University of Dortmund, both in Germany. In
	2002, he was appointed the chair of Computer
	Graphics and Visualization at TUM. His research
	interests comprise scalable data visualization and
	simulation algorithms, GPU computing, real-time
	rendering of large data, and uncertainty visualization.
\end{IEEEbiography}
\vfill
%
%
%




\newpage
\appendices
\renewcommand\thefigure{\thesection.\arabic{figure}}    
\setcounter{figure}{0}
\setcounter{equation}{0}
\section{Quantitative Assessment of Image Quality}
\label{app:quantitative}



PSNR measures the ratio between the power of a signal and the power of noise distorting the signal in decibel (dB). It is computed on a logarithmic scale as
\begin{equation}
\text{PSNR}(O_t^{R}, O_t^{G}) = -10 \log_{10}(||O_t^{R}-O_t^{G}||_2^2),
\end{equation}
where $O_t^{R}$ and $O_t^{G}$ are the rendering results and the ground truth image, respectively.
The PSNR for the ground truth is clamped to a high value of $60$dB.

SSIM is a perception-based model to measure the similarity of two images wrt. structural information.
In contrast to PSNR, SSIM highlights regions of structural (dis-) similarity and enables a detailed analysis of rendering stability and approximation errors wrt. transparency and fragment depth.
SSIM is defined as
\begin{equation}
\text{SSIM}(O_t^{R}, O_t^{G}) = \frac{(2\mu_{R}\mu_{G}+c_1)(2\sigma_{{R},{G}}+c_2)}{(\mu_{R}^2+\mu_{G}^2+c_1)(\sigma_{R}^2+\sigma_{G}^2+c_2)} ,
\end{equation}
where $\mu_{R}$ and $\mu_{G}$ are the average values of $O_t^{R}$ and $O_t^{G}$, $\sigma_{R}^2$ and $\sigma_{G}^2$ are the variances of $O_t^{H}$ and $O_t^{G}$, $\sigma_{{R},{G}}$ is the covariance between $O_t^{H}$ and $O_t^{G}$, and $c_1$ and $c_2$ are two small constants to avoid division by zero.
We compute SSIM pixel-wise using a Gaussian sub-window of 11x11 to weight surrounding pixels.
The SSIM value of an image (1 for the ground truth) is the mean over all pixel-wise SSIM values.

\begin{figure*}[!htb]	
	\centering
	\begin{overpic}[width=0.245\textwidth,clip,trim=0mm 0mm 0mm 10mm]
		{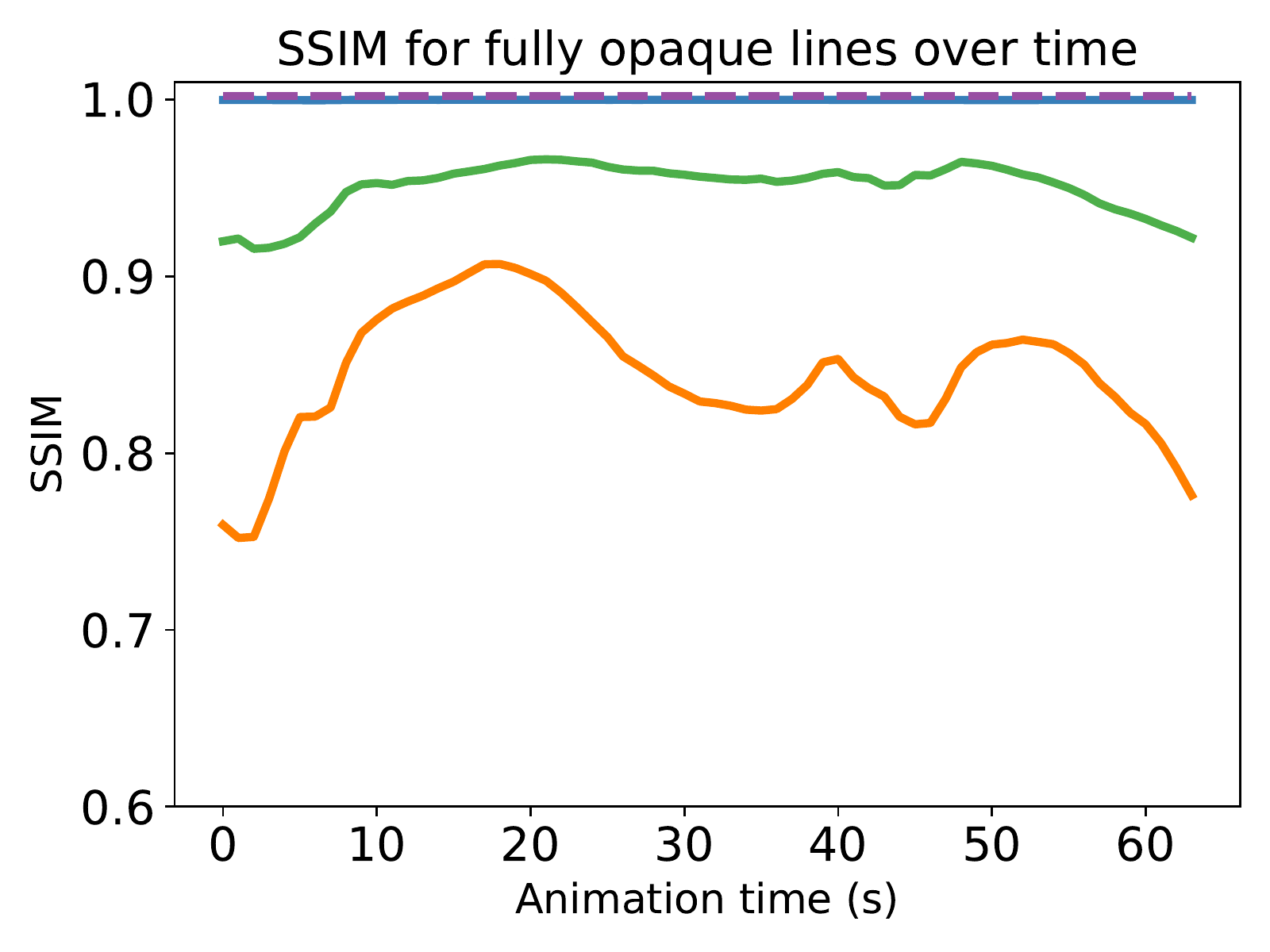}
	\end{overpic}%
	\begin{overpic}[width=0.245\textwidth,clip,trim=0mm 0mm 0mm 10mm]
		{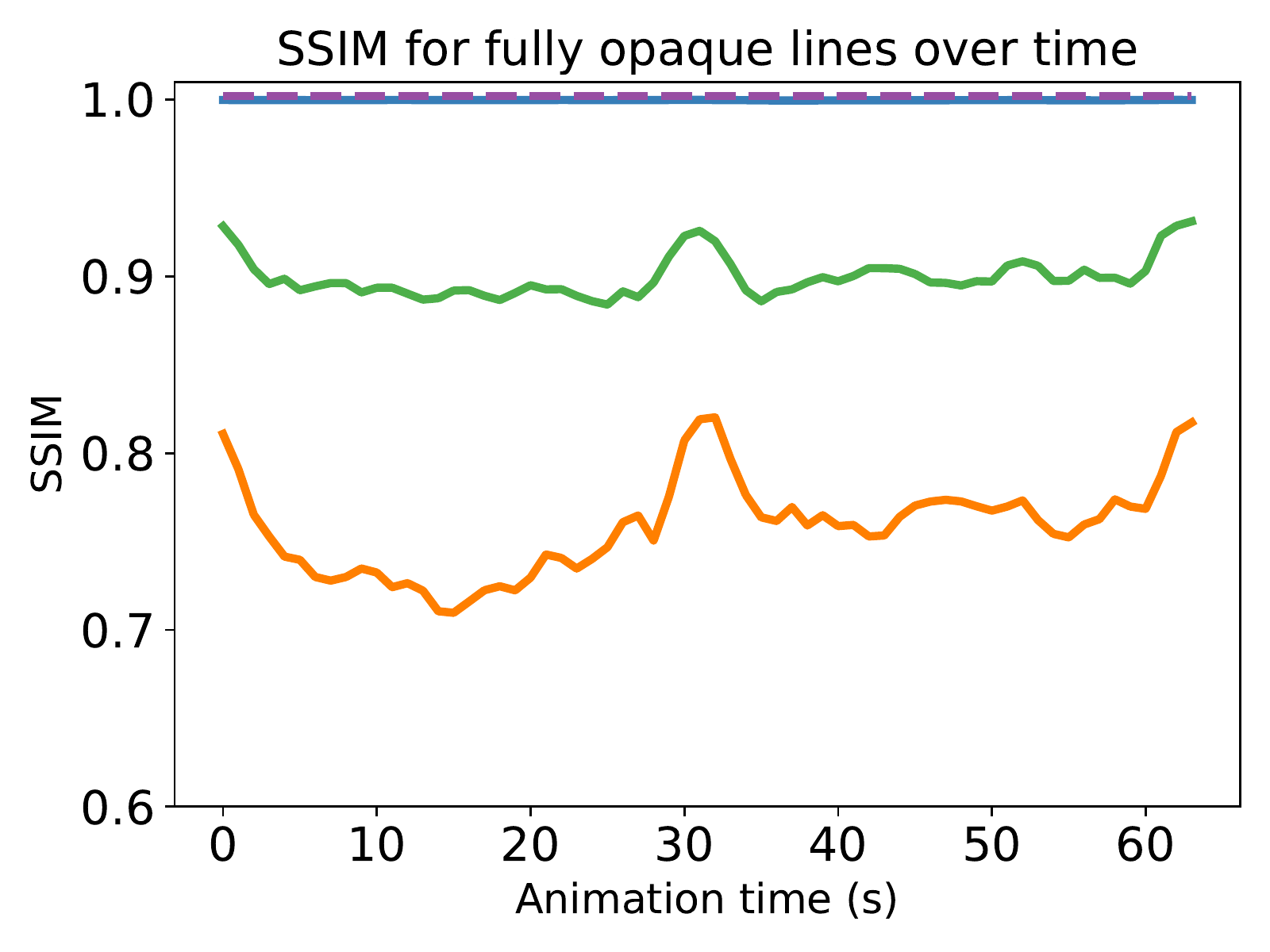}
	\end{overpic}%
	\begin{overpic}[width=0.245\textwidth,clip,trim=0mm 0mm 0mm 10mm]
		{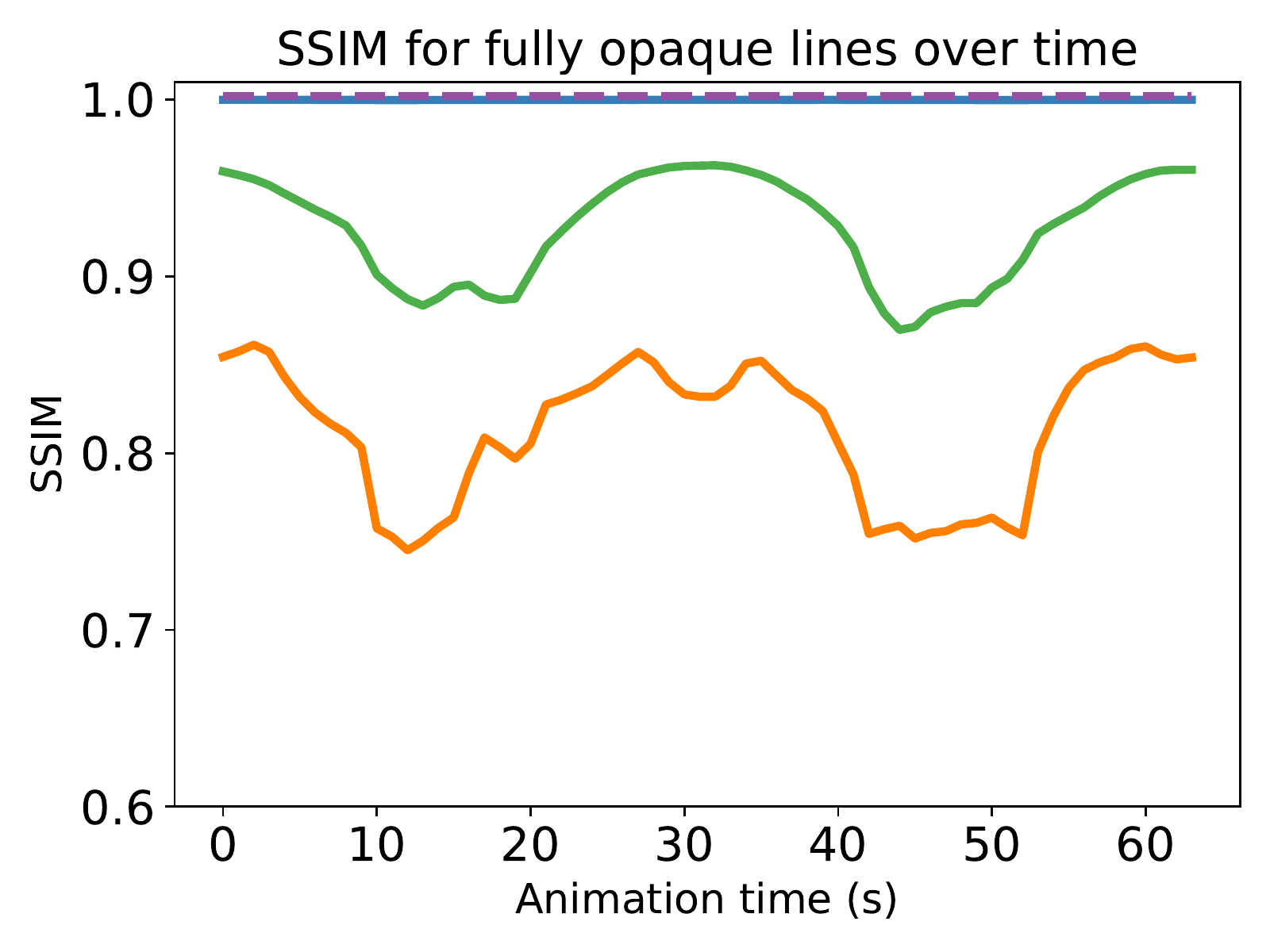}
	\end{overpic}%
	\begin{overpic}[width=0.245\textwidth,clip,trim=0mm 0mm 0mm 10mm]
		{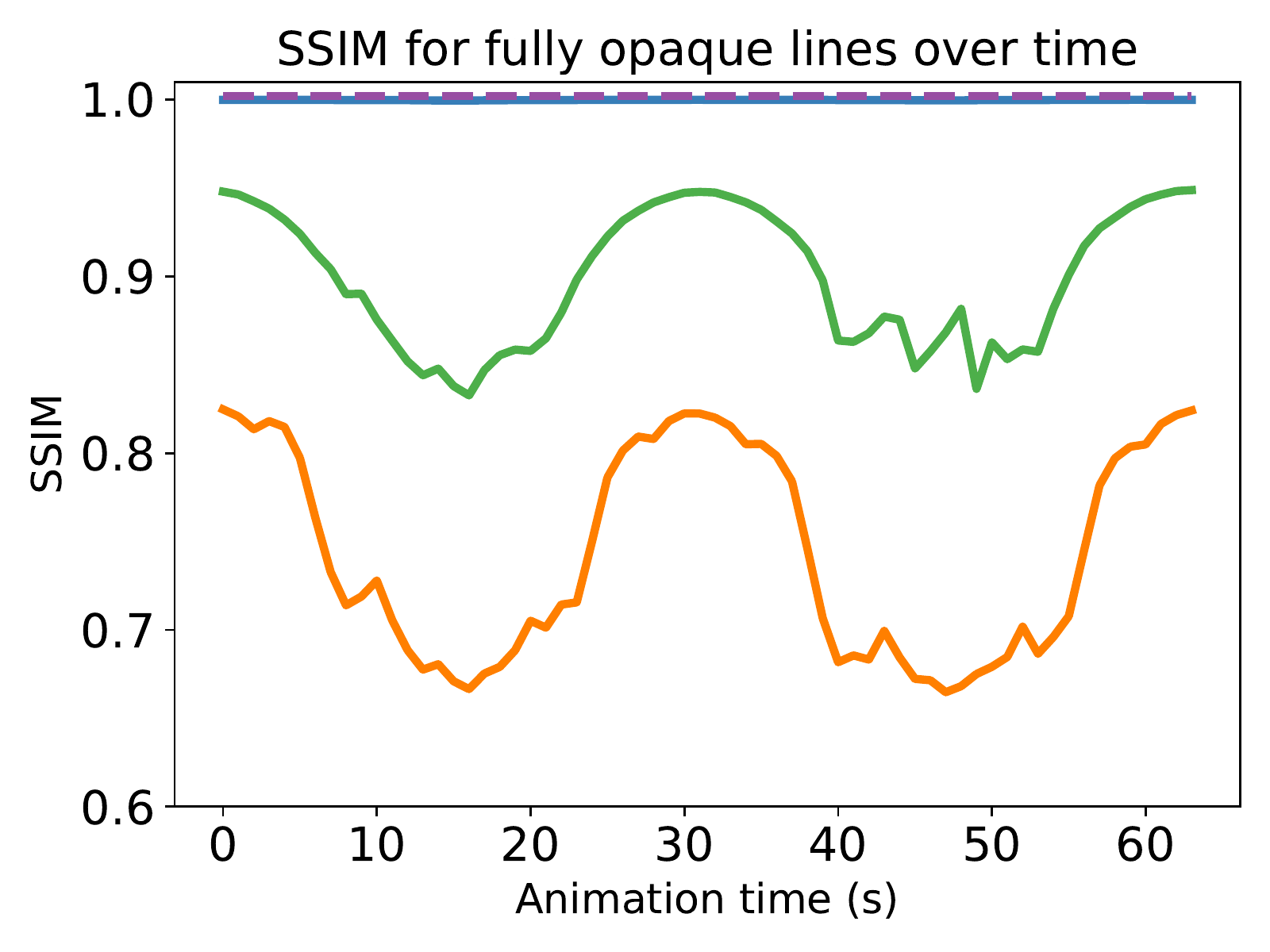}
	\end{overpic}
	\begin{overpic}[width=0.245\textwidth,clip,trim=0mm 0mm 0mm 10mm]
		{figures/tables/tf_quality_semitransparent/error_metric_SSIM_res720p_Aneurysm.pdf}
	\end{overpic}%
	\begin{overpic}[width=0.245\textwidth,clip,trim=0mm 0mm 0mm 10mm]
		{figures/tables/tf_quality_semitransparent/error_metric_SSIM_res720p_ConvectionRolls.pdf}
	\end{overpic}%
	\begin{overpic}[width=0.245\textwidth,clip,trim=0mm 0mm 0mm 10mm]
		{figures/tables/tf_quality_semitransparent/error_metric_SSIM_res720p_Turbulence.pdf}
	\end{overpic}%
	\begin{overpic}[width=0.245\textwidth,clip,trim=0mm 0mm 0mm 10mm]
		{figures/tables/tf_quality_semitransparent/error_metric_SSIM_res720p_UCLA400k.pdf}
	\end{overpic}
	\vspace{12pt}%
	\begin{overpic}[width=0.245\textwidth,clip,trim=0mm 0mm 0mm 10mm]
		{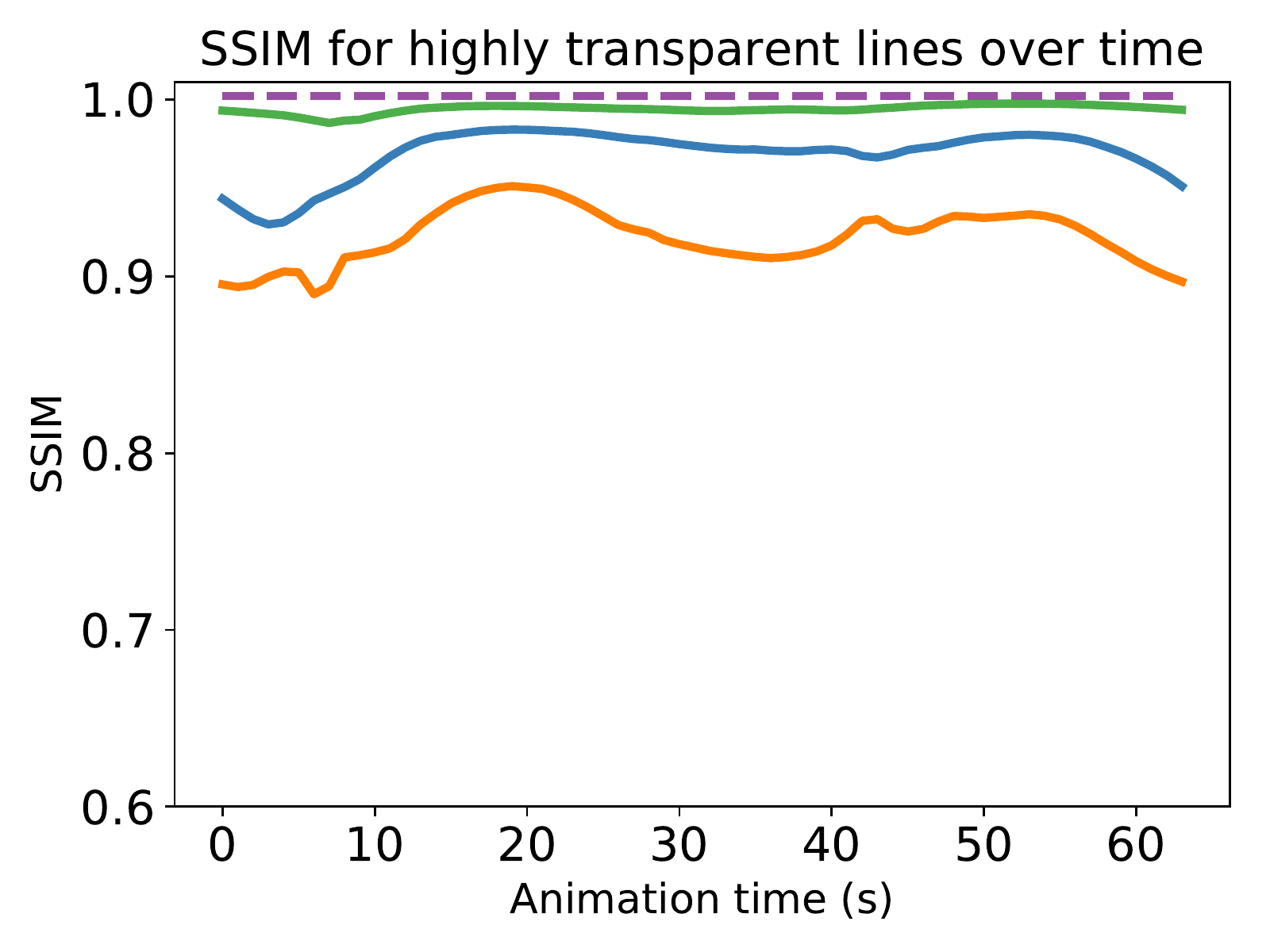}
	\end{overpic}%
	\begin{overpic}[width=0.245\textwidth,clip,trim=0mm 0mm 0mm 10mm]
		{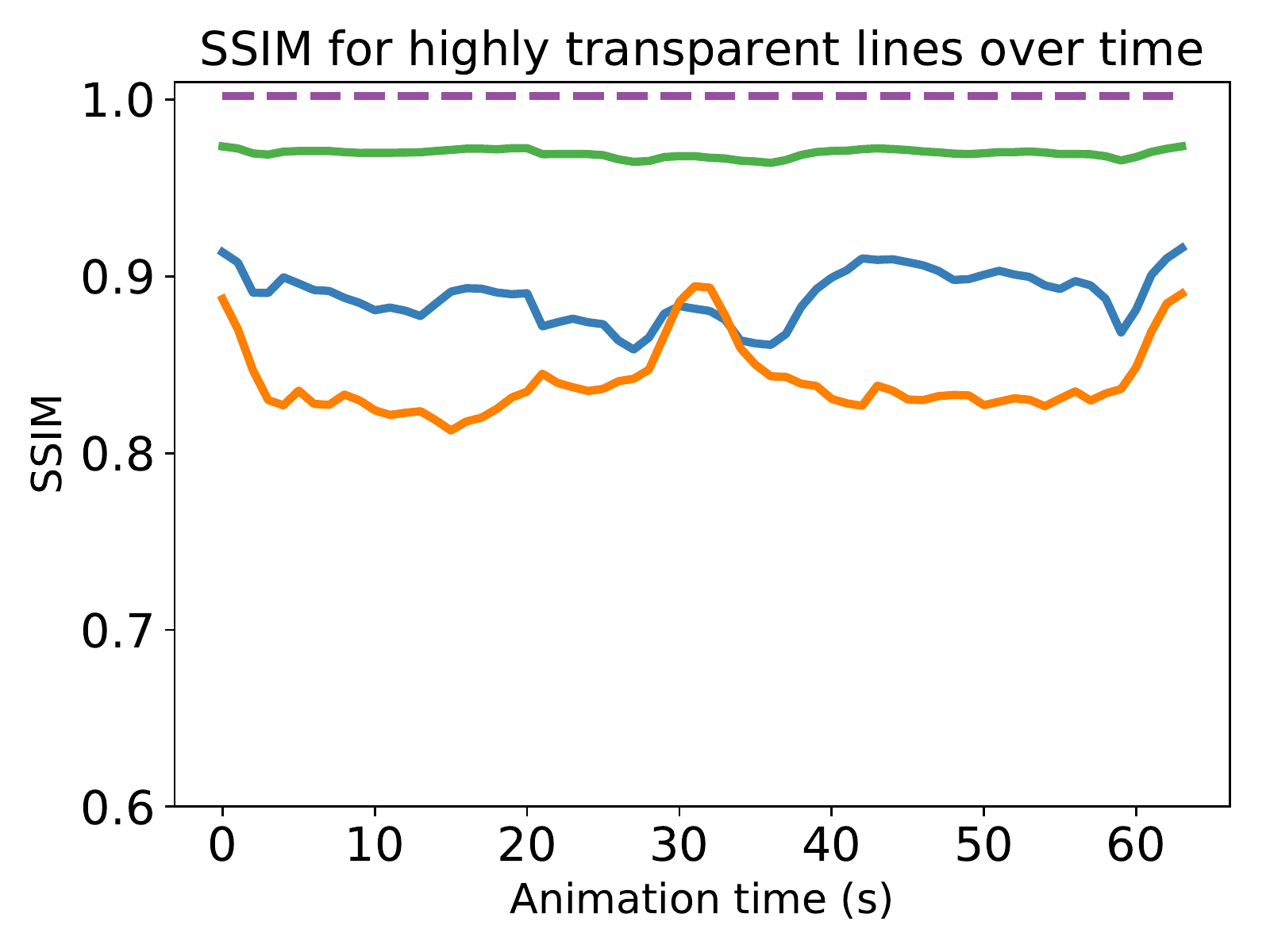}
	\end{overpic}%
	\begin{overpic}[width=0.245\textwidth,clip,trim=0mm 0mm 0mm 10mm]
		{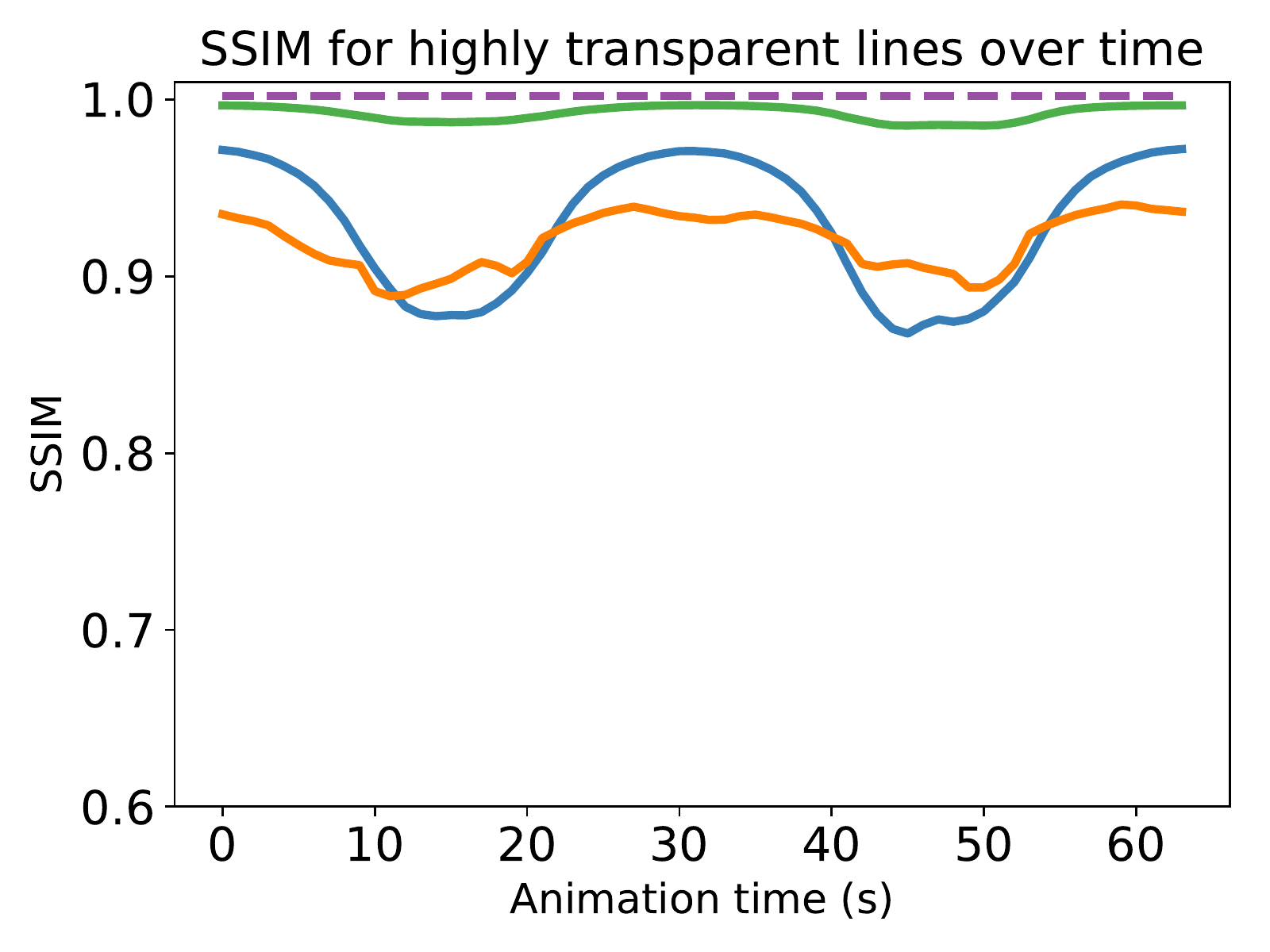}
	\end{overpic}%
	\begin{overpic}[width=0.245\textwidth,clip,trim=0mm 0mm 0mm 10mm]
		{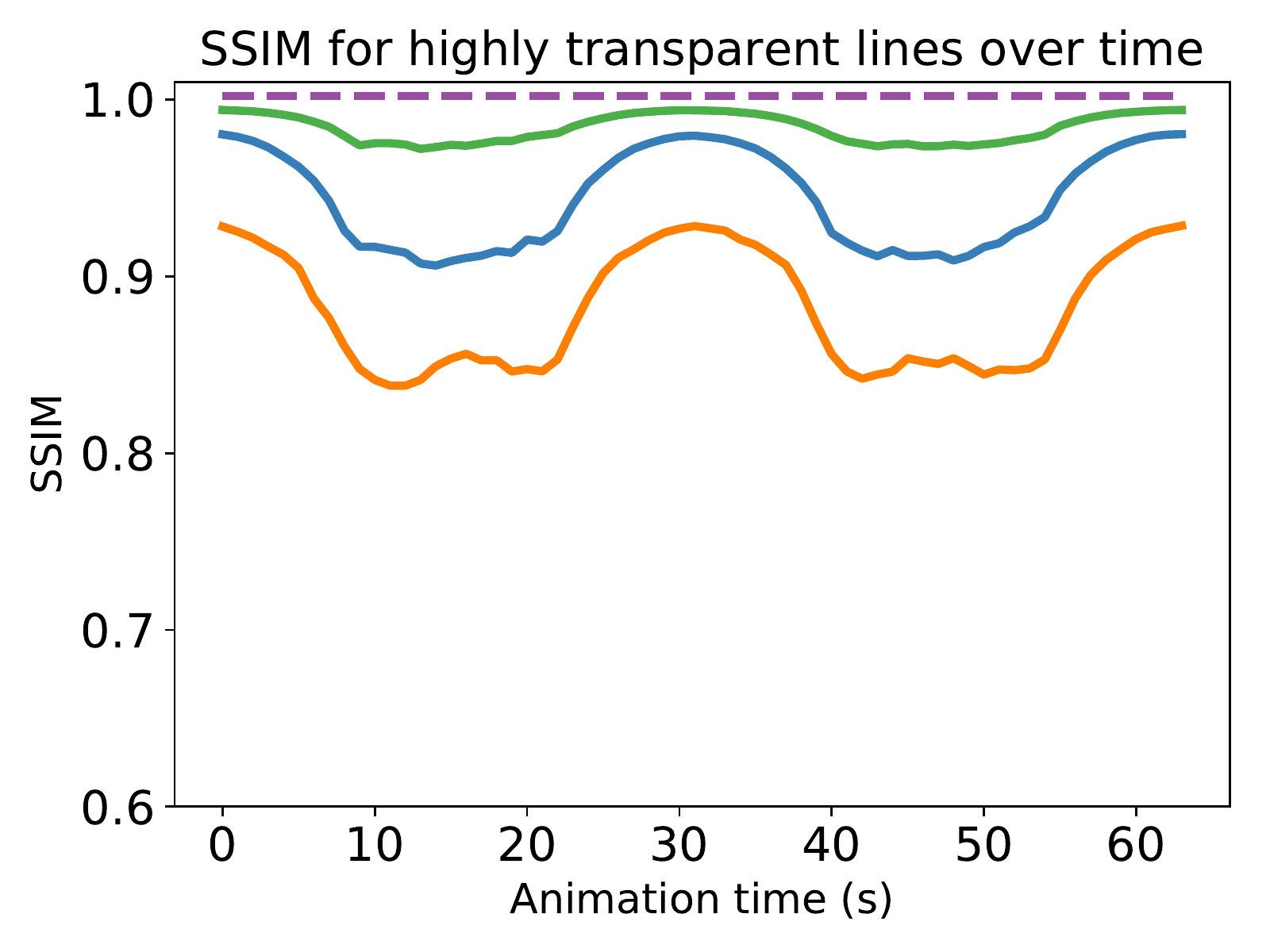}
	\end{overpic}
	\begin{overpic}[width=0.245\textwidth,clip,trim=0mm 0mm 0mm 10mm]
		{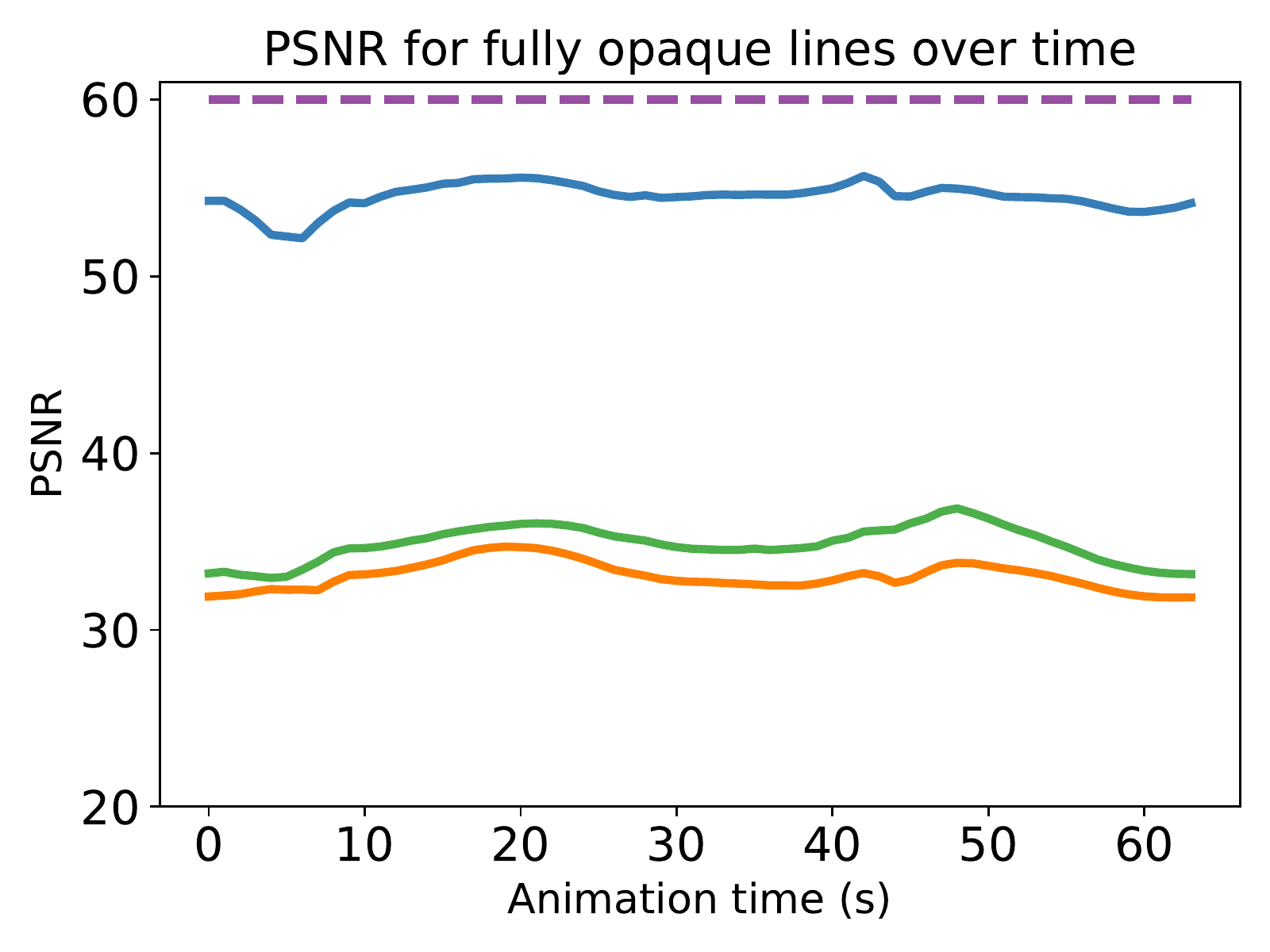}
	\end{overpic}%
	\begin{overpic}[width=0.245\textwidth,clip,trim=0mm 0mm 0mm 10mm]
		{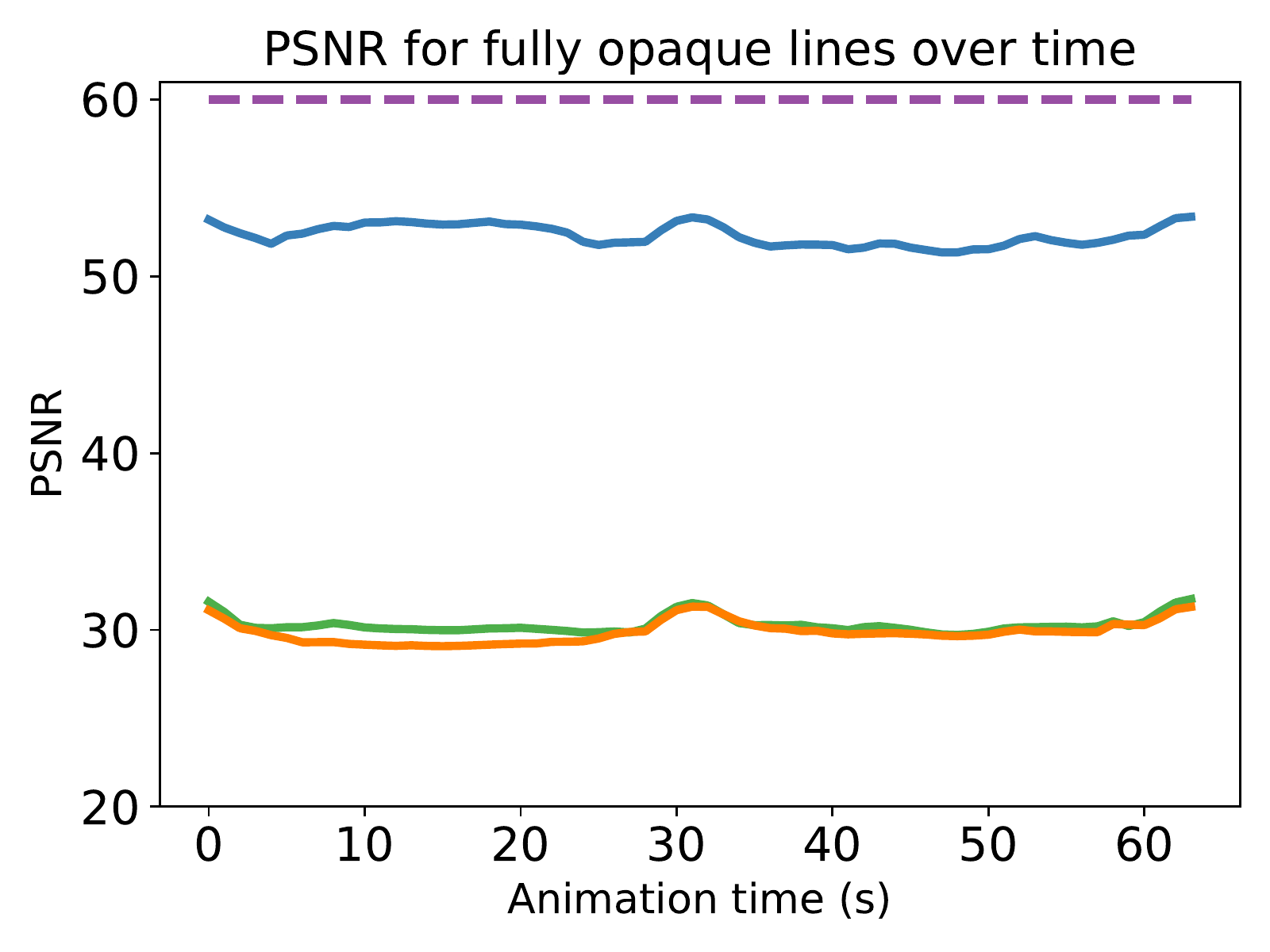}
	\end{overpic}%
	\begin{overpic}[width=0.245\textwidth,clip,trim=0mm 0mm 0mm 10mm]
		{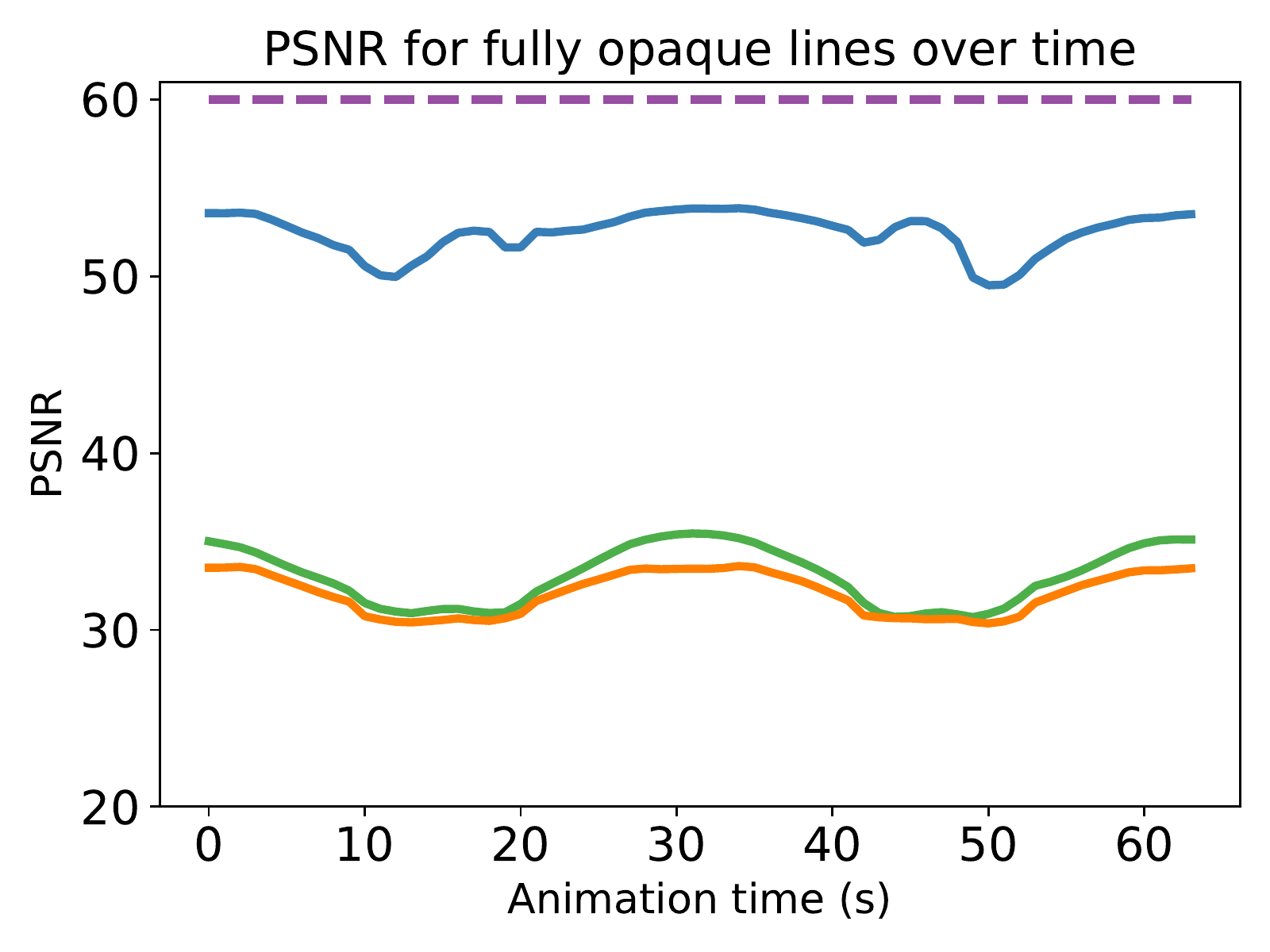}
	\end{overpic}%
	\begin{overpic}[width=0.245\textwidth,clip,trim=0mm 0mm 0mm 10mm]
		{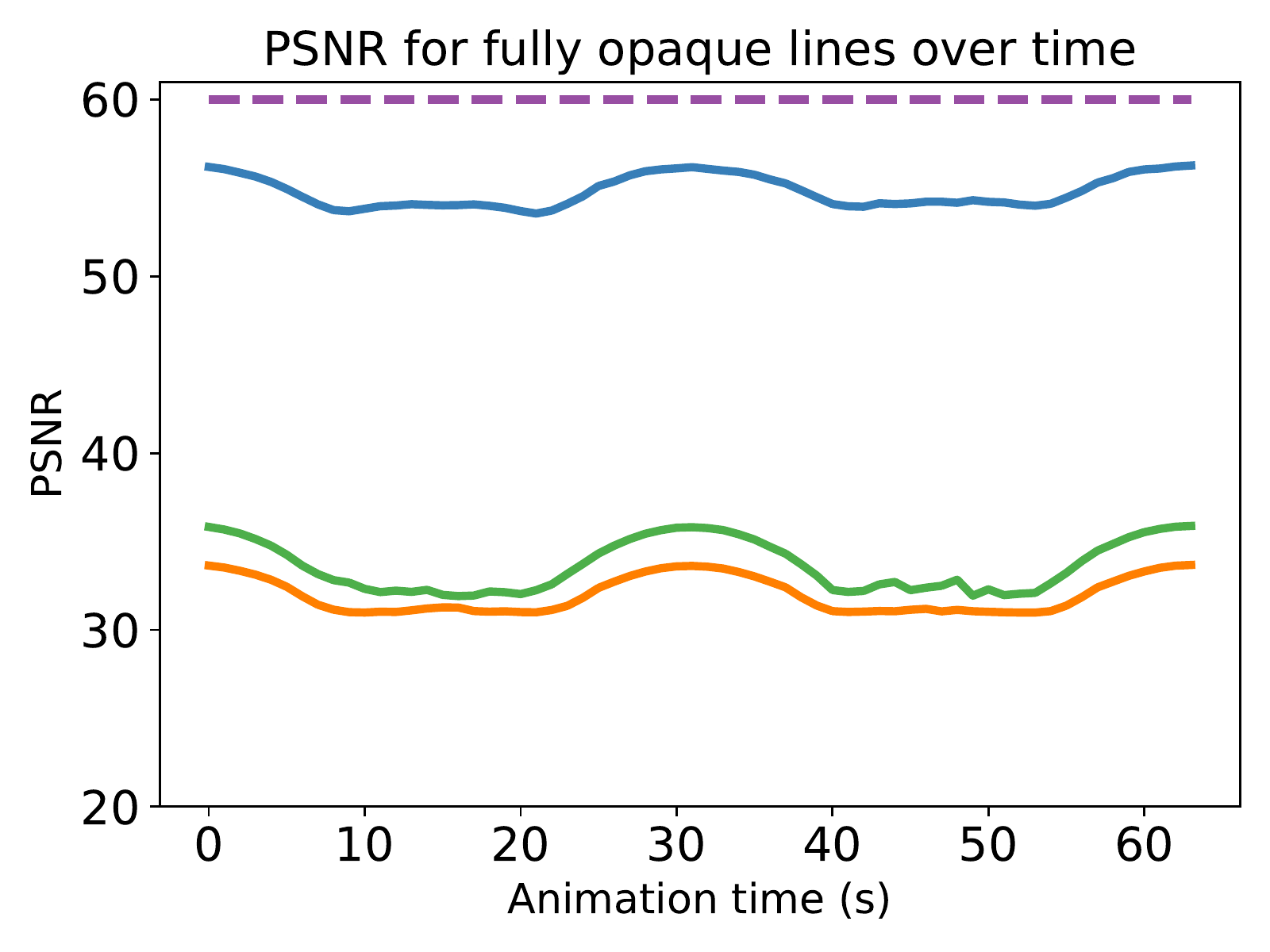}
	\end{overpic}
	\begin{overpic}[width=0.245\textwidth,clip,trim=0mm 0mm 0mm 10mm]
		{figures/tables/tf_quality_semitransparent/error_metric_PSNR_res720p_Aneurysm.pdf}
	\end{overpic}%
	\begin{overpic}[width=0.245\textwidth,clip,trim=0mm 0mm 0mm 10mm]
		{figures/tables/tf_quality_semitransparent/error_metric_PSNR_res720p_ConvectionRolls.pdf}
	\end{overpic}%
	\begin{overpic}[width=0.245\textwidth,clip,trim=0mm 0mm 0mm 10mm]
		{figures/tables/tf_quality_semitransparent/error_metric_PSNR_res720p_Turbulence.pdf}
	\end{overpic}%
	\begin{overpic}[width=0.245\textwidth,clip,trim=0mm 0mm 0mm 10mm]
		{figures/tables/tf_quality_semitransparent/error_metric_PSNR_res720p_UCLA400k.pdf}
	\end{overpic}
	\begin{overpic}[width=0.245\textwidth,clip,trim=0mm 0mm 0mm 10mm]
		{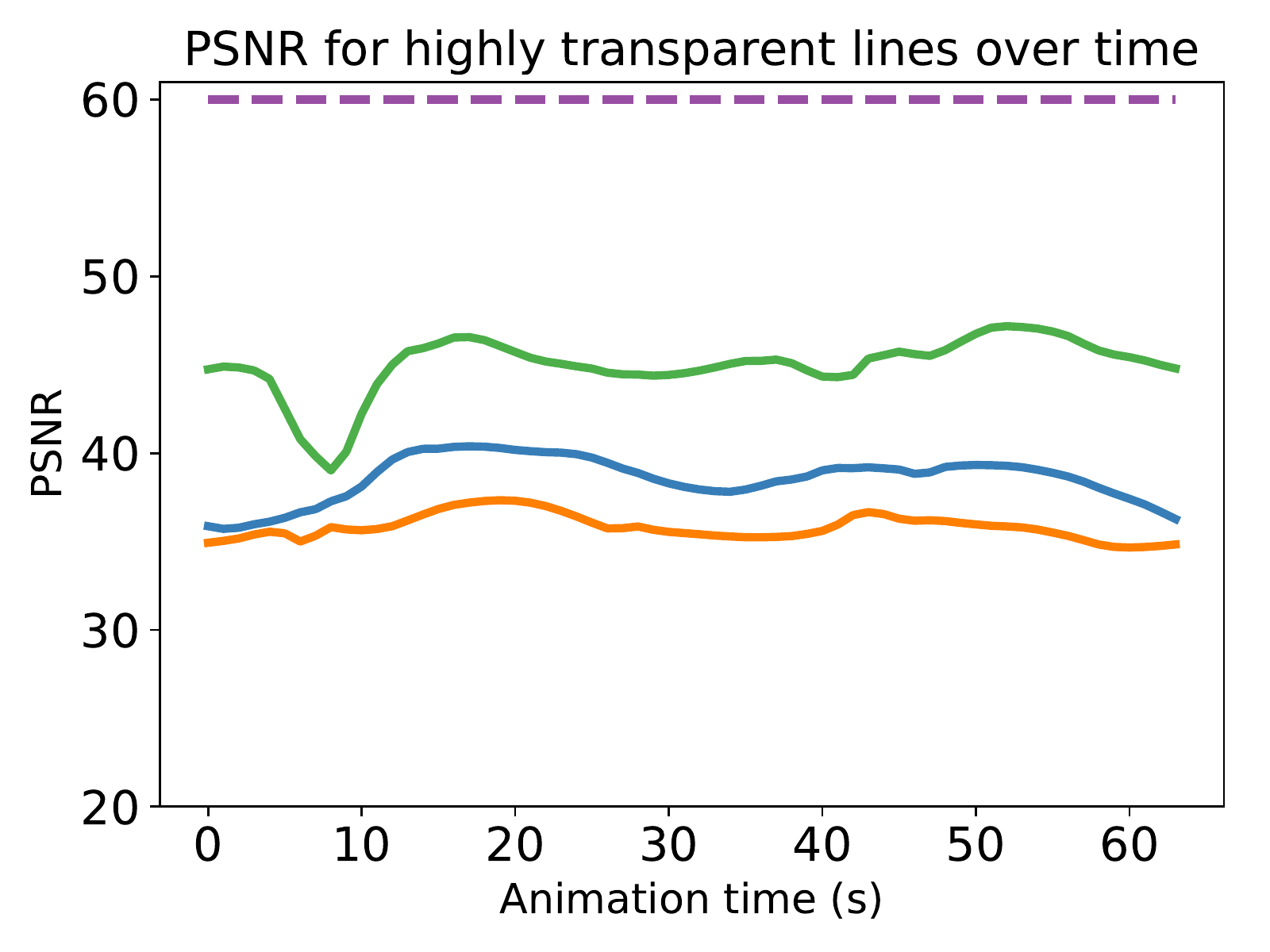}
		\put(38,18){\scriptsize\textbf{\textcolor{gray}{ANEURYSM}}}
	\end{overpic}%
	\begin{overpic}[width=0.245\textwidth,clip,trim=0mm 0mm 0mm 10mm]
		{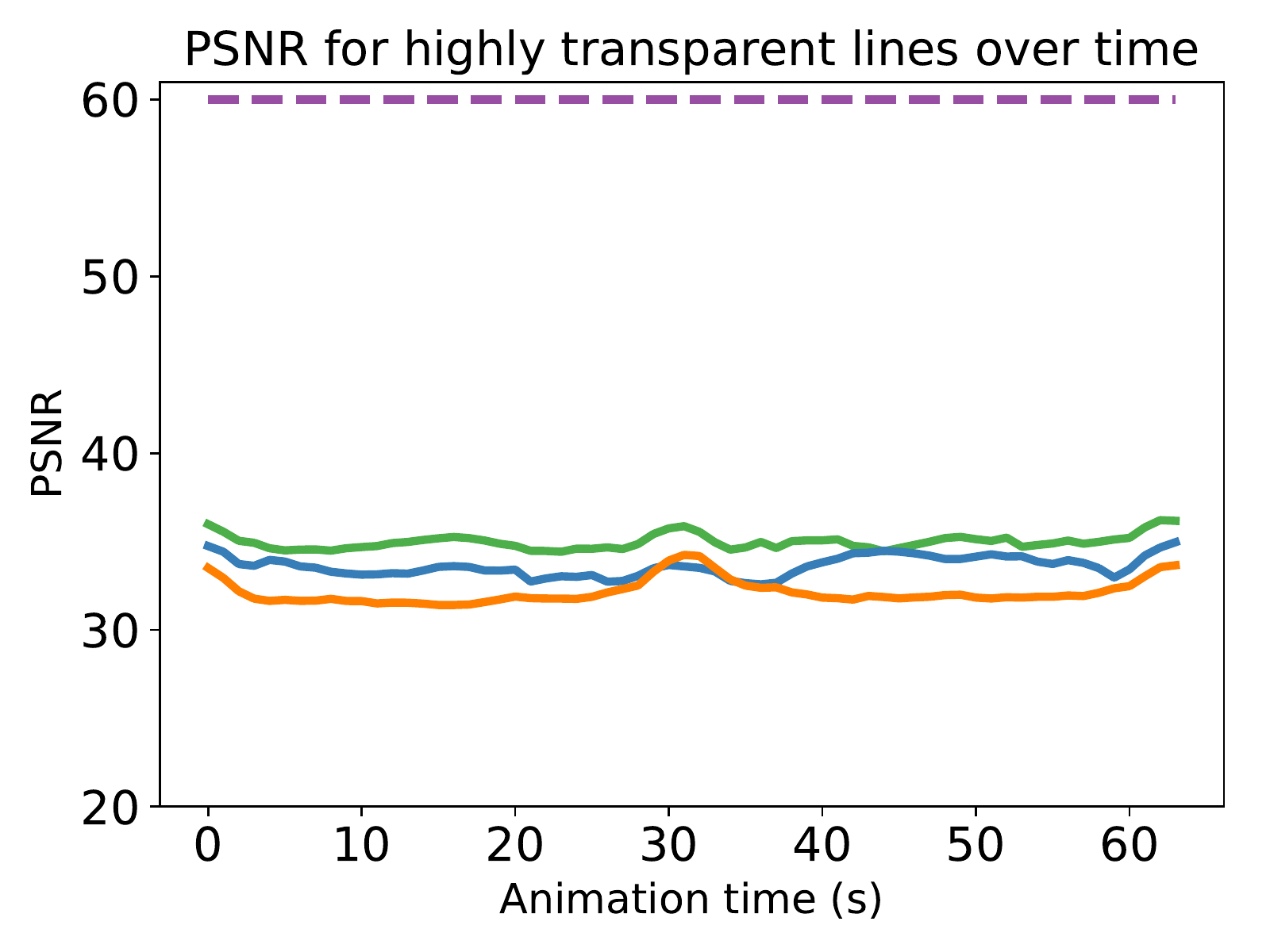}
		\put(34,18){\scriptsize\textbf{\textcolor{gray}{CONVECTION}}}
	\end{overpic}%
	\begin{overpic}[width=0.245\textwidth,clip,trim=0mm 0mm 0mm 10mm]
		{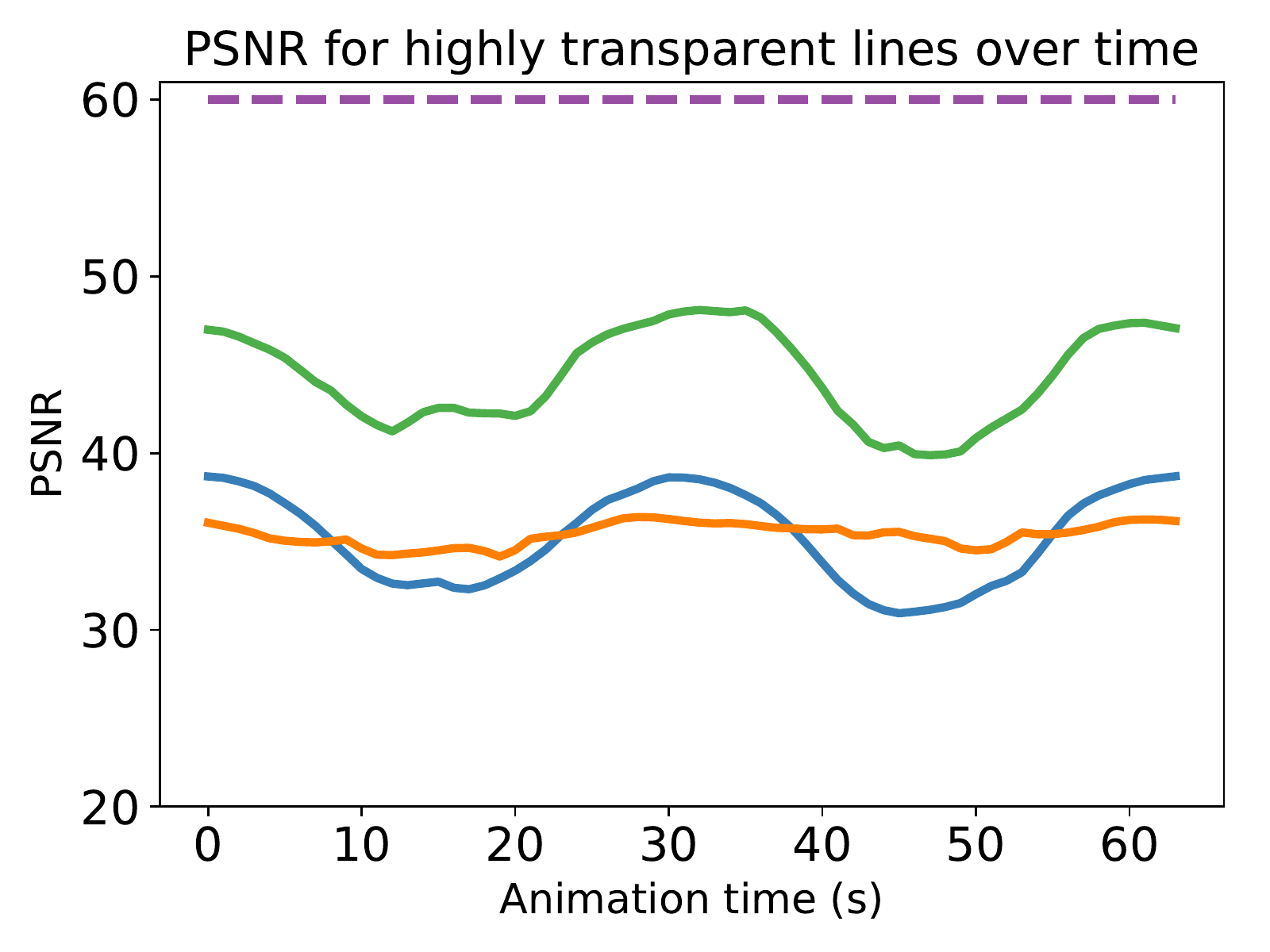}
		\put(36,18){\scriptsize\textbf{\textcolor{gray}{TURBULENCE}}}
	\end{overpic}%
	\begin{overpic}[width=0.245\textwidth,clip,trim=0mm 0mm 0mm 10mm]
		{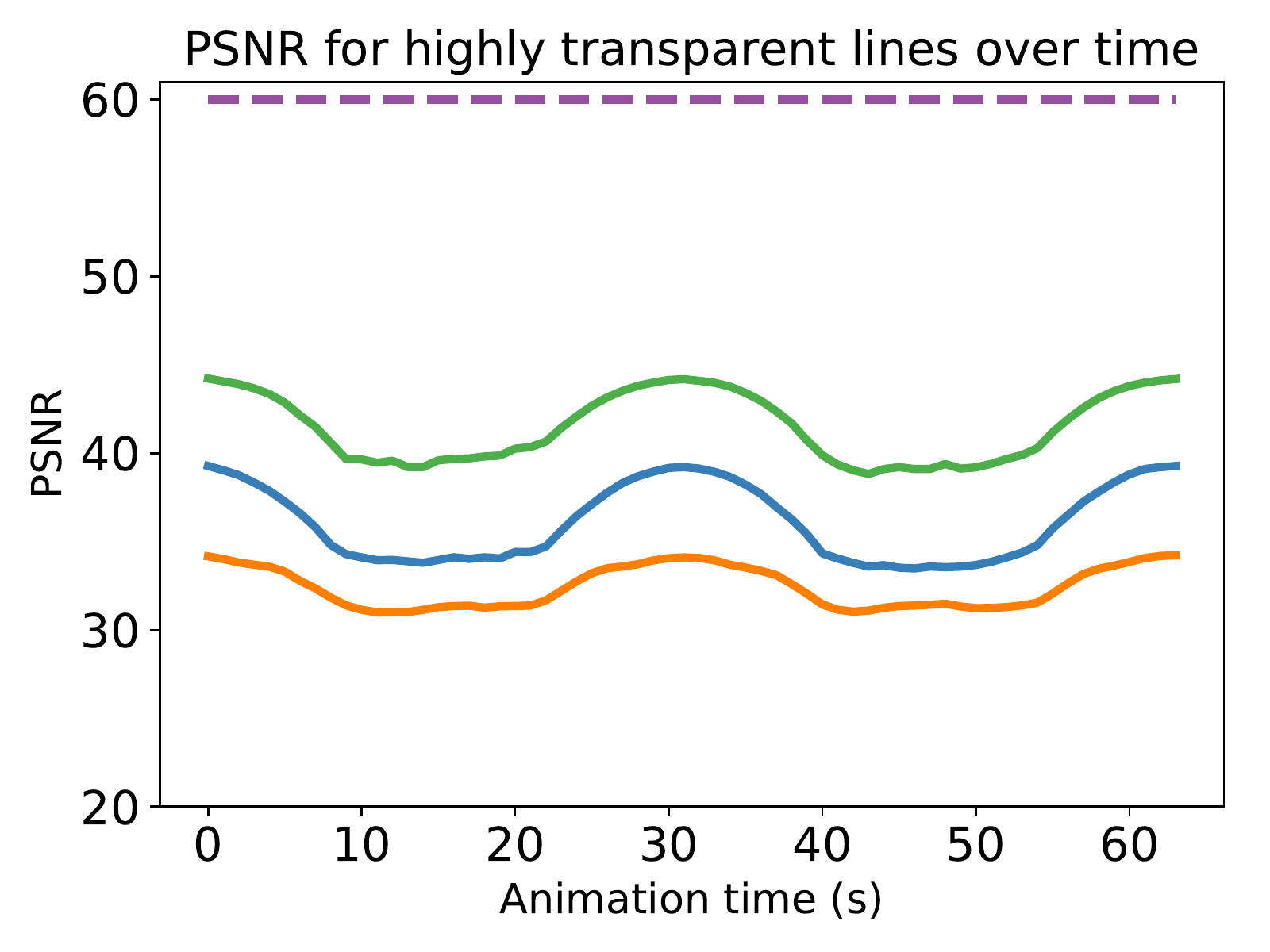}
		\put(42,18){\scriptsize\textbf{\textcolor{gray}{CLOUDS}}}
	\end{overpic}
	\caption{Error metrics of all techniques using opaque lines (1st and 4th row), transparent lines with opaque features of interest (2nd and 5th row), and highly transparent lines (3rd and 6th row). Columns are for ANEURYSM, CONVECTION, TURBULENCE, and CLOUDS, respectively. First three rows show SSIM for each technique, the last three rows show PSNR. A higher value is better. }
	\label{fig:quality_appendix}
	\vspace{10em}
\end{figure*}

The SSIM and PSNR plots in Fig.~\ref{fig:quality_appendix} indicate that MBOIT and MLABDB consistently achieve high image quality. For the smaller data sets, the quality of both techniques doesn't seem to be strongly view dependent, yet during close-ups (cf. frames $10 - 20$ and $40 - 50$) the quality decreases significantly when rendering large line sets like TURBULENCE and CLOUDS.
When zooming into the data, artifacts along a few pixels are now spread across an ever larger area in pixel space, resulting in decreasing SSIM and PSNR values.

For highly opaque lines, MLABDB always comes very close to the ground truth.
This result shows that discarding all transparent fragments behind opaque fragments, similar to early ray termination, helps to improve image quality and stability. 
Nevertheless, bucketing does not seem to work well when high transparency is used, since the correspondence between opacity and depth distance is increasingly lost (see 3rd and 6th rows in Fig.~\ref{fig:quality_appendix}).
In this case, MLABDB operates like MLAB, inheriting the weaknesses mentioned in Sec.~\ref{sec:quality}.

Wrt. SSIM, MBOIT performs slightly worse than MLABDB for lines with low and medium transparency due to either transparency over- or underestimation (cf. first two rows for PSNR and SSIM in Fig.~\ref{fig:quality_appendix}). 
However, for highly transparent lines, MBOIT can accurately simulate the transmittance function for each data set. 
Furthermore, it shows low variation in image quality during animations. 

Regarding the PSNR values, less noise is introduced by MLABDB for highly opaque lines due to bucketing, whereas more noise is produced by MBOIT and VRC with consistently lower PSNR values for all data sets (cf. first two rows for PSNR in Fig.~\ref{fig:quality_appendix}).
Whereas VRC has low PSNR values due to line inaccuracies and silhouette errors, low PSNR values for MBOIT are due to pixel-wise color distortions, originating from high variation in the transmittance, making it harder for MBOIT to properly reconstruct the transmittance.
On the other hand, with increasing transparency and smoother transfer functions, the quality of MBOIT highly improves, and it is able to outperform MLABDB for such settings (compare green PSNR curves for each column in Fig.~\ref{fig:quality_appendix} with best PSNR values in the sixth row).
For high transparency, PSNR also indicates that merge heuristics in MLABDB perform worse and produce high noise in the image due to incorrect blending and color inconsistencies.
These errors are even more pronounced in close-up views where more pixels are affected by incorrect blending, yielding worst PSNR values for TURBULENCE throughout all techniques.


%
The quality of VRC is consistently below that of MBOIT and MLABDB, showing even more severe variations when zooming into the data sets. 
The reason lies in the line discretization used by VRC, which introduces inaccuracies especially along the line silhouettes, i.e., a line is either missed or erroneously hit. 
In both cases, the pixel value can be very different from the ground truth, decreasing the measured image quality. 
Even though fewer lines are seen when zooming into the data, the differences now affect more and more pixels, and thus increasingly affect the image quality. 
With higher line transparency, the line contours are also increasingly blurred out, rendering these artifacts less pronounced and resulting in higher SSIM and PSNR.
In the following, we show images of all rendering techniques using high transparency or semi-transparent rendering settings for ANEURYSM (cf. Fig.~\ref{fig:error_aneursym_appendix}), CONVECTION (Fig.~\ref{fig:error_convectionrolls_appendix}), TURBULENCE (Fig.~\ref{fig:error_turbulence_appendix}), and CLOUDS (Fig.~\ref{fig:error_clouds_appendix}).
The figures are composed of rendering results from Depth Peeling (ground truth) and each approximate technique (MLABDB, MBOIT, and VRC).

Per-pixel absolute error images are included to further highlight differences between each approximate technique and the ground truth image.
Pixel-wise errors are computed using the absolute difference between the ground truth and the rendering for each channel and converted to an inverted gray scale image afterwards.
Here, white color means no difference, and black represents highest possible error in the image.

Furthermore, we show the depth complexity of each view to depict the number of layers per pixel for each scene, where black represents 0 fragments and bright cyan is the maximum number of layers in the current view.
This is achieved by counting the number of total fragments per pixel using atomic operations on the GPU.
For our data set and a viewport of 1920$\times$1080, the total number of fragments varied from 30 million (CONVECTION) to 280 million (CLOUDS) fragments during animation.
The maximum number of layers ranged from $140$ (ANEURYSM, CONVECTION) to $1000$ (TURBULENCE) and even $9000$ (CLOUDS).

Furthermore, we compute error metrics between the rendered image of each approximate technique and the ground truth in scenes with opaque, semi-transparent, and highly transparent render settings for all four data sets used in the paper. 
The plots in Fig.~\ref{fig:quality_appendix} show SSIM (first three rows) and PSNR (last three rows) values of renderings over time produced during a pre-recorded flight with two zoom-ins.

\input{content/error_aneurysm_appendix}%
\input{content/error_convectionrolls_appendix}%
\input{content/error_turbulence_appendix}%
\input{content/error_clouds_appendix}%
\clearpage
\setcounter{figure}{0} 
\section{Surface and Point Data}
\label{app:isosurface_points}
Besides line rendering, we additionally demonstrate the application of the approximate techniques MLABDB and MBOIT to render transparent surface and point models.
Fig.~\ref{fig:error_isosurface_appendix} shows an isosurface in a Richtmyer-Meshkov instability\footnote{R. Cohen, W. Dannevik, A. Dimits, D. Eliason, A. Mirin, Y. Zhou, D. Porter, and P. Woodward, ``Three-dimensional simulation of a richtmyer-meshkov instability with a two-scale initial perturbation,'' Physics of Fluids, vol. 14, no. 10, pp. 3692–3709, 1 2002.} extracted with VTK and comprised of 516 million triangles. Constant transparency is used for all triangles.
Fig.~\ref{fig:error_points_appendix} shows images of 55 million particles from a coal particle combustion simulation in Uintah\footnote{M. Berzins, J. Luitjens, Q. Meng, T. Harman, C. A. Wight, and J. R.Peterson, ``Uintah: A scalable framework for hazard analysis,'' in Proceedings of the 2010 TeraGrid Conference, ser. TG ’10. New York, NY, USA: ACM, 2010, pp. 3:1–3:8} rendered with screen-oriented 2D splats. The velocity of each particle is mapped to transparency and color (from beige to red to blue, with blue being the highest).

In all figures, we show the rendering outcome of DP (ground truth), MLABDB, and MBOIT.
In addition, we highlight the differences between the approximation and ground truth with (a) detailed views below all rendering algorithms, (b) the number of fragments per pixel (depth complexity) to relate per-pixel errors to the complexity of the scene, and (c) per-pixel errors from the ground truth.

\input{content/error_isosurface_appendix}
For both data sets, MLABDB and MBOIT produce similar artifacts as for line sets, discussed in Appendix~\ref{app:quantitative} and Sec.~\ref{sec:error_cases}.
As seen in Fig.~\ref{fig:error_isosurface_appendix}, MBOIT leads to blur effects, especially transitions between neighboring isosurfaces vanish with higher transparency. 
MLABDB, on the other hand, is better at preserving small geometric details. 
For instance, it retains the sharpness of (small) surface contours in regions of high surface variation with multiple transparent layers.

However, for our point cloud data set with a myriad of transparent surfaces --- in our test more than $1000$ transparent layers --- MLABDB is not able to fully reconstruct the correct color due to accumulated errors caused by incorrect fragment merges, leading to blurry or wrongly colored features and misinterpretation of the data.
This effect is even more prominent when mapping attributes with more than two colors (mapping from beige to red to blue, cf. Fig.~\ref{fig:error_points_appendix}).
Note that the depth complexity and per-pixel error images also clearly demonstrate that approximation errors of MLABDB increase with more layers.
Since MLABDB is not order-independent, it also fails to preserve the correct visibility order of fragments, causing hidden features to suddenly appear in the final image (cf. yellow rectangles in Fig.~\ref{fig:error_points_appendix}).

MBOIT, in contrast, can properly handle transmittance and color in this scenario and, besides small approximation errors, is able to preserve the correct visual appearance of features in the data (compare the detailed views in Fig.~\ref{fig:error_points_appendix}).
In both detailed views, MBOIT is able to reconstruct the sharpness of red colored particles (highlighted in the purple view) and does not cause hidden interior features to incorrectly become visible on the particle combustion data set (yellow view).

In summary, these findings reflect the results observed when rendering large line sets such as TURBULENCE and CLOUDS, which can be taken from green and blue curves in PSNR and SSIM plots (for all transparent rendering settings) and per-data image comparisons in Appendix~\ref{app:quantitative}.

\input{content/error_points_appendix}

\end{document}